\documentclass[11pt,a4paper]{article}
\pdfoutput=1

\usepackage{jcappub}

\usepackage{bm}
\usepackage{amsfonts}
\usepackage{amsmath}
\usepackage{subfigure}
\usepackage{float}

\newcommand{\be}{\begin{equation}}
\newcommand{\ee}{\end{equation}}
\newcommand{\bea}{\begin{eqnarray}}
\newcommand{\eea}{\end{eqnarray}}

\begin{document}



\title{Fitting functions on the cheap: the relative nonlinear matter power spectrum}

\author[a]{Steen~Hannestad} 
\author[b]{and Yvonne Y.~Y.~Wong}

\affiliation[a]{Department of Physics and Astronomy, University of Aarhus, DK-8000 Aarhus C, Denmark}
\affiliation[b]{School of Physics, The University of New South Wales, Sydney NSW 2052, Australia}

\emailAdd{sth@phys.au.dk, yvonne.y.wong@unsw.edu.au}

\abstract{We propose an alternative approach to the construction of fitting functions to the nonlinear matter power spectrum extracted from $N$-body simulations  based on the relative matter power spectrum~$\delta(k,a)$, defined as the fractional deviation in the absolute matter power spectrum produced by a target cosmology away from a reference $\Lambda$CDM prediction.   From the computational perspective, $\delta(k,a)$ is fairly insensitive to the specifics of the simulation settings, and numerical convergence at the 1\%-level can be readily achieved without the need for huge computing capacity.  
Furthermore, with the $w$CDM class of models tested, $\delta(k,a)$ exhibits several interesting properties that enable a piece-wise construction of the full fitting function, whereby component fitting functions are sought for single-parameter variations and then multiplied together to form the final product.  Then, to obtain 1\%-accurate absolute power spectrum predictions for any target cosmology only requires that the community as a whole invests in producing {\it one single} ultra-precise reference $\Lambda$CDM absolute power spectrum, to be combined with the fitting function to produce the desired result. To illustrate the power of this approach, we have constructed the fitting function  {\sc RelFit}  using only  five relatively inexpensive $w$CDM simulations (box length $L=256 \, h^{-1}$Mpc,  $N=1024^3$ particles, initialised at $z_{i}=49$).  In a 6-parameter space spanning $\{\omega_{\rm m},A_{s},n_{s},w,\omega_{\rm b},h\}$, the output relative power spectra of {\sc RelFit} are consistent with the predictions of the {\sc CosmicEmu} emulator to 1\% or better for a wide range of cosmologies up to $k \simeq 10$/Mpc.
Thus, our approach could provide an inexpensive and democratically accessible route to fulfilling the 1\%-level accuracy demands of the upcoming generation of large-scale structure probes, especially in the exploration of ``non-standard'' or ``exotic'' cosmologies on nonlinear scales.  
}

\maketitle


\section{Introduction}

The upcoming generation of  large-scale structure surveys such as the ESA {\sc Euclid} mission~\cite{euclid} and the Large Synoptic Survey Telescope (LSST)~\cite{lsst} have the potential to measure cosmological observables at an unprecedented level of precision.  In terms of the matter power spectrum, the measurement uncertainty is expected to be at the 1\% level down to length scales corresponding  to wavenumbers $k \sim O(5) \, h$/Mpc. Such high precisions in turn put heavy demands on theoretical calculations of the observables.

On large scales where perturbations are expected to remain well below $O(1)$, linear perturbation theory can easily satisfy the 1\% precision requirement.  Likewise, perturbative methods can be extended to higher orders on weakly nonlinear scales ($k \sim 0.05 \to 0.1 \, h$/Mpc at scale factor $a=1$),  and much effort has been devoted recently towards improving the convergence  of these computations (see, e.g.,~\cite{Bernardeau:2013oda}).  Calculations in the fully nonlinear scales, i.e,. $k \gtrsim O(0.1) \, h$/Mpc at $a= 1$, however,  belong in the domain of  numerical simulations.

However, simulations are inherently computationally expensive, and it is currently not economical to run full simulations for more than a select $O(10 \to 100)$ parameter combinations ``representative'' of a large cosmological parameter space.  In fact, achieving the required 1\%~precision for even one single set of cosmological parameters is a computational challenge that necessitates the use of some of the largest computing facilities in the world~\cite{Heitmann:2008eq,Schneider:2015yka}.  As an example, each cosmology in the Mira--Titan suite of $w$CDM simulations is realised by two high-resolution simulations with 30 billion+ and 60 billion+ particles each, plus 16 lower-resolution 100-million-particle runs~\cite{Heitmann:2015xma,Lawrence:2017ost}. Only a select few researchers in the world have access to the requisite computing power to carry out such calculations {\it en masse}.

Currently, in order to explore large parameter spaces with parameter combinations running into $O(10^5)$---as is required in a typical Markov Chain Monte Carlo parameter estimation analysis---the favoured approach is to  employ fitting functions such as {\sc Halofit}~\cite{halofit,Takahashi:2012em} or  {\sc HMCode}~\cite{Mead:2015yca,hmcode2016} that have been calibrated against simulation results.  Alternatively, one can interpolate between a set of simulations spanning the parameter spaces of interest, such as the emulator approach of~\cite{Heitmann:2013bra,Heitmann:2008eq, Heitmann:2009cu,Lawrence:2009uk}.  
However, as the accuracy of any fitting or interpolation function is contingent upon  there being sufficient calibrators to fairly sample the parameter space {\it and} the calibrating simulations {\it themselves} having the required level of precision, the burden is again back on the simulations {\it and} the  same select few research groups that have the computing resources to supply these calculations.
Such a strong reliance on computing resources clearly poses severe limitations on the participation of the wider scientific community, especially in the  exploration of  ``exotic'' cosmologies such as decaying or interacting dark matter (e.g.,~\cite{Dakin:2019dxu,Diacoumis:2018ezi}), or dark energy perturbations~(e.g., \cite{Dakin:2019vnj})  on nonlinear scales.

In this paper  we put forward a different approach to constructing fitting functions to the nonlinear  matter power spectrum that will alleviate to a large extent the precision burden on the calibrating simulations and potentially democratise the exploration of precision cosmology on nonlinear scales: Instead of the usual practice of fitting or interpolating directly the  {\it absolute} simulated matter power spectrum $P({\bf \Theta}; k;a)$ for a select few cosmological parameter combinations ${\bf \Theta}$, we propose to construct a fitting function to a set of spectra $\delta({\bf \Theta}; {\bf \Theta}_0;k;a)$, defined as 
\begin{equation}
\label{eq:relative}
\delta({\bf \Theta}; {\bf \Theta}_0; k;a)  \equiv  \frac{P({\bf \Theta}; k;a)-P({\bf \Theta}_0; k;a)}{P({\bf \Theta}_0; k;a)}
\end{equation}
{\it relative} to the absolute matter power spectrum of a reference cosmological model, $P({\bf \Theta}_0; k;a)$.  
As we shall demonstrate, there are a number of reasons why  fitting the relative power spectra may be superior to fitting their absolute counterparts:
\begin{enumerate}
\item From the computational perspective, relative power spectra can be calculated much more precisely than absolute power spectra from $N$-body simulations using the same box size and number of particles~\cite{McDonald:2005gz}.    This is because many systematic uncertainties are multiplicative and affect all simulations in the same way; taking the ratio of two simulation results therefore enable these uncertainties to cancel to a large extent.  Indeed, the use of ratios to ``get around'' systematic uncertainties that may not be completely well understood  is a well-known technique used in many areas of physics, e.g., collider phenomenology~\cite{Aad:2014rta}, precision cosmology~\cite{Song:2008qt}, and neutrino physics~\cite{Villante:1998pe,Ren:2017xov}.

An immediate corollary of this observation is that a nominal accuracy goal can be achieved at much a lower computational cost using relative power spectrum simulations than  their absolute counterparts.
Once a fitting function to $\delta = \delta({\bf \Theta}; {\bf \Theta}_0;k;a)$  is available as a function of the underlying cosmology,  to obtain  an accurate estimation of a target $P({\bf \Theta}; k;a)$  for {\it any} parameter combination requires only that we perform {\it one single} ultra-high precision simulation of the reference cosmological model to establish~$P({\bf \Theta}_0; k;a)$ and then combine this result with the fitting function.  In this regard, our proposal parallels  the ``halo model reaction'' approach of~\cite{Cataneo:2018cic,Giblin:2019iit} and the {\sc CosmicEFT} approach of~\cite{Cataneo:2016suz}, wherein the equivalent of $\delta$ is computed using semi-analytical methods such as the halo model and effective field theory.

\item The present generation of linear cosmological probes, e.g., measurements of the cosmic microwave background (CMB) temperature and polarisation anisotropies by the Planck mission~\cite{Ade:2015xua,Aghanim:2018eyx}, already constrains cosmology to the extent that variations in the absolute power spectra are typically $\lesssim 10\%$.  This means that any fitting function to $\delta$ need only be calibrated to at most $\sim 10$\%-precision in order to reproduce a target  $P({\bf \Theta}; k;a)$  with $\lesssim 1$\%-level error  (assuming, of course, that an ultra-precise reference $P({\bf \Theta}_0; k;a)$   is available), and the smaller $\delta$ is the laxer the calibration precision requirement. This is a  trivial demand in comparison with the 1\%~calibration precision required of fitting functions designed to {\it directly} reproduce~$P({\bf \Theta}; k;a)$.  

\item  Since typically $|\delta| \lesssim 0.1$, it is strongly suggestive that the relative matter power spectrum~$\delta$ may be computable perturbatively from similarly small deviations in the {\it linear} power spectrum away from the reference cosmology.  Indeed, we find that~$\delta$ can be related to relative changes in, e.g., the linear growth function, the primordial power spectrum, etc., in a remarkably cosmology-independent way.  This attractive feature enables a multiplicative construction of the full fitting function, whereby component fitting functions are sought for variations of cosmological model parameters (or their proxies such as the linear growth function) one at a time and the full fitting function pasted together via a simple multiplication of the components.

\end{enumerate}

\begin{table*}[t]
	\begin{center}
		{\footnotesize
			\hspace*{0.0cm}\begin{tabular}
				{lcc} \hline \hline
				Parameter & Symbol & Value \\ \hline
				Total physical matter density & $\omega_{\rm m}$ & 0.1422 \\ 
				Physical baryon density &	$\omega_{\rm b}$ & 0.0221 \\
				Physical neutrino density & $\omega_{\nu}$ & 0 \\
				Spatial curvature &	$\Omega_{\rm k}$ & 0 \\
				Effective number of neutrinos & $N_{\rm eff}$ & 3.04 \\
				Dark energy equation of state parameter &	$w$ & $-1$ \\
				Dimensionless Hubble parameter &	$h$ & 0.673  \\
				Primordial scalar fluctuation amplitude at $k_{\rm piv}=0.05$/Mpc &	$10^9 A_s$ & $2.198$ \\
				Scalar spectral index & $n_s$ & 0.96 \\
				Running of the scalar spectral index &	$n_{\rm run}$ & 0 \\
				Tensor-to-scalar ratio &	$r$ & 0 \\
				Optical depth to reionisation &	$\tau$ & 0.09 \\
				\hline \hline
			\end{tabular}
		}
	\end{center}
	\caption{Cosmological parameter values of the reference $\Lambda$CDM model.\label{tab:params}}
\end{table*}

The paper is organised as follows.  We begin in section~\ref{sec:convergence} with a discussion of the convergence of the absolute and the relative matter power spectrum, using cosmologies with a non-canonical dark energy equation of state  parameter as an example. Section~\ref{sec:prelim} examines the properties of the relative power spectrum under single- and multi-parameter variations, through which we motivate a strategy for the construction of a fitting function for~$\delta$. We propose specific functional forms for the fitting function  in section~\ref{sec:fitting} , which we then calibrate against $N$-body simulations to produce  {\sc RelFit}.  Comparisons of the predictions of {\sc RelFit} and other approaches are  presented in the same section.
Section~\ref{sec:conc} contains our conclusions.   Throughout the work we use as our reference cosmology~${\bf \Theta}_0$ a $\Lambda$CDM model with parameter values given in table~\ref{tab:params},
roughly comparable to the  best-fit  of the Planck 2015 CMB data~\cite{Ade:2015xua}.  Where confusion is unlikely to arise, we shall sometimes omit writing out the dependences of the absolute and relative matter power spectra on $k$ and/or $a$.  


\section{Numerical convergence of the absolute and the relative spectrum}
\label{sec:convergence}

Many factors may influence the numerical convergence of a simulation result.  Chief amongst these are the simulation box size and the number of particles employed to sample the cosmological fluid (i.e., cold dark matter in a $\Lambda$CDM-type cosmology) phase space.  Other important factors include the redshift at which a simulation is initialised, and the gravitational softening length adopted in the simulation to prevent spurious relaxation.
In this section we examine the extent to which numerical convergence of the absolute and  relative power spectra depends on these factors, using a series of $N$-body simulations performed with the {\sc Gadget-2} code~\cite{Springel:2005mi}.  The specifics of each simulation  are summarised in table~\ref{tab:runs1}.

\begin{table*}[t]
	\begin{center}
		{\footnotesize
			\hspace*{0.0cm}\begin{tabular}
				{lcccccccc} \hline \hline
				Run & $L (h^{-1}$Mpc) & $N$ & $z_i$ & $r_s(h^{-1}$kpc) & $\omega_{\rm m}$ & $10^9 A_s$ & $n_s$ & $w$ \\       \hline
				{\tt Ref} & 320 & $1282^3$ & 49 & 6 & 0.1422 & 2.198 & 0.96 & $-1.00$ \\
				{\tt Ref$w2$} & 320 & $1282^3$ &  49 & 6 & 0.1422 & 2.198 & 0.96 & $-0.85$   \\
				{\tt Ref2} & 960 & $1282^3$ & 49& 6  & 0.1422 & 2.198 & 0.96 & $-1.00$ \\
				{\tt Ref2$w2$} & 960 & $1282^3$ &  49 & 6 & 0.1422 & 2.198 & 0.96 & $-0.85$   \\
				{\tt Ref3} & 1920 & $1282^3$ & 49 & 6 & 0.1422 & 2.198 & 0.96 & $-1.00$ \\
				{\tt Ref3$w2$} & 1920 & $1282^3$ & 49 & 6 & 0.1422 & 2.198 & 0.96 & $-0.85$ \\ \hline
				{\tt 1024Ref-512} & 512 & $1024^3$ &  49 & 6 & 0.1422 & 2.198 & 0.96 & $-1.00$  \\
				{\tt 1024$w2$-512} & 512 & $1024^3$ &  49 & 6  & 0.1422 & 2.198 & 0.96 & $-0.85$  \\
				{\tt 1024Ref} & 256 & $1024^3$ &  49 & 6 & 0.1422 & 2.198 & 0.96 & $-1.00$  \\
				{\tt 1024$w2$} & 256 & $1024^3$ &  49 & 6 & 0.1422 & 2.198 & 0.96 & $-0.85$  \\
				{\tt 1024Ref-128} & 128 & $1024^3$ &  49 & 6 & 0.1422 & 2.198 & 0.96 & $-1.00$ \\
				{\tt 1024$w2$-128} & 128 & $1024^3$ &  49 & 6 & 0.1422 & 2.198 & 0.96 & $-0.85$ \\
				{\tt 768Ref-512} & 512 & $768^3$ &  49 & 6 & 0.1422 & 2.198 & 0.96 & $-1.00$  \\
				{\tt 768$w2$-512} & 512 & $768^3$ &  49 & 6 & 0.1422 & 2.198 & 0.96 & $-0.85$  \\
				{\tt 768Ref-256} & 256 & $768^3$ &  49 & 6 & 0.1422 & 2.198 & 0.96 & $-1.00$ \\
				{\tt 768$w2$-256} & 256 & $768^3$ &  49 & 6 & 0.1422 & 2.198 & 0.96 & $-0.85$  \\
				{\tt 768Ref-128} & 128 & $768^3$ &  49 & 6 & 0.1422 & 2.198 & 0.96 & $-1.00$  \\
				{\tt 768$w2$-128} & 128 & $768^3$ &  49 & 6 & 0.1422 & 2.198 & 0.96 & $-0.85$  \\
				{\tt 512Ref-512} & 512 & $512^3$ &  49 & 6 & 0.1422 & 2.198 & 0.96 & $-1.00$  \\
				{\tt 512$w2$-512} & 512 & $512^3$ &  49 & 6 & 0.1422 & 2.198 & 0.96 & $-0.85$  \\
				{\tt 512Ref-256} & 256 & $512^3$ &  49 & 6 & 0.1422 & 2.198 & 0.96 & $-1.00$ \\
				{\tt 512$w2$-256} & 256 & $512^3$ &  49 & 6 & 0.1422 & 2.198 & 0.96 & $-0.85$  \\
				{\tt 512Ref-128} & 128 & $512^3$ &  49 & 6 & 0.1422 & 2.198 & 0.96 & $-1.00$  \\
				{\tt 512$w2$-128} & 128 & $512^3$ &  49 & 6 & 0.1422 & 2.198 & 0.96 & $-0.85$  \\   \hline
				{\tt 1024Ref-256-$z_i$29} & 256 & $1024^3$ &  29 & 6 & 0.1422 & 2.198 & 0.96 & $-1.00$  \\
				{\tt 1024$w2$-256-$z_i$29} & 256 & $1024^3$ &  29 & 6 & 0.1422 & 2.198 & 0.96 & $-0.85$  \\	
				{\tt 1024Ref-256-$r_s$12} & 256 & $1024^3$ &  49 & 12 & 0.1422 & 2.198 & 0.96 & $-1.00$  \\
				{\tt 1024$w2$-256-$r_s$12} & 256 & $1024^3$ &  49 & 12 & 0.1422 & 2.198 & 0.96 & $-0.85$  \\	
				\hline
				{\tt pkdgravRef} & 384 & $1024^3$ & 49 & 6 & 0.1422 & 2.198 & 0.96 & $-1.00$ \\
					{\tt pkdgrav$w2$} & 384 & $1024^3$ & 49 & 6 & 0.1422 & 2.198 & 0.96 & $-0.85$ \\
				\hline \hline
			\end{tabular}
		}
	\end{center}
	\caption{Simulations used in section~\ref{sec:convergence}: $L$ is the simulation box length, $N$ the number of simulation particles, $z_i$ the initial redshift, $r_s$ the gravitational softening length, and $\{\omega_{\rm m}, A_s, n_s, w\}$ are cosmological model parameters described in table~\ref{tab:params}.  All except the last two simulations have been performed using {\sc Gadget}-2/{\sc Camb}; the last two are outputs of {\sc Pkdgrav3}/{\sc Class}.~\label{tab:runs1}}
\end{table*}

For each simulation we employ initial conditions generated via the Zel'dovich approximation from linear transfer functions outputted by {\sc Camb}~\cite{Lewis:1999bs}. We include baryons in the computation of the linear transfer function required for initial condition generation, but do not distinguish baryons from cold dark matter in the actual simulations. The latter is certainly an oversimplification in precision calculations of an {\it absolute} power spectrum, but can be expected to be a reasonable approximation in the case of a  {\it relative} power spectrum.


\subsection{Box size and  number of particles}

It is well known that numerical convergence of the absolute matter power spectrum requires simulations in large boxes with many particles. If the box size is too small sample (cosmic) variance becomes a serious issue.  Increasing the box size however  requires that we also up the number of particles in order to suppress shot noise on small scales. These issues have been discussed in detail in, e.g., a series of papers related to the {\sc Coyote} simulations (e.g.,~\cite{Heitmann:2008eq}) and more recently in~\cite{Schneider:2015yka}.  The general conclusion is that to achieve an absolute power spectrum calculation at the 1\% level of precision requires box lengths exceeding $L=500 \, h^{-1}$Mpc and particle numbers of order $N=10000^3$.

\begin{figure}[t]
\begin{center}
\includegraphics[width=15.4cm]{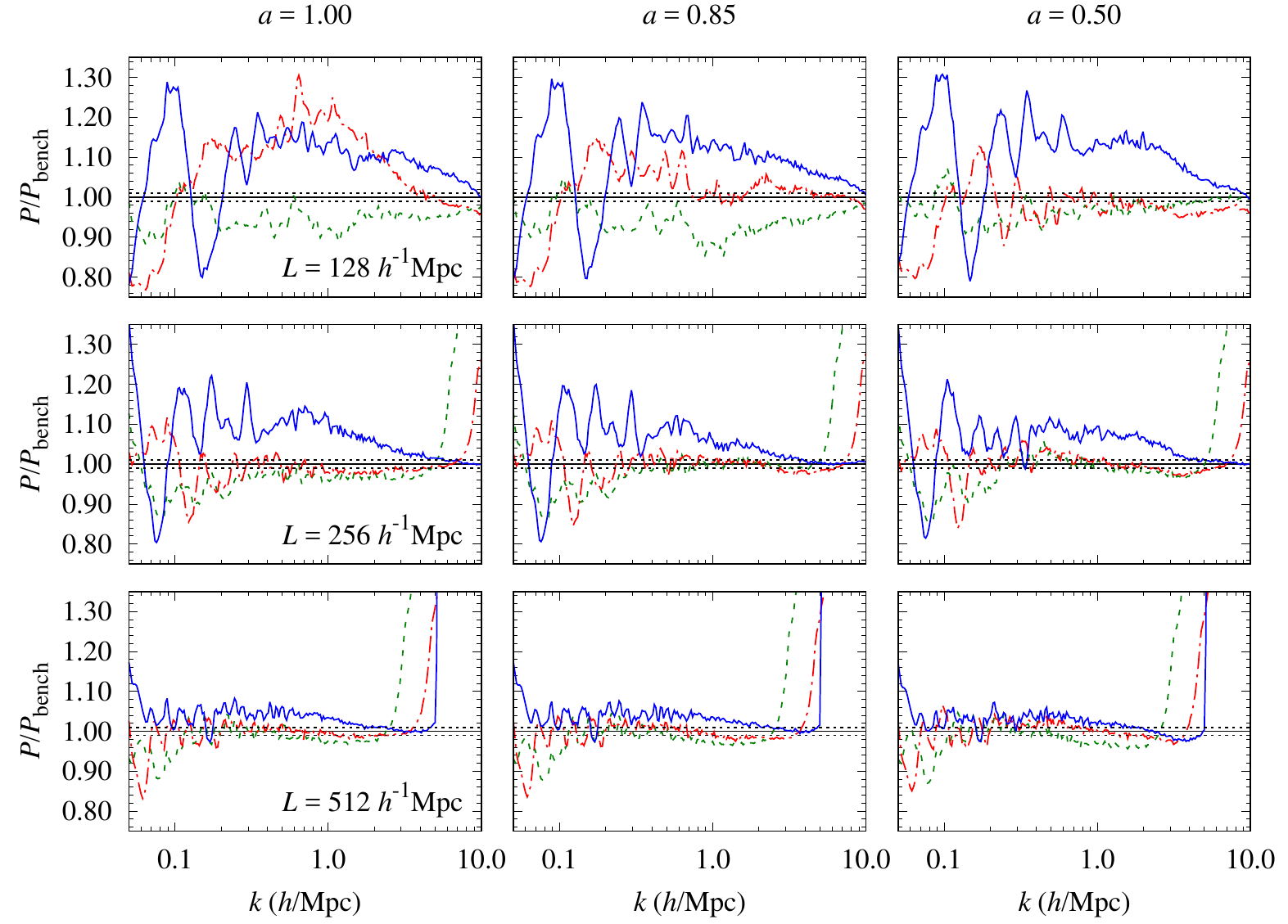}
\end{center}
\caption{Absolute matter power spectrum at $a=1.00,0.85,0.50$ for the reference $\Lambda$CDM model (parameter values in table~\ref{tab:params}), computed using different box sizes and particle numbers. The green/dashed, red/dot-dash, and blue/solid lines represent respectively $N=512^3, 768^3, 1024^3$.  All spectra have been normalised to a benchmark matter power spectrum $P_{\rm bench}$  constructed from the  {\tt Ref}, {\tt Ref2}, and {\tt Ref3} simulations (see table~\ref{tab:runs1}). 
\label{fig:abs}}
\end{figure}

\begin{figure}[t]
	\begin{center}
		\includegraphics[width=15.4cm]{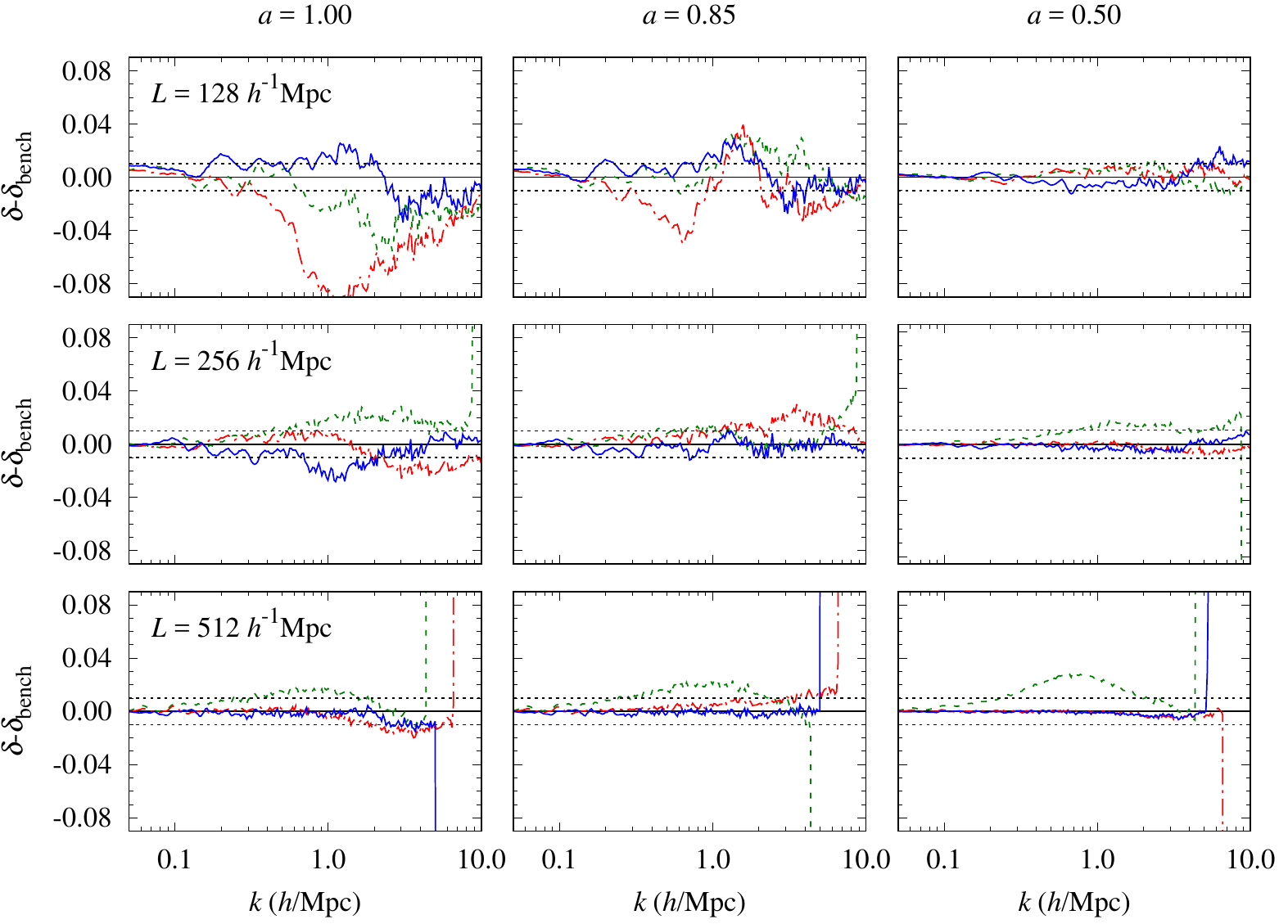}
	\end{center}
	\caption{Relative matter power spectrum at $a=1.00, 0.85,0.50$  between two cosmologies with $w=-0.85$ and $w = -1$ (all other parameters fixed at the values in table~\ref{tab:params}), computed using 
		different  box sizes and particle numbers. The green/dashed, red/dot-dash, and blue/solid lines represent respectively $N=512^3, 768^3, 1024^3$, and we have subtracted away the benchmark $\delta_{\rm bench}$ constructed from the {\tt Ref$w$2}, {\tt Ref2$w$2}, {\tt Ref3$w$2}, {\tt Ref}, {\tt Ref2}, and {\tt Ref3} simulations (see table~\ref{tab:runs1}).\label{fig:rel}}
\end{figure}

To illustrate the {\it lack of convergence} of the absolute matter power spectrum $P({\bf \Theta}; k;a)$, we show in figure~\ref{fig:abs} $P({\bf \Theta}_0; k;a)$
constructed from various reference $\Lambda$CDM simulations using different box sizes and particle numbers (but keeping for now the initialisation redshift and softening length fixed at $z_i=49$ and  $r_s = 6\, h^{-1}$kpc respectively) summarised in table~\ref{tab:runs1}. These are normalised to a benchmark power spectrum $P_{\rm bench}({\bf \Theta}_0; k;a)$,
constructed from amalgamating the power spectra extracted from three ``high-quality'' runs---{\tt Ref3} at $k < 1 \, h$/Mpc, {\tt Ref2} in the range $1 \, h$/Mpc$ < k < 3 \, h$/Mpc, and {\tt Ref} at $k > 3 \, h$/Mpc.  Clearly, no single simulation is able to converge to the benchmark at better than 10\% across the entire $k$-range, and convergence worsens as the scale factor $a$ approaches unity.

In contrast, the relative change in the matter power spectrum between two cosmologies with different parameter values, $\delta = \delta({\bf \Theta},{\bf \Theta}_0;k;a)$ as defined in equation~(\ref{eq:relative}), is much less susceptible to  sample variance,  {\it provided} the two simulations used to construct~$\delta$ have been run under identical conditions  and initialised  with {\it identical phases} in the density field.  We emphasise that these requirements of identical simulations settings are crucial, as it is precisely this sameness that ensures two simulations suffer largely the same systematic effects that eventually cancel out when forming a ratio, leaving a $\delta$ that is ultimately relatively insensitive to the simulation settings.
A similar observation has also been made in~\cite{McDonald:2005gz}.

This relative insensitivity to the simulation settings also means that numerical  convergence in~$\delta$ can be achieved using much smaller boxes and hence smaller numbers of simulation particles than in the case of the absolute power spectrum.  
Figure~\ref{fig:rel} illustrates this point by way of the relative change in power $\delta({\bf \Theta},{\bf \Theta}_0;k;a)$ between two cosmological models specified respectively by the parameter values
\begin{equation}
\begin{aligned}
\label{eq:cosmologies}
{\bf \Theta} &= \{\theta_w=\bar{\theta}_{w}; w=-0.85 \}, \\
{\bf \Theta}_0 &= \{\theta_w= \bar{\theta}_{w}; w = -1 \},
\end{aligned}
\end{equation}
where $w$ denotes the dark energy equation of state parameter, and $\theta_w=\bar{\theta}_{w}$ stipulates that all other model parameters {\it besides} $w$ are to be held fixed at their reference values $\bar{\theta}_{w}$ given in table~\ref{tab:params}.  As in figure~\ref{fig:abs}, the relative power spectra here have been constructed from the simulations of table~\ref{tab:runs1} using different combinations of box sizes and particle numbers,  and for clarity we have subtracted away the benchmark relative power spectrum~$\delta_{\rm bench}$ constructed from the ``high-quality'' {\tt Ref$w$2}, {\tt Ref2$w$2}, {\tt Ref3$w$2}, {\tt Ref}, {\tt Ref2}, and {\tt Ref3} runs of  table~\ref{tab:runs1}.

\begin{figure}[t]
	\begin{center}
		\includegraphics[width=11cm]{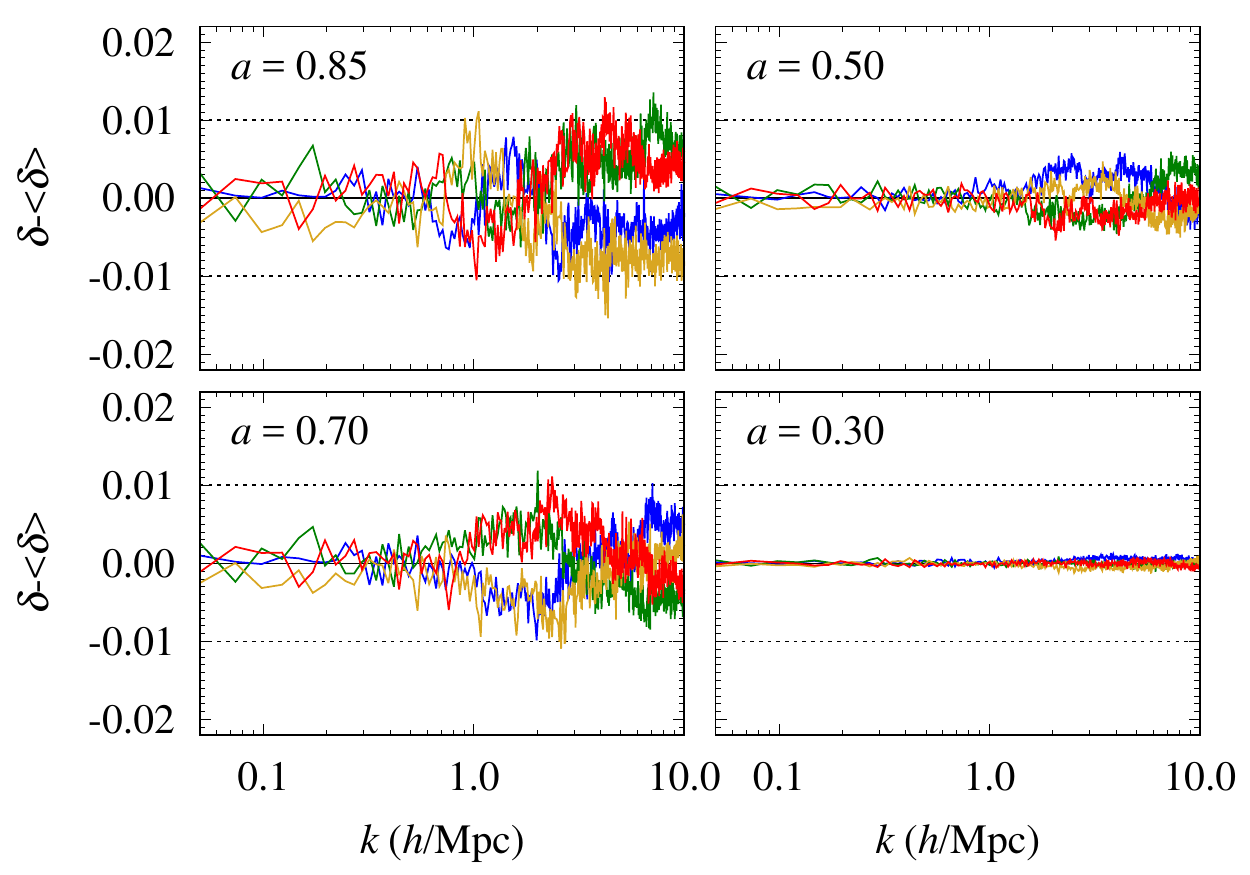}
	\end{center}
	\caption{Deviations of four realisations (i.e., initialised with four different sets of phases) of the relative matter power spectrum between two cosmologies with $w=-0.85$ and $w = -1$ from the ensemble average~$\langle \delta \rangle$.  All simulations used  $N=1024^3$ particles in a box of side length $L=256 \, h^{-1}$Mpc.
		\label{fig:seeds}}
\end{figure}

Clearly, independently of the number of simulation particles employed, sample variance dominates when the box size is too small, but becomes manageable once  the box length reaches $L=256 \, h^{-1}$Mpc.   In terms of particle numbers, we find $N = 1024^3$ to be sufficient to eliminate to a large extent shot noise in boxes of side length $L  \geq 256 \, h^{-1}$Mpc, enabling numerical convergence at the $0.01$ level down to wavenumbers close to the Nyquist frequency at $a=0.85$ and better than $0.005$ at $a=0.50$; even at $a=1$, convergence at the (not unacceptable) 0.02 level is possible for a large range of wavenumbers.
Importantly, these conclusions are independent of the choice of initial phases, as demonstrated in figure~\ref{fig:seeds}, where we have re-simulated the relative power spectrum of the two cosmologies of equation~(\ref{eq:cosmologies}) using four different sets of initials seeds for the setting $L=256 \, h^{-1}$Mpc and  $N = 1024^3$, and plotted their deviations from the ensemble average~$\langle \delta \rangle$.

Note that the alternative choice of $L = 512 \,h^{-1}$Mpc and $N = 1024^3$ could even enable the attainment of 0.01 numerical convergence  at $a = 1$, as shown in figure~\ref{fig:rel}. The downside, however, is that such a setting yields power spectrum predictions only up to $k = 5 \,h$/Mpc, and to achieve a better resolution in $L = 512 \, h^{-1}$Mpc boxes would require a computing capacity beyond our current means.  Henceforth, we shall adopt the setting   $L=256 \, h^{-1}$Mpc and $N=1024^3$,  a fair compromise between computing power and the accuracy demands of future large-scale structure probes,%
\footnote{A scale factor $a=0.85$ corresponds to a redshift $z=0.176$, reasonably low relative to the median redshift  $z_{\rm m}=0.8 \to 0.9$  of the {\sc Euclid} and LSST galaxy redshift surveys~\cite{euclid,lsst}.  Similarly, while cosmic shear is in principle sensitive to the matter distribution at $z=0$, in practice the lensing weights are dominated by structures at roughly half the source-to-observer comoving distance;  for a shear tomographic bin at  $z=0.5\to 1.0$, such as used in the {\sc Euclid} parameter sensitivity forecast~\cite{euclid}, the weight peaks at  $z\sim 0.3$.  This encourages us to think that 0.01 numerical convergence of the matter power spectrum down to $z=0.176$ may suffice.}
and restrict our attention to  $a \leq 0.85$.


\subsection{Initial redshift and gravitational softening}

\begin{figure}[t]
	\begin{center}
		\includegraphics[width=11.5cm]{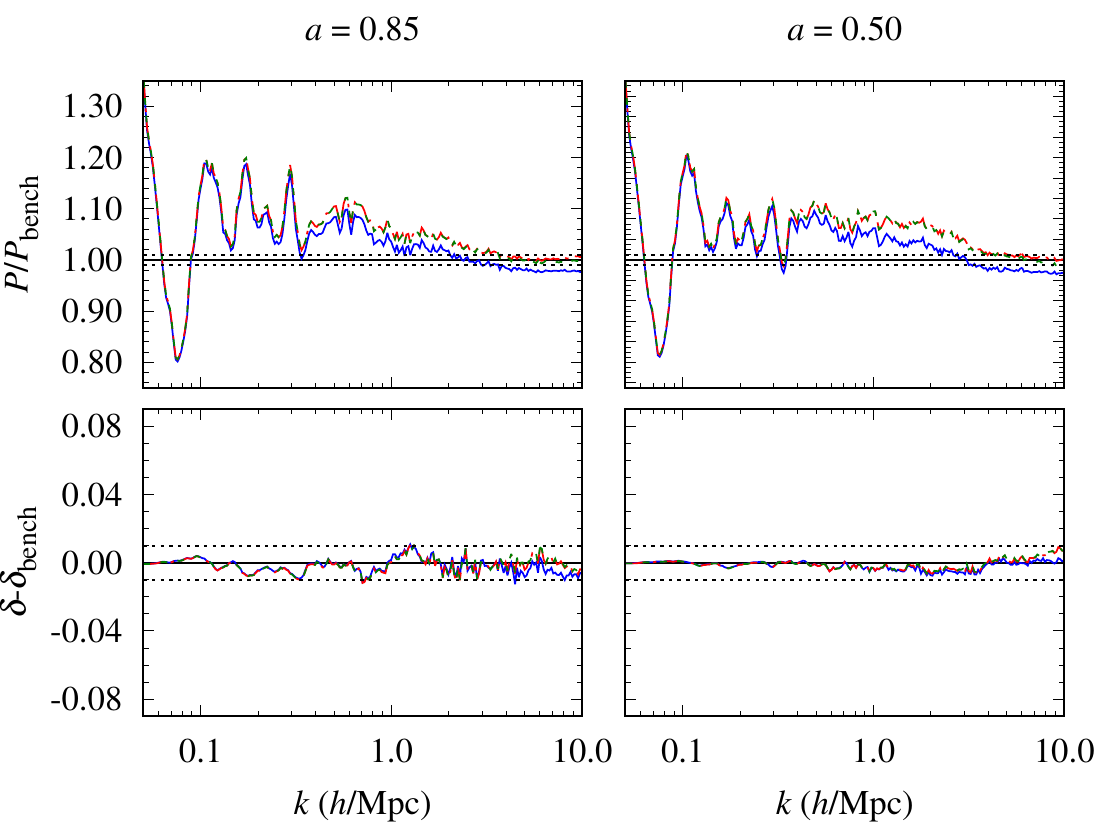}
	\end{center}
	\caption{Absolute matter power spectrum of the reference $\Lambda$CDM model (top) and relative matter power spectrum between the $w=-0.85$ and the $w=-1$ cosmologies (bottom) at $a=0.85, 0.50$, computed using different initial redshifts $z_i$ and gravitational softening length~$r_s$.  The red/dot-dash lines represent the default choice of $z_i=49$ and  $r_s = 6 \, h^{-1}$kpc, the blue/solid lines denote variation from the default to $z_i=29$, and the green/dashed lines variation to $r_s=12 \, h^{-1}$kpc.  All spectra have been normalised to the benchmark $P_{\rm bench}$ or $\delta_{\rm bench}$.\label{fig:zini}}
\end{figure}

We consider also the sensitivity of the absolute and relative matter power spectrum to the simulation initial redshift $z_i$ and gravitational softening length $r_s$, and vary these simulation parameters from the default $z_i=49$ and $r_s = 6 \, h^{-1}$kpc to $z_i=29$ and $r_s = 12 \, h^{-1}$kpc respectively.  The results at $a=0.85,0.50$ are shown in figure~\ref{fig:zini}.

Evidently, changing the gravitational softening length has no discernible effect on the relative power spectrum at either $a=0.85$ or $a=0.50$, and alters the absolute power spectrum only at the percent level at $k=10\, h$/Mpc.  On the other hand, with initial conditions set by the Zel'dovich approximation, both initialisation redshifts tested are clearly too low to achieve reasonable accuracy for the absolute power spectrum because of long-lived transients (although the problem of transients can be avoided by adopting 2LPT initial conditions~\cite{Crocce:2006ve}).  The relative power spectrum, however, appears to be largely insensitive to  $z_i$ within the 0.01 accuracy requirement.

Of course the case of varying  {\it only} $w$ away from its reference $\Lambda$CDM value is particularly benevolent in the sense that even for $w=-0.85$ the evolution history of the density perturbations at $z \gg 1$ is essentially identical to the $w=-1$ case.  This means that any transient excited as a result of the initialisation procedure must be identical in both cases, and cancel out exactly when we form the relative power spectrum.


\subsection{Code comparison}

 Lastly, we compare the stability of the relative matter power spectrum predictions  with respect to the numerical codes used to generate the results.  This we achieve by performing a new set of simulations using the {\sc Pkdgrav3} $N$-body code~\cite{Potter:2016ttn} with initial conditions generated from the linear transfer function output of {\sc Class}~\cite{Blas:2011rf} using the method outlined in \cite{Dakin:2017idt}. (Recall that our main suite of simulation results have been generated with {\sc Gadget}-2 and {\sc Camb}.)  The simulation specifics are summarised in the last two lines of table~\ref{tab:runs1}.  

\begin{figure}[t]
	\begin{center}
		\includegraphics[width=11.5cm]{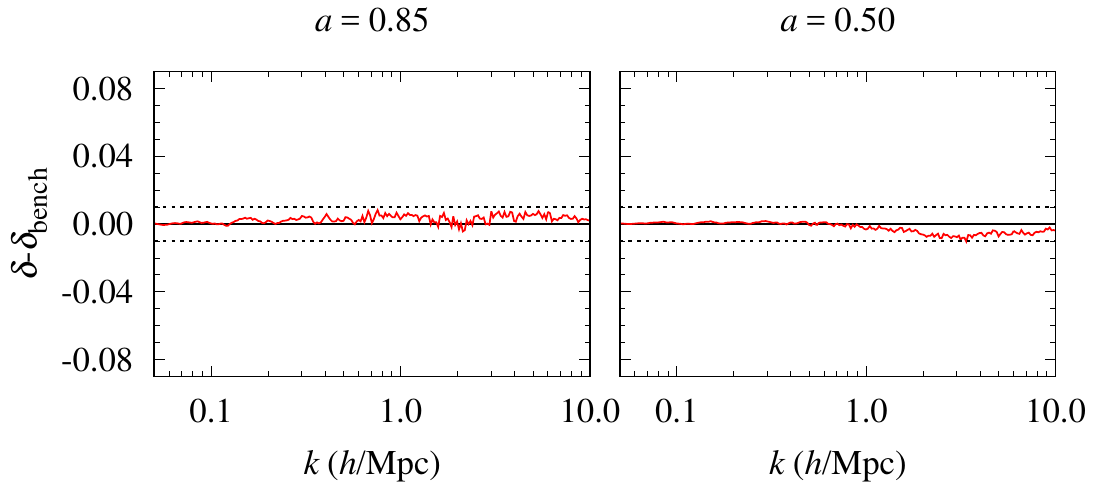}
	\end{center}
	\caption{Relative matter power spectrum between the $w=-0.85$ and the $w=-1$ cosmologies at $a=0.85, 0.50$, computed using {\sc Pkdgrav3} initialised with the linear transfer function outputs of {\sc Class}.  All spectra have been normalised to the  benchmark~$\delta_{\rm bench}$ computed using {\sc Gadget}-2/{\sc Camb}.~\label{fig:code}}
\end{figure}

Figure~\ref{fig:code} shows the relative matter power spectrum between the $w=-0.85$ and the $w=-1$ cosmologies at $a=0.85,0.50$
computed in this manner relative to the {\sc Gadget}-2/{\sc Camb} benchmark~$\delta_{\rm bench}$.  Clearly, the discrepancy between the two different code outputs is well within our 0.01 accuracy tolerance and comparable to the deviations one would expect from initialising the simulations with different sets of random phases (see figure~\ref{fig:seeds}).  We therefore conclude that the relative power spectrum is largely insensitive to $N$-body code from which it is generated or to the linear Boltzmann solver that provides the initial conditions.


\section{Properties of the relative power spectrum}
\label{sec:prelim}

Having established the advantage of the relative matter  power spectrum $\delta$ over its absolute counterpart in terms of numerical convergence, we now examine its properties more closely, in order to devise a fitting strategy and eventually a functional form that can directly fit $\delta$.

\begin{table*}[t]
\begin{center}
{\footnotesize
  \hspace*{0.0cm}\begin{tabular}
  {lccccccccc} \hline \hline
  Run & $L (h^{-1}$Mpc) & $N$ & $z_i$ & $r_s(h^{-1}$kpc)  & $\omega_{\rm m}$ & $10^9 A_s$ & $n_s$ & $w$ &Cal  \\       \hline
	{\tt 1024Ref} & 256 & $1024^3$ &  49 & 6 & 0.1422 & 2.198 & 0.96 & $-1.00$& *  \\
	{\tt 1024$w1$} & 256 & $1024^3$ &  49  & 6 & 0.1422 & 2.198 & 0.96 & $-0.92$ &   \\
	{\tt 1024$w2$} & 256 & $1024^3$ &  49  & 6 & 0.1422 & 2.198 & 0.96 & $-0.85$ & * \\
	{\tt 1024$w3$} & 256 & $1024^3$ &  49  & 6 & 0.1422 & 2.198 & 0.96 & $-0.75$  &\\
	{\tt 1024$w4$} & 256 & $1024^3$ &  49  & 6 & 0.1422 & 2.198 & 0.96 & $-1.15$ & * \\
	{\tt 1024$\omega_{{\rm m},l}$} & 256 & $1024^3$ &  49  & 6 & 0.1381 & 2.198 & 0.96 & $-1.00$ & * \\
	{\tt 1024$\omega_{{\rm m},h}$} & 256 & $1024^3$ &  49  & 6 & 0.1361 & 2.198 & 0.96 & $-1.00$  & * \\
	{\tt 1024$\omega_{{\rm m},l}w1$} & 256 & $1024^3$ &  49  & 6 & 0.1381 & 2.198 & 0.96 & $-0.92$  &\\
	{\tt 1024$\omega_{{\rm m},h}w1$} & 256 & $1024^3$ &  49  & 6 & 0.1461 & 2.198 & 0.96 & $-0.92$  & \\
	{\tt 1024$\omega_{{\rm m},l}w2$} & 256 & $1024^3$ &  49  & 6 & 0.1381 & 2.198 & 0.96 & $-0.85$  & \\
	{\tt 1024$\omega_{{\rm m},h}w2$} & 256 & $1024^3$ &  49  & 6 & 0.1461 & 2.198 & 0.96 & $-0.85$  & \\
	{\tt 1024$n_{s,l}$} & 256 & $1024^3$ &  49  & 6 & 0.1422 & 2.198 & 0.93 & $-1.00$ & *  \\
	{\tt 1024$n_{s,h}$} & 256 & $1024^3$ &  49  & 6 & 0.1422 & 2.198 & 0.98 & $-1.00$ & * \\
	{\tt 1024$n_{s,l}w2$} & 256 & $1024^3$ &  49  & 6 & 0.1422 & 2.198 & 0.93 & $-0.85$ & \\
	{\tt 1024$n_{s,h}w2$} & 256 & $1024^3$ &  49  & 6 & 0.1422 & 2.198 & 0.98 & $-0.85$  & \\
	{\tt 1024$A_{s,l}$} & 256 & $1024^3$ &  49  & 6 & 0.1422 & 2.100 & 0.96 & $-1.00$ & * \\
	{\tt 1024$A_{s,h}$} & 256 & $1024^3$ &  49 & 6  & 0.1422 & 2.300 & 0.96 & $-1.00$ & * \\
	{\tt 1024$A_{s,l}w2$} & 256 & $1024^3$ &  49  & 6 & 0.1422 & 2.100 & 0.96 & $-0.85$  & \\
	{\tt 1024$A_{s,h}w2$} & 256 & $1024^3$ &  49  & 6 & 0.1422 & 2.300 & 0.96 & $-0.85$  & \\
	{\tt 1024$\omega_{{\rm m},l}A_{s,l}w2$} & 256 & $1024^3$ &  49  & 6 & 0.1381 & 2.100 & 0.96 & $-0.85$ & \\
	{\tt 1024$\omega_{{\rm m},h}A_{s,h}w2$} & 256 & $1024^3$ &  49  & 6 & 0.1461 & 2.300 & 0.96 & $-0.85$ & \\
  \hline \hline
  \end{tabular}
  }
  \end{center}
    \caption{Simulations discussed  in sections~\ref{sec:prelim}, a subset of which---indicated by an asterisk---will be used in section~\ref{sec:fitting} to calibrate our fitting function: $L$ is the simulation box length, $N$ the number of simulation particles, $z_i$ the initial redshift, $r_s$ the gravitational softening length,  and $\{\omega_{\rm m}, A_s, n_s, w\}$ are cosmological model parameters described in table~\ref{tab:params}.
    	All simulations have been performed using {\sc Gadget}-2/{\sc Camb}.~\label{tab:runs2}}
\end{table*}

To this end we have performed a suite of  simulations summarised in table~\ref{tab:runs2}  using {\sc Gadget}-2/{\sc Camb}, varying the parameter values of $\omega_{\rm m}$, $A_s$,  $n_s$, and $w$ away from their reference $\Lambda$CDM values one at a time as well as in combination in the ranges
\begin{equation}
\begin{aligned}
\label{eq:calibrationmodels}
0.1381 & \leq \omega_{\rm m} \leq 0.1461, \\ 
2.1 & \leq 10^9 A_s \leq 2.3, \\
0.93 & \leq n_s \leq 0.98, \\
-1.15& \leq w \leq -0.75.
\end{aligned}
\end{equation}
In terms of measurement uncertainties, our choice of $\omega_{\rm m}$ values spans a range comparable to $9.2$ times the standard deviation inferred from the 2018 Planck+external data combination%
\footnote{For $\omega_{\rm m}$, $A_s$, and $n_s$, this means the 2018 Planck TT+TE+EE+lowE+lensing+BAO combination~\cite{Aghanim:2018eyx}, while for $w$ the set includes also SNe.} 
in a vanilla 6-parameter $\Lambda$CDM fit 
($\sigma(\omega_{\rm  m})=0.00087$)~\cite{Aghanim:2018eyx}; $6.6$ times for the primordial fluctuation amplitude ($\sigma (10^9 A_s )= 0.030$); 
$13$ for the spectral index  ($\sigma(n_s) = 0.0038$); and  $12$ for a time-independent dark energy equation of state parameter
 ($\sigma(w) = 0.032$).
While these ranges differ between parameters in terms of the number of standard deviations,  the effects the parameter variations produce on the  nonlinear matter power spectrum are of very similar magnitudes---typically no more than $20$\% at $a=0.85$, as shown in figures~\ref{fig:w} to \ref{fig:asomegamns}.

Two interesting properties of $\delta({\bf \Theta}; {\bf \Theta}_0; k;a)$  can be discerned from our simulation set: an approximate universality and multiplicability.  We discuss these properties in detail below, and propose how they can be jointly exploited as a strategy for constructing a fitting function to any general $\delta({\bf \Theta} ,{\bf \Theta}_0; k;a)$ in a multivariate parameter space.


\subsection{Approximate universality: varying one parameter at a time}
\label{sec:universality}

\begin{figure}[t]
	\begin{center}
		\includegraphics[width=15.4cm]{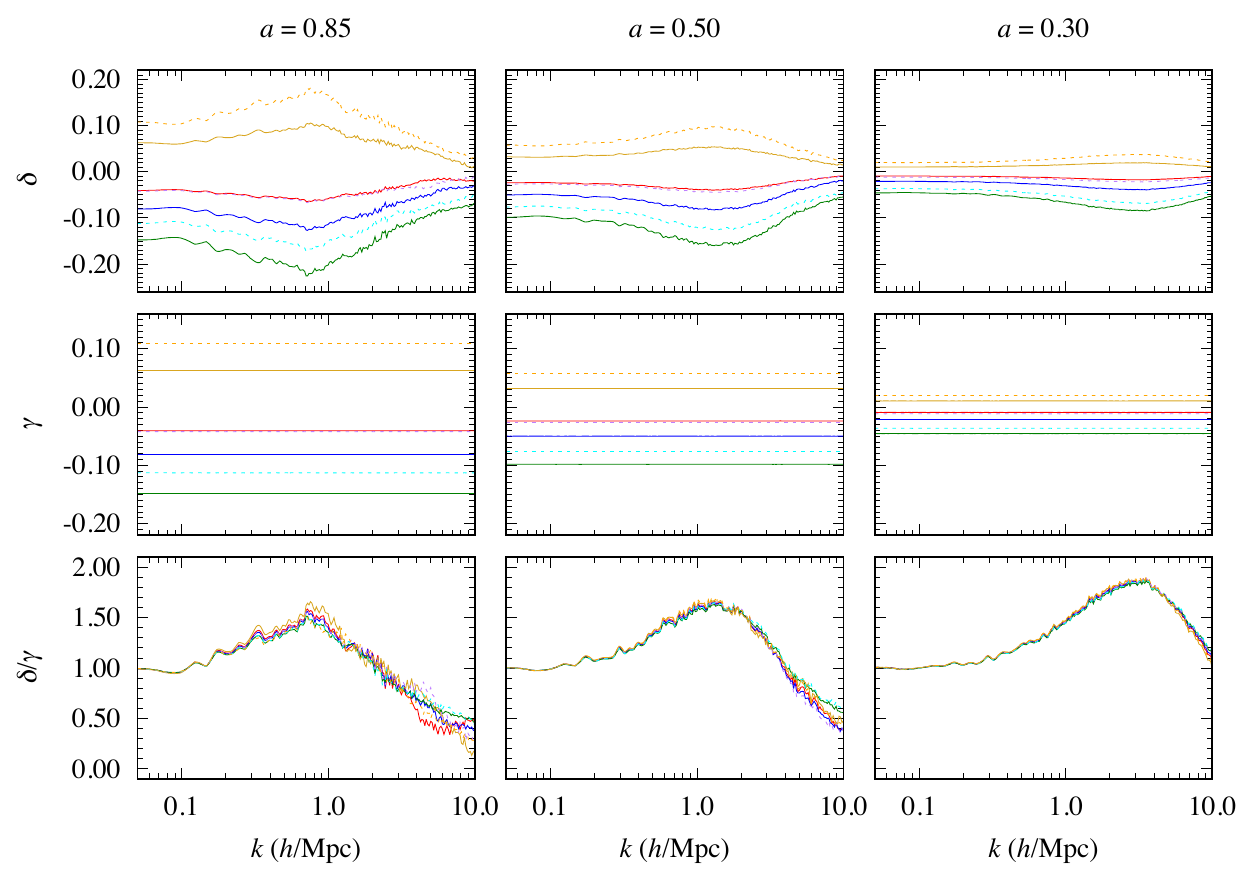}
	\end{center}
	\caption{Relative matter  power spectra~$\delta$ (top panels), their linear counterparts $\gamma$ (middle), and the ratios of the two~$\delta/\gamma$ (bottom) of several target and reference cosmologies~${\bf \Theta}$ and ${\bf \Theta}_0$ at, from left to right, $a=0.85,0.50, 0.30$.  In each case, ${\bf \Theta}$ and ${\bf \Theta}_0$ differ from one another only in the choice of the dark energy equation of state parameter~$w$.
		Solid lines denote target cosmologies with $w=-0.75,-0.85,-0.92,-1.15$, respectively, relative to a reference cosmology with $w=-1$, while dashed lines represent $w=-0.75,-0.85,-1.15$  relative to $w=-0.92$. In all cases, all non-$w$ cosmological parameters~$\theta_w$ have been held fixed at their reference $\Lambda$CDM values~$\bar{\theta}_w$ given in table~\ref{tab:params}.\label{fig:w}}
\end{figure}

Consider figure~\ref{fig:w}. The solid lines in the top panels show the relative matter power spectra $\delta=\delta({\bf \Theta}, {\bf \Theta}_0; k;a)$ at $a = 0.85,0.50, 0.30$ of four  target cosmological models described by ${\bf \Theta} = \{\theta_w=\bar{\theta}_{w};w= -0.73,-0.85,-0.92,-1.15 \}$,
where all non-$w$ model parameters  $\theta_w$ are held fixed at their reference values $\bar{\theta}_{w}$, relative to the canonical reference $\Lambda$CDM model ${\bf \Theta}_0 = \{\theta_w=\bar{\theta}_{w};w =-1 \}$ of table~\ref{tab:params}.   The middle panels show the same cosmological models in a similar construct $\gamma=\gamma({\bf \Theta}, {\bf \Theta}_0; k;a)$, defined as
 \begin{equation}
\gamma({\bf \Theta}, {\bf \Theta}_0; k;a) \equiv   \frac{P_{\rm L}({\bf \Theta}; k;a)-P_{\rm L}({\bf \Theta}_0; k;a)}{P_{\rm L}({\bf \Theta}_0; k;a)},
\end{equation}
i.e., akin to $\delta({\bf \Theta}, {\bf \Theta}_0; k;a )$, but with the target  and reference absolute  power spectra $P({\bf \Theta})$ and $P({\bf \Theta}_0)$ replaced with their linear counterparts $P_{\rm L}({\bf \Theta})$ and $P_{\rm L}({\bf \Theta}_0)$ outputted by {\sc Camb}~\cite{Lewis:1999bs}.
The bottom panels show the ratios $\delta/\gamma$.

An immediately notable feature in  figure~\ref{fig:w} is that despite their differences in~$\delta$ and~$\gamma$, at each scale factor $a$ and over a wide range of wavenumbers~$k$, all four target cosmologies return a functional form for the ratio $\delta/\gamma$ that is quantitatively remarkably independent of the chosen value of~$w$; the function tends to unity in the linear regime, 
peaks at an $a$-dependent~$k_{\rm peak}$, and  drops off to zero at large $k$ values.  At $k \lesssim 4 \ h$/Mpc the agreement between models is always better  than 10\%.  This  ``approximate universality'' of $\delta/\gamma$ likewise holds for a reference~$w$ value different  from the canonical choice of  $-1$ in  ${\bf \Theta}_0$, as demonstrated by the dashed lines in figure~\ref{fig:w}  (which feature $w=-0.92$ in ${\bf \Theta}_0$), 
 provided of course that we choose the same reference~$w$ for both $P({\bf \Theta}_0)$ and  $P_{\rm L}({\bf \Theta}_0)$  in the construction of~$\delta/\gamma$.
 
\begin{figure}[t]
	\begin{center}
		\includegraphics[width=15.4cm]{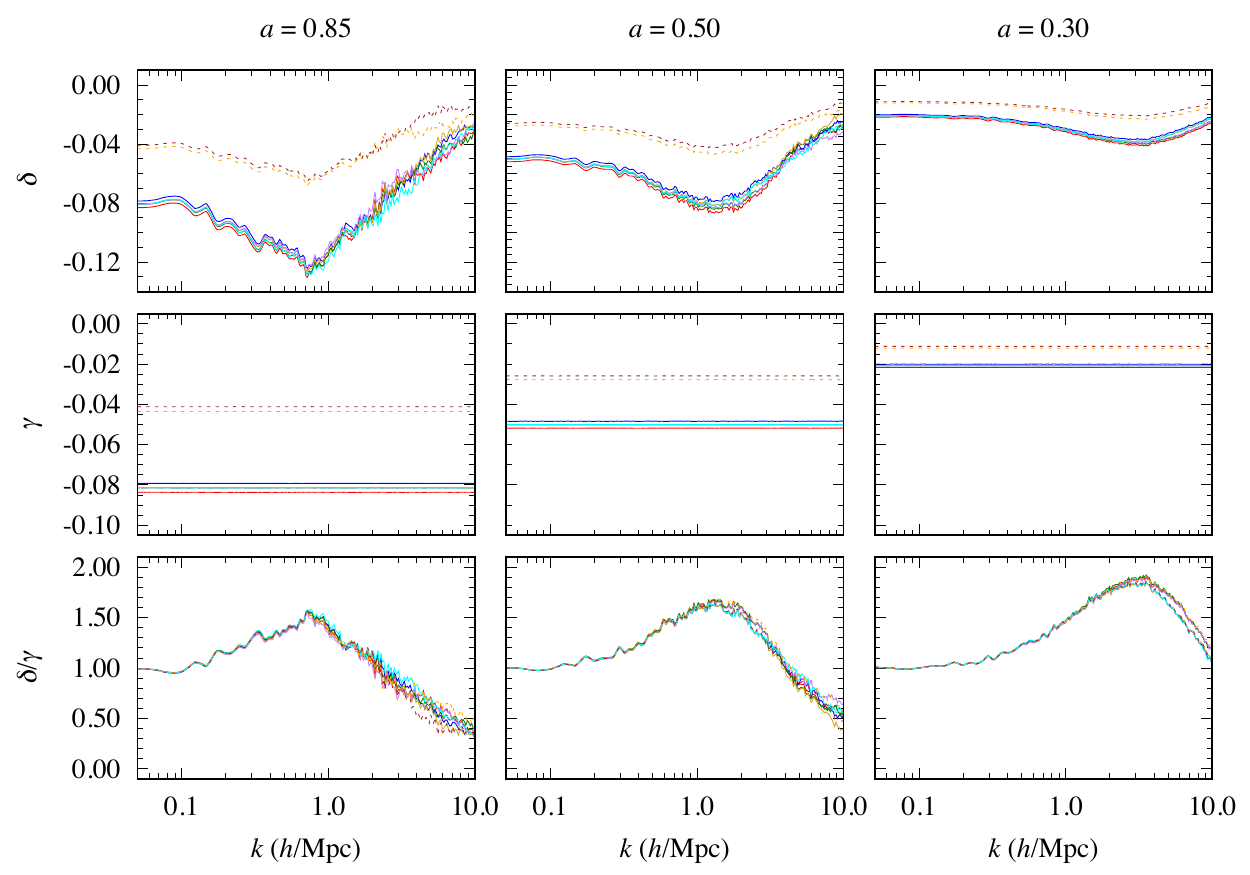}
	\end{center}
	\caption{Relative matter power spectra~$\delta$ (top panels), their linear counterparts $\gamma$ (middle), and the ratios of the two~$\delta/\gamma$ (bottom) of several target and reference cosmologies~${\bf \Theta}$ and ${\bf \Theta}_0$ at, from left to right, $a=0.85,0.50,0.30$.  Similarly to figure~\ref{fig:w}, in each case ${\bf \Theta}$ and ${\bf \Theta}_0$ differ from one another only in the choice of the dark energy equation of state parameter~$w$.  Contrary to figure~\ref{fig:w}, however, some non-$w$ cosmological parameters~$\theta_w$ common between  ${\bf \Theta}$ and ${\bf \Theta}_0$ have had their numerical values altered from their canonical  values~$\bar{\theta}_w$.
		The solid lines represent six target cosmologies with $w=-0.85$ relative to a $w=-1$ reference model, wherein the common parameters $A_s,\omega_{\rm m}, n_s$ between the target and the reference have been changed, one at a time, from their canonical  values to $10^9 A_s = \{2.100,2.300\}$, $\omega_{\rm m}=\{0.1381,0.1461\}$, and $n_s = \{-0.93, -0.96\}$. The dashed lines denote two target cosmologies with $w=-0.85$ relative to a $w=-0.92$ reference, and variations to $\omega_{\rm m}=\{0.1381,0.1361\}$.
		\label{fig:wtheta}}
\end{figure}

Approximate universality in $\delta/\gamma$ extends also to the case in which we employ a set of the non-$w$ cosmological parameters~$\theta_w$ different from~$\bar{\theta}_{w}$, again on the understanding that whatever values we choose for~$\theta_w$  in the construction of~$\delta/\gamma$
are held constant across the four target and reference absolute power spectra, $P({\bf \Theta}), P({\bf \Theta}_0), P_{\rm L}({\bf \Theta})$, and $P_{\rm L}({\bf \Theta}_0)$. This is illustrated in figure~\ref{fig:wtheta} by the solid lines, representing $\delta, \gamma$, and $\delta/\gamma$ constructed from a selection of target and reference cosmologies
from the simulations of table~\ref{tab:runs2}, where ${\bf \Theta} = \{\theta_w=\bar{\theta}_{w};w= -0.85 \}$, ${\bf \Theta}_0 = \{\theta_w=\bar{\theta}_{w}; w= -1 \}$, and $\theta_w \neq \bar{\theta}_{w}$.   In the same figure, the cosmological models represented by the dashed lines  feature in addition a non-canonical reference~$w$ value in ${\bf \Theta}_0$ (in this instance, $w=-0.92$); again, their respective $\delta/\gamma$ conforms to the same approximately universal form already observed amongst the solid lines as well as in figure~\ref{fig:w}.

\begin{figure}[t]
\begin{center}
\includegraphics[width=15.4cm]{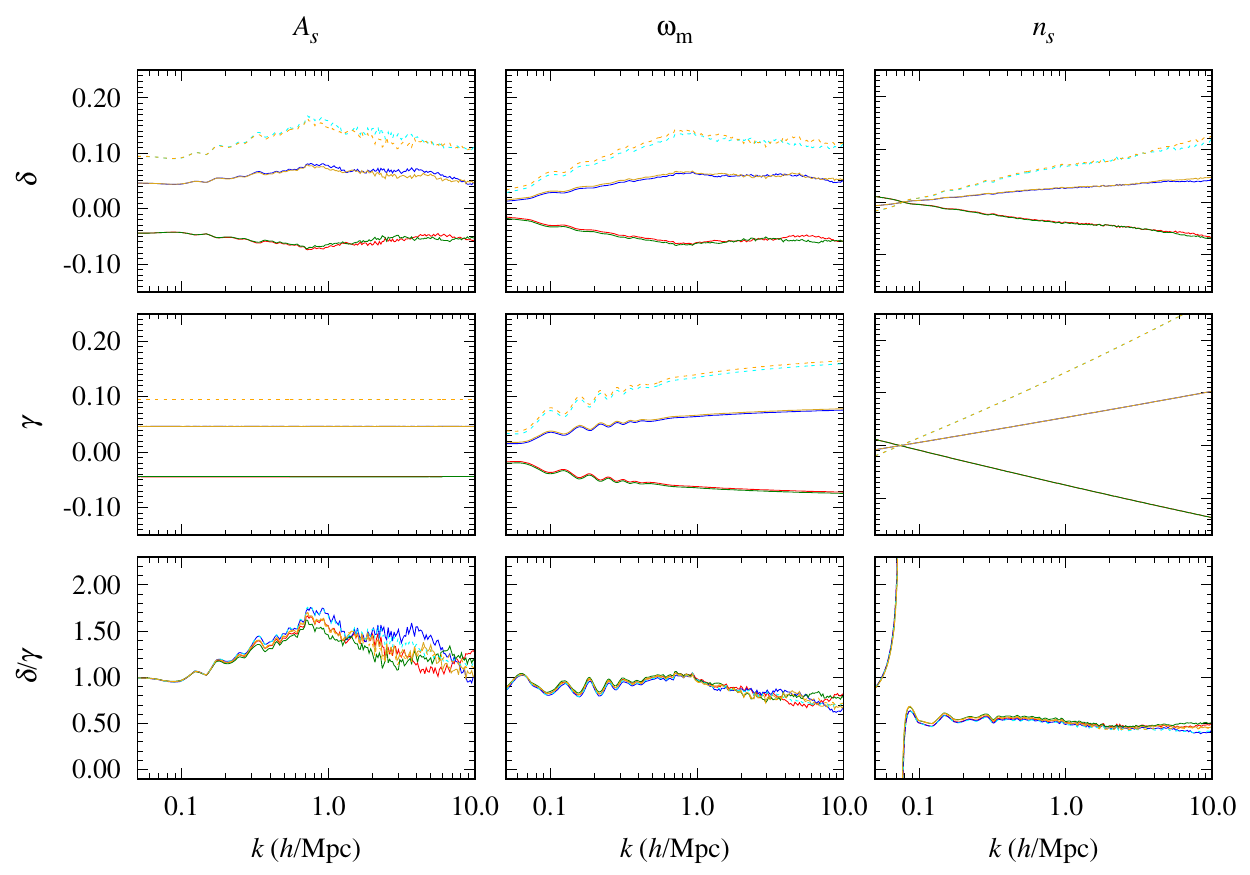}
\end{center}
\caption{Relative matter power spectra~$\delta$ (top panels), their  linear counterparts $\gamma$ (middle), and the ratios of the two~$\delta/\gamma$ (bottom) of several target and reference cosmologies~${\bf \Theta}$ and ${\bf \Theta}_0$ at  $a=0.85$, where, from left to right, ${\bf \Theta}$ and ${\bf \Theta}_0$ differ from one another only in the choice of~$X= A_s,\omega_{\rm m}, n_s$.  Solid lines denote a canonical choice for $X$ in the reference cosmology~${\bf \Theta}_0$, while dashed lines represent a non-canonical option.  The non-$X$ parameters may or may not satisfy $\theta = \bar{\theta}$.  See text for details of the models.\label{fig:asomegamns}.}
\end{figure}

\begin{figure}[t]
\begin{center}
\includegraphics[width=15.4cm]{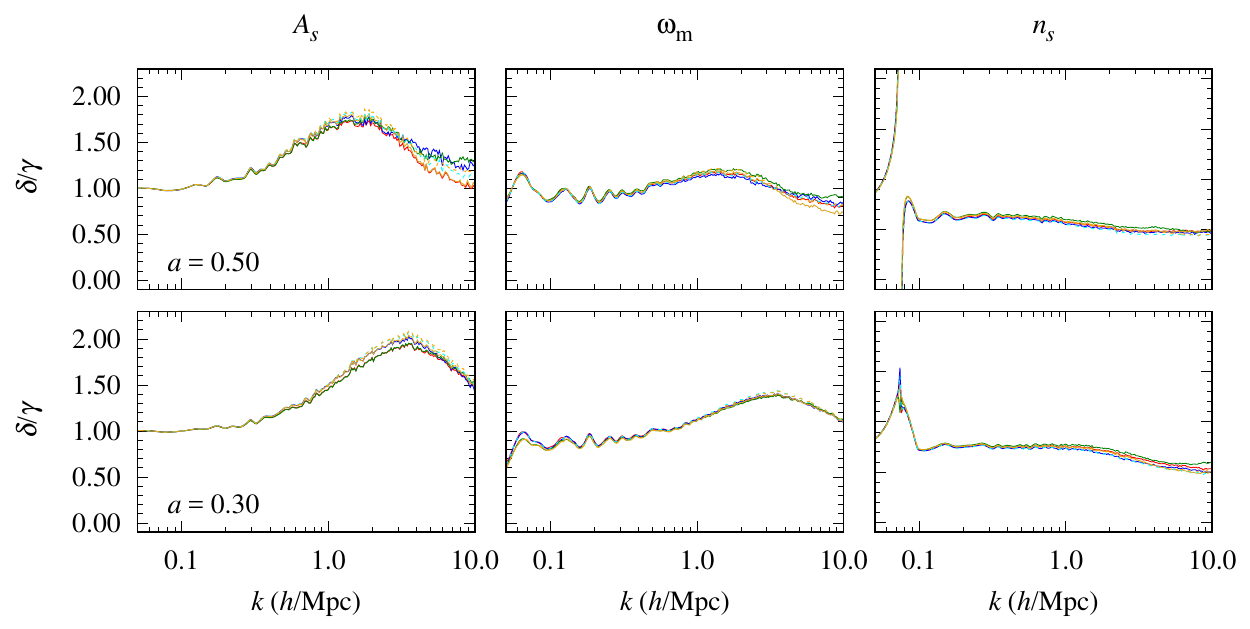}
\end{center}
\caption{Same as figure~\ref{fig:asomegamns},  but only the ratios~$\delta/\gamma$ at  $a=0.50$ (top panels) and at $a=0.30$ (bottom).\label{fig:asomegamnshighz}}
\end{figure}

So far we have  discussed the approximate universality of $\delta/\gamma$ exclusively in the context wherein the target and reference cosmologies, ${\bf \Theta}$ and ${\bf \Theta}_0$, differ only by  their~$w$ parameter value.  To further test the hypothesis of a $\delta/\gamma$ approximate universality under variation of {\it any} one cosmological parameter besides~$w$, 
we show in figure~\ref{fig:asomegamns} $\delta, \gamma$, and $\delta/\gamma$ for three families of relative power spectra at $a=0.85$ described by
\begin{enumerate}
\item[1a.] Solid:  ${\bf \Theta} = \{\theta_{A_s}; 10^9 A_s=2.100, 2.300\}$, ${\bf \Theta}_0 = \{\theta_{A_s}; 10^9 A_s=2.198 \}$;
\item[b.] Dashed: ${\bf \Theta} = \{\theta_{A_s}; 10^9 A_s=2.300  \}$, ${\bf \Theta}_0 = \{\theta_{A_s}; 10^9 A_s=2.100  \}$;
\item[2a.] Solid:  ${\bf \Theta} = \{\theta_{\omega_{\rm m}}; \omega_{\rm m} =0.1381,0.1461\}$, ${\bf \Theta}_0 = \{\theta_{\omega_{\rm m}}; \omega_{\rm m}=0.1422 \}$;
\item[b.] Dashed: ${\bf \Theta} = \{\theta_{\omega_{\rm m}}; \omega_{\rm m} =0.1461\}$, ${\bf \Theta}_0 = \{\theta_{\omega_{\rm m}}; \omega_{\rm m}=0.1381 \}$;
\item[3a.] Solid: ${\bf \Theta} = \{\theta_{n_s}; n_s=0.93,0.98\}$, ${\bf \Theta}_0 = \{\theta_{n_s}; n_s=0.96 \}$;
\item[b.] Dashed: ${\bf \Theta} = \{\theta_{n_s}; n_s=0.98\}$, ${\bf \Theta}_0 = \{\theta_{n_s}; n_s=0.93 \}$.
\end{enumerate}
See also figure~\ref{fig:asomegamnshighz} for $\delta/\gamma$ of these models at $a=0.50,0.30$.
Here, the convention $\theta_X$ again denotes all model parameters other than $X$, and we consider both $\theta_X=\bar{\theta}_X$ and $\theta_X \neq \bar\theta_X$ selected from the simulations of table~\ref{tab:runs2}.  Again, the close similarity of $\delta/\gamma$ {\it within each family} is unmistakable.
 In the case of variations in $A_s$ and $\omega_{\rm m}$, we see that the $a$-dependent locations of the peaks are similar to $k_{\rm peak}$ previously identified for variations in $w$.

Note that in the case of variation of~$\omega_{\rm m}$, $\delta/\gamma$ exhibits prominent oscillations at $k \lesssim 1\ h$/Mpc.  Oscillations arise in the first place from a phase difference in the baryon acoustic oscillations between cosmologies with different matter densities, and can already be seen in both~$\delta$ and~$\gamma$.   Nonlinear evolution additionally alters the amplitudes and phases of these oscillations, so that a residual survives in $\delta/\gamma$.

A final remark concerns the singularities in $\delta/\gamma$ under variation of $n_s$  observed in figures~\ref{fig:asomegamns} and~\ref{fig:asomegamnshighz}.  These are artefacts following from our choice of pivot scale $k_{\rm piv} = 0.05$/{\rm Mpc} for the primordial power spectrum~${\cal P_R}(k)=A_s (k/k_{\rm piv})^{n_s-1}$.  In fact, 
a singularity will arise in~$\delta/\gamma$ whenever the linear power spectra of the target and reference cosmologies cross over.  
A judicious choice of $k_{\rm piv}$, e.g., $k_{\rm piv}=0.002$/Mpc, would have confined such cross-overs to scales outside of the range of interest and facilitated the task of finding a fitting function.
However, as we shall discuss in section~\ref{sec:further}, rather than re-running simulations with a different~$k_{\rm piv}$, it transpires that for power-law primordial power spectra the remedy is very simple.

Then, to summarise section~\ref{sec:universality}, for a family of relative matter  power spectra described by the target and reference cosmological model parameters ${\bf \Theta} = \{\theta_{X}; X \}$ and ${\bf \Theta}_0 = \{\theta_X; \bar{X} \}$, the ratio of the relative (nonlinear) power spectrum to the relative linear power spectrum, $\delta/\gamma$, is, at each scale factor $a$ and over a wide range of wavenumbers~$k$, largely independent of the values of $\theta_X$, $X$, and $\bar{X}$. 
We shall denote this approximately  universal ratio $(\delta/\gamma)_X$.


\subsection{Multiplicability: varying two or more parameters at a time}
\label{sec:multiplicability}

Consider now three target cosmological models specified respectively by the parameters
\begin{equation}
\begin{aligned}
\label{eq:targets}
{\bf \Theta}_2 & = \{\theta_{w,A_s}=\bar{\theta}_{w,A_s}; w, A_s \}, \\
{\bf \Theta}_{1a} & =  \{\theta_{w}= \bar{\theta}_{w}; w\}, \\
{\bf \Theta}_{1b} & =  \{\theta_{A_s}= \bar{\theta}_{A_s}; A_s\}, 
\end{aligned}
\end{equation}
where $\theta_{X,Y}$ denotes all model parameters besides $X$ and $Y$, $\bar{\theta}_{X,Y}$ their reference values in table~\ref{tab:params}, and our canonical reference model  is again defined by ${\bf \Theta}_0  =  \{\theta_w = \bar{\theta}_w; w=-1\}$.  

From the definition~(\ref{eq:relative}) it is easy to establish that the three target cosmologies must have relative power spectra $\delta$ satisfying at all times  the general relations
\begin{equation}
\begin{aligned}
\label{eq:general}
1+ \delta({\bf \Theta}_2, {\bf \Theta}_0) & = \left[1+ \delta({\bf \Theta}_2, {\bf \Theta}_{1a}) \right] \left[1+ \delta({\bf \Theta}_{1a}, {\bf \Theta}_0) \right] \\
& = \left[1+ \delta({\bf \Theta}_{1b}, {\bf \Theta}_0) \right]  \left[1+ \delta({\bf \Theta}_2, {\bf \Theta}_{1b}) \right] ,
\end{aligned}
\end{equation}
irrespective of our exact choice of model parameter values.  For the {\it particular} target cosmologies~(\ref{eq:targets}) under consideration, the corresponding relative linear power spectra $\gamma$ also happen to obey
\begin{equation}
\begin{aligned}
\label{eq:linearrelations}
\gamma({\bf \Theta}_2, {\bf \Theta}_{1a}) &= \gamma({\bf \Theta}_{1b}, {\bf \Theta}_0),\\
\gamma({\bf \Theta}_{2}, {\bf \Theta}_{1b}) &= \gamma({\bf \Theta}_{1a}, {\bf \Theta}_0)
\end{aligned}
\end{equation}
because of the especially simple and, importantly, {\it separable} effects variations of $w$ and $A_s$ induce on the absolute linear power spectrum,   in the sense that $P_{\rm L}({\bf \Theta})$ is a separable function of $w$, $A_s$, and $\theta_{w,A_s}$:
\begin{equation}
\label{eq:separable}
P_{\rm L}({\bf \Theta}= \{\theta_{w,A_s}; w, A_s \}) = f(w) g(A_s) h(\theta_{w,A_s}).
\end{equation}
 It then follows straightforwardly from the approximate universality of $\delta/\gamma$ 
 discussed in section~\ref{sec:universality} that $\delta({\bf \Theta}_2, {\bf \Theta}_{1a})\simeq  \delta({\bf \Theta}_{1b}, {\bf \Theta}_0)$ and 
$\delta({\bf \Theta}_{2}, {\bf \Theta}_{1b}) \simeq \delta({\bf \Theta}_{1a}, {\bf \Theta}_0)$, and hence 
\begin{equation}
\label{eq:approximate}
1+ \delta({\bf \Theta}_2, {\bf \Theta}_0)  \simeq   \left[1+ \delta({\bf \Theta}_{1b}, {\bf \Theta}_0) \right]  \left[1+ \delta({\bf \Theta}_{1a}, {\bf \Theta}_0) \right]
\end{equation}
as an approximation to equation~(\ref{eq:general}).

\begin{figure}[t]
\begin{center}
\includegraphics[width=15.4cm]{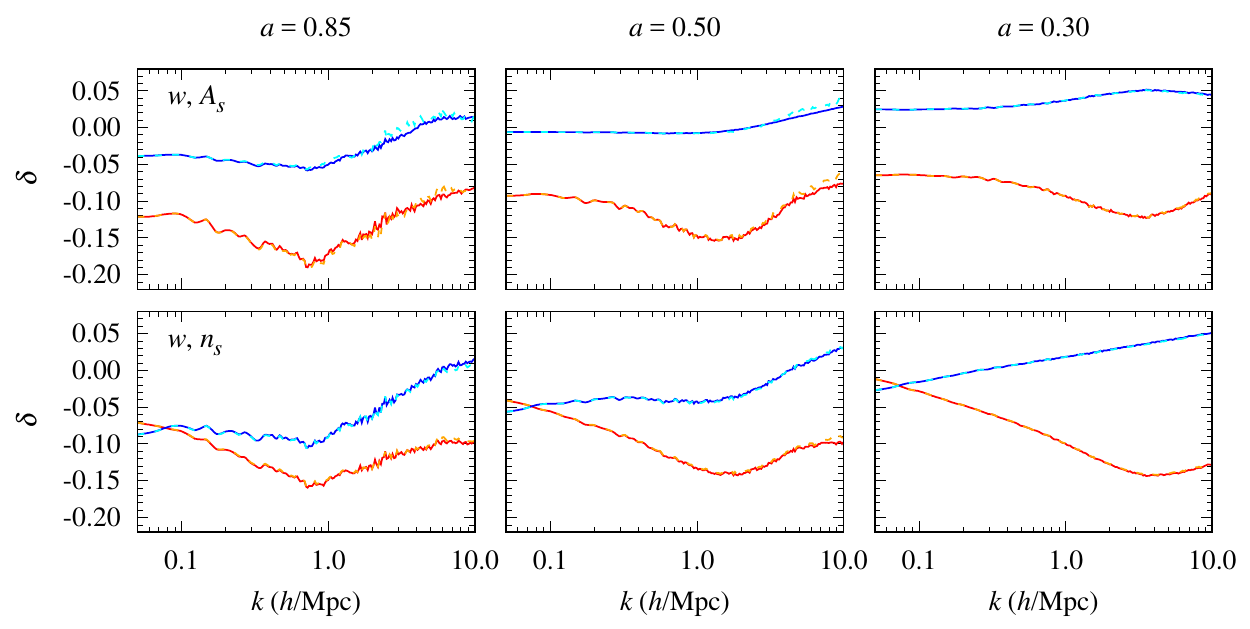}
\end{center}
\caption{{\it Top}: Exact relative matter  power spectrum for the simultaneous variation of two parameters $A_s$ and $w$  (solid) and its approximate form constructed from single-parameter variations via equation~(\ref{eq:approximate}) (dashed) at, from left to right, $a=0.85,0.50, 0.30$.  The two target cosmologies shown correspond respectively to $\{w=-0.85, 10^9 A_s=2.100\}$ (red/orange) and $\{w=-0.85, 10^9 A_s=2.300\}$ (blue/cyan). The reference cosmology~${\bf \Theta}_0$ is the canonical $\Lambda$CDM model of~table~\ref{tab:params}.
{\it Bottom}: Same as the top panels, but variations in $A_s$ in the target cosmologies have been replaced with variations in $n_s$, specifically, $n_s=0.93$ (red/orange) and $n_s=0.98$ (blue/cyan). \label{fig:multiplyasw}}
\end{figure}

The top panels of figure~\ref{fig:multiplyasw}  demonstrate the remarkable correspondence between the exact $\delta({\bf \Theta}_2, {\bf \Theta}_0)$  and its approximation constructed from $\delta({\bf \Theta}_{1b}, {\bf \Theta}_0)$ and  $\delta({\bf \Theta}_{1a}, {\bf \Theta}_0)$ via equation~(\ref{eq:approximate}) for $w=-0.85$ and $10^9 A_s = 2.100, 2.300$; at all scale factors and for the entire range of wavenumbers under consideration, the approximation is able to reproduce the exact relative matter power spectrum to $0.01$ or better.  The bottom panels provide a second example of this excellent correspondence for the target cosmologies ${\bf \Theta}_2  = \{\theta_{w,n_s}=\bar{\theta}_{w,n_s}; w=-0.85, n_s=0.93,0.98 \}$, ${\bf \Theta}_{1a}  =  \{\theta_{w}= \bar{\theta}_{w}; w=-085\}$, and ${\bf \Theta}_{1b}  =  \{\theta_{n_s}= \bar{\theta}_{n_s}; n_s=0.93,0.98\}$ (for which the equivalents of equations~(\ref{eq:linearrelations}), (\ref{eq:separable}), and hence~(\ref{eq:approximate}) also  hold).

Naturally, alternatively to equation~(\ref{eq:approximate}), the approximate universality of $\delta/\gamma$ under variation of one parameter means that we could also have approximated 1+$\delta({\bf \Theta}_2, {\bf \Theta}_0)$ using  instead $\left[1+\delta({\bf \Theta}_{2}, {\bf \Theta}_{1b}) \right] \left[ 1+\delta({\bf \Theta}_{2}, {\bf \Theta}_{1a}) \right]$---or, indeed, any other combination of two relative power spectra in which we vary only one parameter at a time and whose linear counterparts equate to the relations~(\ref{eq:linearrelations})---with similarly good although not identical results to  figure~\ref{fig:multiplyasw}.  The essence of equation~(\ref{eq:approximate}), however, lies in its suggestion that the multiplicative nature of the relative power spectrum and the approximate-universal form $(\delta/\gamma)_X$ under variation of  $X$ may be jointly exploited as a relatively simple strategy for constructing a fitting function to any general $\delta({\bf \Theta} ,{\bf \Theta}_0)$ in a multivariate parameter space.

Furthermore, the condition of separability~(\ref{eq:separable}) implies that the natural division of cosmological models into families (for the purpose of finding the approximate-universal forms~$(\delta/\gamma)_X$) is not in terms of the model parameters {\it per se}, but rather their linear ``proxies''---the linear transfer function~$T$, the linear growth function~$D$, etc.---that naturally cast the absolute linear power spectrum in power-law $w$CDM-type cosmologies  into a separable function:
\begin{equation}
\label{eq:linearpowerspectrum}
P_{\rm L}( {\bf \Theta}; k;a) = {\cal N} (A_s, \omega_{ \rm m})  \left( \frac{k}{k_{\rm piv}} \right)^{n_s-1} D^2(w, \omega_{\rm m},h;a) T^2(\omega_{\rm m}, \omega_{\rm b}; k)
\end{equation}
following the  textbook convention of~\cite{Dodelson:2003ft}, where ${\cal N} \equiv A_s/\omega_{\rm m}^2$ is the overall normalisation of the linear matter power spectrum up to an irrelevant multiplicative constant.

Then, for the parameter variations represented by the independent parameters of equation~(\ref{eq:linearpowerspectrum}),
 it follows from the same reasoning of approximate universality and multiplicability that a multivariate relative power spectrum may be most conveniently approximated by
\begin{equation}
\begin{aligned}
\label{eq:form}
&1+ \delta({\bf \Theta}, {\bf \Theta}_0; k;a) \\
&  \simeq   \left[1+  (\delta/\gamma)_{\cal N}  \frac{\Delta {\cal N}}{\bar{\cal N} }\right] 
 \left[1+  (\delta/\gamma)_{n_s} \frac{\Delta {\cal Q}}{\bar{\cal Q}} \right]
  \left[1+  (\delta/\gamma)_{D}  \frac{\Delta D^2}{\bar{D}^2} \right] 
  \left[1+  (\delta/\gamma)_{T}   \frac{\Delta T^2}{\bar{T}^2}  \right],
  \end{aligned}
\end{equation}
where $\Delta X \equiv X-\bar{X}$ with the revised understanding that $X$ may be a model parameter or a linear proxy, $(\delta/\gamma)_X$ is the approximate-universal form of $\delta/\gamma$ under variation of~$X$ alone,  ${\cal Q} \equiv  (k/k_{\rm piv})^{n_s-1}$, and 
\begin{equation}
\begin{aligned}
{\bf \Theta} &= \{\theta_{w,\omega_{\rm m}, \omega_{\rm b},h,A_s,n_s}=\bar{\theta}_{w,\omega_{\rm m},\omega_{\rm b}, h, A_s,n_s}; {\cal N} (A_s, \omega_{ \rm m}), n_s, D(w,\omega_{\rm m},h;a),T(\omega_{\rm m},\omega_{\rm b};k) \}, \\
{\bf \Theta}_0 &= \{\theta_{w, \omega_{\rm m},\omega_{\rm b},h,A_s,n_s}= \bar{\theta}_{w, \omega_{\rm m},\omega_{\rm b},h,A_s,n_s}; \bar{\cal N},
 \bar{n}_s, \bar{D}, \bar{T} \},
\end{aligned}
\end{equation}
with  $\bar{\cal N} \equiv {\cal N} (\bar{A}_s,\bar{\omega}_{ \rm m})$,  $\bar{D}\equiv D(\bar{w},\bar{\omega}_{\rm m},\bar{h};a)$, and  $\bar{T}\equiv T(\bar{\omega}_{\rm m},\bar{\omega}_{\rm b};k )$,
specify the target and the reference cosmology respectively.


\subsection{Further remarks}
\label{sec:further}

Equation~(\ref{eq:form}) serves as a starting point for the construction of a fitting function of the relative power spectrum; Three more remarks are in order before we proceed.

\paragraph{Remark 1: Fitting functions}  The salient feature of equation~(\ref{eq:form}) is that the cosmological dependence of the relative power spectrum has been largely subsumed by the linear quantities~$\Delta X/\bar{X}$.  Thus, the task of finding a full fitting function for~$\delta({\bf \Theta}, {\bf \Theta}_0; k;a)$  boils down at the most elementary level to writing down a cosmology-independent functional form in terms of the wavenumber~$k$ and scale factor~$a$ for each of the four familial approximate-universal forms~$(\delta/\gamma)_X$.  For fixed values of~$a$ this is a trivial exercise. A more useful endeavour would be to model the approximate-universal forms' dependence on the scale factor~$a$, to be pursued in section~\ref{sec:fitting}.

In a more sophisticated model one could of course also incorporate the small, cosmology-dependent deviations from the approximate-universal forms that inevitably creep in at large wavenumbers.  We do not however see this as a necessary step at this stage: the one-parameter approximate-universal forms~$(\delta/\gamma)_X$ are in the worst case $10 \to 20$\% ``off'' at $k \gtrsim 4 \ h$/Mpc (see figures~\ref{fig:w} to~\ref{fig:asomegamnshighz}), while the linear deviations~$\Delta X/\bar{X}$ are typically $O(0.1)$.  Thus,  barring an unfortunate add-up of errors, 
we can be confident that $\delta$ can be reproduced to $\pm 0.01 \to 0.02$ up to $k \sim 10 \ h$/Mpc.

\paragraph{Remark 2: Varying the matter density} 
Equation~(\ref{eq:form}) is amenable to further algebraic manipulation, a property that is especially useful in those cases where a cosmological model parameter controls more than one linear proxy.  The case in point  is the physical matter density $\omega_{\rm m}$, the only parameter that controls the linear transfer function~$T$ in the cosmologies under consideration.  Because $\omega_{\rm m}$ affects also the linear growth function~$D$ and the normalisation ${\cal N}$, it is {\it a priori} not possible to  establish the approximate-universal form~$(\delta/\gamma)_T$ directly from a set of $N$-body simulations such as detailed in table~\ref{tab:runs2} that uses $\omega_{\rm m}$ as a base parameter.

 However, equation~(\ref{eq:form}) permits us to write
\begin{equation}
\label{eq:matteronly}
 1+  (\delta/\gamma)_{\omega_{\rm m}}  \gamma_{\omega_{\rm m}}
 \simeq 
  \left[1+  (\delta/\gamma)_{\cal N}  \frac{\Delta {\cal N}_1}{\bar{\cal N}} \right] 
 \left[1+  (\delta/\gamma)_{D}  \frac{\Delta D_1^2}{\bar{D}^2} \right] 
  \left[1+  (\delta/\gamma)_{T}   \frac{\Delta T^2}{\bar{T}^2}  \right],
\end{equation}
where 
\begin{equation}
\begin{aligned}
\Delta {\cal N}_1 &\equiv {\cal N}_1 - \bar{\cal N} \equiv {\cal N} (\bar{A}_s, \omega_m)- \bar{\cal N},\\
\Delta D_1^2 &\equiv D_1^2 - \bar{D}^2  \equiv D^2(\bar{w}, \omega_{\rm m},\bar{h};a)- \bar{D}^2,
\end{aligned}
\end{equation}
and $\gamma_{\omega_{\rm m}} \equiv \gamma({\bf \Theta}=\{\bar{\theta}_{\omega_{\rm m}}; \omega_{\rm m}\},
{\bf \Theta}_0=\{\bar{\theta}_{\omega_{\rm m}};\bar{ \omega}_{\rm m}\})$ denotes the  relative linear matter power spectrum under variations in~$\omega_{\rm m}$ alone.   Then, solving for $(\delta/\gamma)_T$ and substituting back into equation~(\ref{eq:form}) itself yields an alternative form
\begin{equation}
\begin{aligned}
\label{eq:formalt}
1+ \delta({\bf \Theta}, {\bf \Theta}_0; k;a)  
\simeq  \frac{ 1+  (\delta/\gamma)_{\cal N}  \frac{\Delta {\cal N}}{\bar{\cal N}} }{ 1+  (\delta/\gamma)_{\cal N}  \frac{\Delta {\cal N}_1}{\bar{\cal N}} }
 \left[1+  (\delta/\gamma)_{n_s} \frac{\Delta {\cal Q}}{\bar{\cal Q}} \right]
   \left[1+  (\delta/\gamma)_{\omega_{\rm m}}  \gamma_{\omega_{\rm m}} \right]
 \frac{ 1+  (\delta/\gamma)_{D}  \frac{\Delta D^2}{\bar{D}^2}  }{ 1+  (\delta/\gamma)_{D}  \frac{\Delta D_1^2}{\bar{D}^2} },
  \end{aligned}
\end{equation}
which has the desirable feature that~$(\delta/\gamma)_{\omega_{\rm m}}$ can be  determined directly from simulations.  Indeed, for the cosmologies of table~\ref{tab:runs2}, equation~(\ref{eq:formalt}) may  be the more convenient albeit less general fitting form than equation~(\ref{eq:form}).

\begin{figure}[t]
\begin{center}
\includegraphics[width=15.4cm]{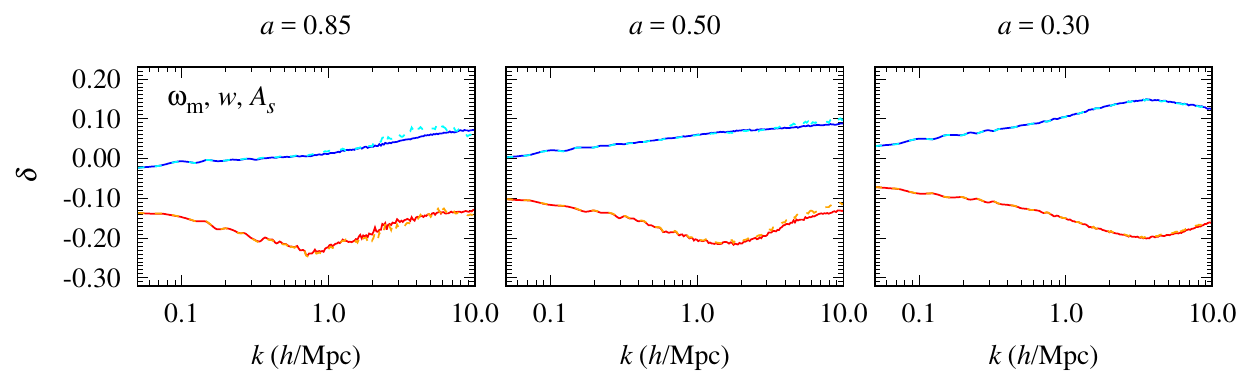}
\end{center}
\caption{Exact relative matter power spectrum for the simultaneous variation of three parameters $A_s$, $w$, and $\omega_{\rm m}$ (solid) and its approximate form constructed from single-parameter variations via equation~(\ref{eq:formalt}) (dashed) at, from left to right, $a=0.85,0.50, 0.30$.  The two target cosmologies shown correspond respectively to $\{\omega_{\rm m}=0.1381, w=-0.85, 10^9 A_s=2.100 \}$ (red/orange) and $\{\omega_{\rm m}=0.1361, w=-0.85, 10^9 A_s=2.300\}$ (blue/cyan).
The reference cosmology is the canonical $\Lambda$CDM model of~table~\ref{tab:params}.\label{fig:multiplyomasw}}
\end{figure}

Figure~\ref{fig:multiplyomasw} shows the exact relative  matter power spectrum from the simultaneous variation of $\{\omega_{\rm m}, w, A_s \}$ and its approximate form constructed from single-parameter variations via equation~(\ref{eq:formalt}), for two target cosmologies  $\{\omega_{\rm m}=0.1461,w= -0.85, 10^9 A_s=2.300 \}$ and 
$\{\omega_{\rm m}=0.1381,w= -0.85, 10^9 A_s=2.100\}$. The agreement is excellent: typically much better than 0.01, and in the worst case $\sim 0.02$ at large $k$ values.

Note that to calculate the linear growth functions $D$ and $D_1$ we have solved numerically the differential equation~\cite{Linder:2003dr}
\begin{equation}
\label{eq:growthde}
g''+  \left[\frac{7}{2}-\frac{3}{2}\frac{w(a)}{1+X(a)} \right] \frac{g'}{a} +\frac{3}{2} \frac{1-w(a)}{1+X(a)} \frac{g}{a^2}=0
\end{equation}
 with the initial conditions $g(a_{\rm ini}) = 1$, $g'(a_{\rm ini}) = 0$, and $a_{\rm ini}=10^{-3}$.  Here, $g \equiv D/a$, the prime $(\cdots)'$ denotes a derivative with respect to the scale factor $a$, $w(a)$ is the dark energy equation of state parameter which may be time-dependent, and  
 \begin{equation}
 \begin{aligned}
 X(a) & = \frac{\omega_{\rm m}}{h^{2}-\omega_{\rm m}} \exp \int^1_a {\rm d} \ln a' \, w(a')\\
 &=\frac{\omega_{\rm m}}{h^{2}-\omega_{\rm m}} \, a^{3(w_0+w_a)} \exp \left[ 3 w_a (1-a)\right],
 \end{aligned}
\end{equation} 
 where the second equality applies in the case $w(a) = w_0 + w_a (1-a)$~\cite{Chevallier:2000qy,Linder:2002et}.
 It is usually understood that the solution to equation~(\ref{eq:growthde}) can be approximated to high accuracy by the integral~\cite{Linder:2005in} 
\begin{equation}
\label{eq:growthformula}
g (a)= \exp \int_0^a d \ln a'  \left[ \Omega(a')^\rho-1 \right],
\end{equation}
 where  $\Omega(a)$ is the reduced matter density at $a$, and $\rho = 0.55 + 0.05 [1+w(z=1)]$ was originally proposed in~\cite{Linder:2005in}.  Indeed, we have checked that for even time-dependent equations of state, the approximate formula~(\ref{eq:growthformula}) is able to reproduce numerical solutions to roughly 1 part in $10^4$, sufficient to approximate the growth function {\it differences} $\Delta D^2/\bar{D}^2$ to $O(0.001)$ accuracy for the models tested.
Nonetheless, we prefer to err on the side of caution and  work directly with the differential equation~(\ref{eq:growthde}).

\paragraph{Remark 3: Varying the scalar spectral index} 

As already pointed out in section~\ref{sec:universality}, because of our choice of pivot scale~$k_{\rm piv}=0.05$/Mpc, variation of the scalar spectral index $n_s$ introduces a singularity in $(\delta/\gamma)_{n_s}$ in the $k$ range of interest.  This singularity can be easily removed by recognising that the relative linear power spectrum $\gamma$ under variation of $n_s$ alone, $\gamma_{n_s} \equiv \Delta {\cal Q}/\bar{\cal Q}$, can be recast as
\begin{equation}
\begin{aligned}
\label{eq:Gamma}
\gamma_{n_s} & =(k/k_{\rm piv})^{\Delta n_s}-1 = \Gamma(k) \ln ( k/k_{\rm piv}),
\end{aligned}
\end{equation}
where $\Delta n_s \equiv n_s - \bar{n}_s$, and 
$\Gamma (k) \equiv \Delta n_s \sum_{i=0}^\infty  \left[\Delta n_s \ln (k/k_{\rm piv})\right]^{i}/(i+1)!$
is always finite and, at leading order, equal to $\Delta n_s$.  It then follows that  the corresponding approximate-universal form is equivalently
\begin{equation}
\label{eq:Gammauniversal}
(\delta/\gamma)_{n_s} = \frac{( \delta /\Gamma)_{n_s}}{\ln( k/k_{\rm piv})},
\end{equation}
and hence $(\delta/\gamma)_{n_s} \, \Delta {\cal Q}/\bar{\cal Q} =   (\delta/\Gamma)_{n_s} \Gamma$,
where the ratio $(\delta/\Gamma)_{n_s}$ must also be approximately universal for all variations of $n_s$, albeit better-behaved than the original $(\delta/\gamma)_{n_s}$.

Then, applying this understanding to equation~(\ref{eq:form}) and its restricted form~(\ref{eq:formalt}), we find respectively
\begin{equation}
\begin{aligned}
\label{eq:formns}
&1+ \delta({\bf \Theta}, {\bf \Theta}_0; k;a)  \\
&\simeq 
\left[ 1+  (\delta/\gamma)_{\cal N}  \frac{\Delta {\cal N}}{\bar{\cal N}} \right]
  \left[1+  (\delta/\Gamma)_{n_s} \Gamma \right]
\left[ 1+  (\delta/\gamma)_{D}  \frac{\Delta D^2}{\bar{D}^2}  \right]
\left[ 1+  (\delta/\gamma)_{T}  \frac{\Delta T^2}{\bar{T}^2}  \right],
  \end{aligned}
  \end{equation}
and
\begin{equation}
\begin{aligned}
\label{eq:formaltns}
1+ \delta({\bf \Theta}, {\bf \Theta}_0; k;a)  
\simeq \frac{ 1+  (\delta/\gamma)_{\cal N}  \frac{\Delta {\cal N}}{\bar{\cal N}} }{ 1+  (\delta/\gamma)_{\cal N}  \frac{\Delta {\cal N}_1}{\bar{\cal N}} }
  \left[1+  (\delta/\Gamma)_{n_s} \Gamma \right]
\left[1+  (\delta/\gamma)_{\omega_{\rm m}}  \gamma_{\omega_{\rm m}} \right]
\frac{ 1+  (\delta/\gamma)_{D}  \frac{\Delta D^2}{\bar{D}^2}  }{ 1+  (\delta/\gamma)_{D}  \frac{\Delta D_1^2}{\bar{D}^2} }.
  \end{aligned}
  \end{equation}
Our fitting function for the relative matter power spectrum,  {\sc RelFit},  will be based  upon these expressions;  we shall determine the functional forms for $(\delta/\Gamma)_{n_s}$ and $(\delta/\gamma)_{X}$ in section~\ref{sec:fitting}.


\section{{\sc RelFit} fitting functions}
\label{sec:fitting}

That the ratio $\delta/\gamma$ should take on an essentially cosmology-independent form under variation of one cosmological model parameter or proxy is perhaps not very surprising  upon scrutiny. As the top and middle panels of figures~\ref{fig:w} to~\ref{fig:asomegamns} demonstrate, the current generation of observations on linear scales already constrains cosmology to the extent that  $\delta, \gamma,  \Delta X/\bar{X} \ll 1$. Such tight constraints imply
 that perturbing $P({\bf \Theta})$ in $X$ around a reference model~${\bf \Theta}_0$ will always yield to leading order in  $\Delta X/\bar{X}$ a linear dependence of~$\delta$ on~$\Delta X/\bar{X}$, i.e.,
 \begin{equation}
 \label{eq:taylorexpansion}
\delta({\bf \Theta}, {\bf \Theta}_0) = \sum_X \left. \frac{\partial \ln P}{\partial \ln X}\right|_{{\bf \Theta}={\bf \Theta}_0} \frac{\Delta X}{\bar{X}},
\end{equation}
regardless of the exact functional dependence of $P({\bf \Theta}; k;a)$ on $X$.  

Furthermore, while the functional derivatives $\left. \partial \ln P/\partial \ln X\right|_{{\bf \Theta}={\bf \Theta}_0}$  
depend in principle on our choice of expansion point~${\bf \Theta}_0$, the correction incurred by choosing a different ${\bf \Theta}_0$ must be $\ll {\cal O}(1)$ if the new expansion point remains within the observationally allowed range.  Thus, in this restricted sense the derivatives $\left. \partial \ln P/\partial \ln X\right|_{{\bf \Theta}={\bf \Theta}_0}$ are essentially ``approximately universal'', and we identify them with the approximate-universal forms~$(\delta/\gamma)_X$ defined in section~\ref{sec:universality}.  Then, to first order in small $\Delta X/\bar{X}$, equations~(\ref{eq:form}) and~(\ref{eq:taylorexpansion}) are the same.

Identifying the approximate-universal forms $(\delta/\gamma)_X$ with finite-difference estimates of the  functional derivatives of  $P({\bf \Theta}; k;a)$  immediately suggests that a reasonable approximation of their functional forms can be established using  as few as two simulations per family~$X$, where $\Delta X/\bar{X}$ should be chosen to be as close to zero as is permitted by the precision limitations.  Then, the full $w$CDM fitting function can in principle be constructed with as few as five simulations in total. Given however that we have already at our disposal a set of some 20 simulations, we opt instead to compute the derivatives based on double-sided estimation, which ups the number of required simulations to nine.

In finding functional forms for $(\delta/\gamma)_X$  we adopt a strictly empirical approach and simply match rational functions to our simulated spectra, irrespective of their limiting behaviours on very small scales.  This also means that  extrapolating {\sc RelFit} to outside the calibration $k$-region may return nonsensical results.
We note however that our simulated $(\delta/\gamma)_X$, especially $X={\cal N}, D$, appear to exhibit small-scale behaviours consistent with the stable clustering ansatz~\cite{Hamilton:1991es}.  This suggests that a physically motivated fitting function could be constructed from, e.g., recasting the well-known Peacock--Dodds fitting formula~\cite{PD} as a fitting function directly for $(\delta/\gamma)_X$. We refer the interested reader to appendix~\ref{sec:pd} for details.


\subsection{Functional forms for $(\delta/\gamma)_X$}
\label{sec:fitforms}

Following the findings of section~\ref{sec:prelim},  we choose as the independent variable in our fitting functions
\begin{equation}
\label{eq:y}
y(k,a) \equiv \frac{k}{k_{\rm peak}(a)},
\end{equation}
where $k_{\rm peak}$ specifies the  locations of the peak features in  $(\delta/\gamma)_{{\cal N}, \omega_{ \rm m},D}$.   Interpolating our simulation outputs at $a= 0.85,0.7,0.5,0.3$, we find $k_{\rm peak}$  to be well described by 
\begin{equation}
\label{eq:kpeak}
k_{\rm peak} (a) = \left[\frac{k_\sigma (a) }{h/{\rm Mpc}} \right]^{0.65} \ h/{\rm Mpc}, 
\end{equation}
with an $a$-dependent $k_\sigma\equiv 1/x$ defined by the condition
\begin{equation}
\label{eq:sigma}
\sigma^2(x=k_\sigma^{-1},a) = \frac{1}{2 \pi^2} \int d \ln k \ k^3 P_{\rm L}({\bf \Theta}_0; k;a) \ e^{-k^2 x^2} = 1
\end{equation}
evaluated for  the reference cosmological model.  For the reference $\Lambda$CDM cosmology of table~\ref{tab:params}, $k_\sigma/(h/{\rm Mpc}) \simeq 0.844,1.11,2.06,7.92$ at $a= 0.85,0.7,0.5,0.3$ respectively.%


\subsubsection{$X={\cal N}, \omega_{\rm m},D,n_s$}
\label{sec:restrictedfit}

We use a subset of the $N$-body simulation results of table~\ref{tab:runs2} to calibrate $(\delta/\gamma)_X$ in the wavenumber range $k = 0.05 \to 10 \ h$/Mpc at  $a=0.85, 0.7, 0.50, 0.30$.   Specifically,  we use relative matter power spectra formed from the  pairs:
\begin{itemize}
\item $X={\cal N}$: \{{\tt 1024}$A_{s,l}$, {\tt 1024Ref}\},  \{{\tt 1024}$A_{s,h}$, {\tt 1024Ref}\}; 
\item 
$X=\omega_{\rm m}$:  \{{\tt 1024}$\omega_{{\rm m},l}$, {\tt 1024Ref}\},  \{{\tt 1024}$\omega_{{\rm m},h}$, {\tt 1024Ref}\};
\item  $X=D$: \{{\tt 1024}$w2$, {\tt 1024Ref}\},   \{{\tt 1024}$w4$, {\tt 1024Ref}\}; 
\item $X=n_s$:  \{{\tt 1024}$n_{s,l}$, {\tt 1024Ref}\}, \{{\tt 1024}$n_{s,h}$, {\tt 1024Ref}\}.
\end{itemize}
At each scale factor~$a$, we construct for each pair the corresponding ratio $\delta/\gamma$ and combine them to form a mean $(\delta/\gamma)_X$ for each family $X$ weighted by the inverse of  the linear relative power spectrum, $|\gamma|^{-1}$,  at that scale factor.

\begin{figure}[t]
\begin{center}
\includegraphics[width=15.4cm]{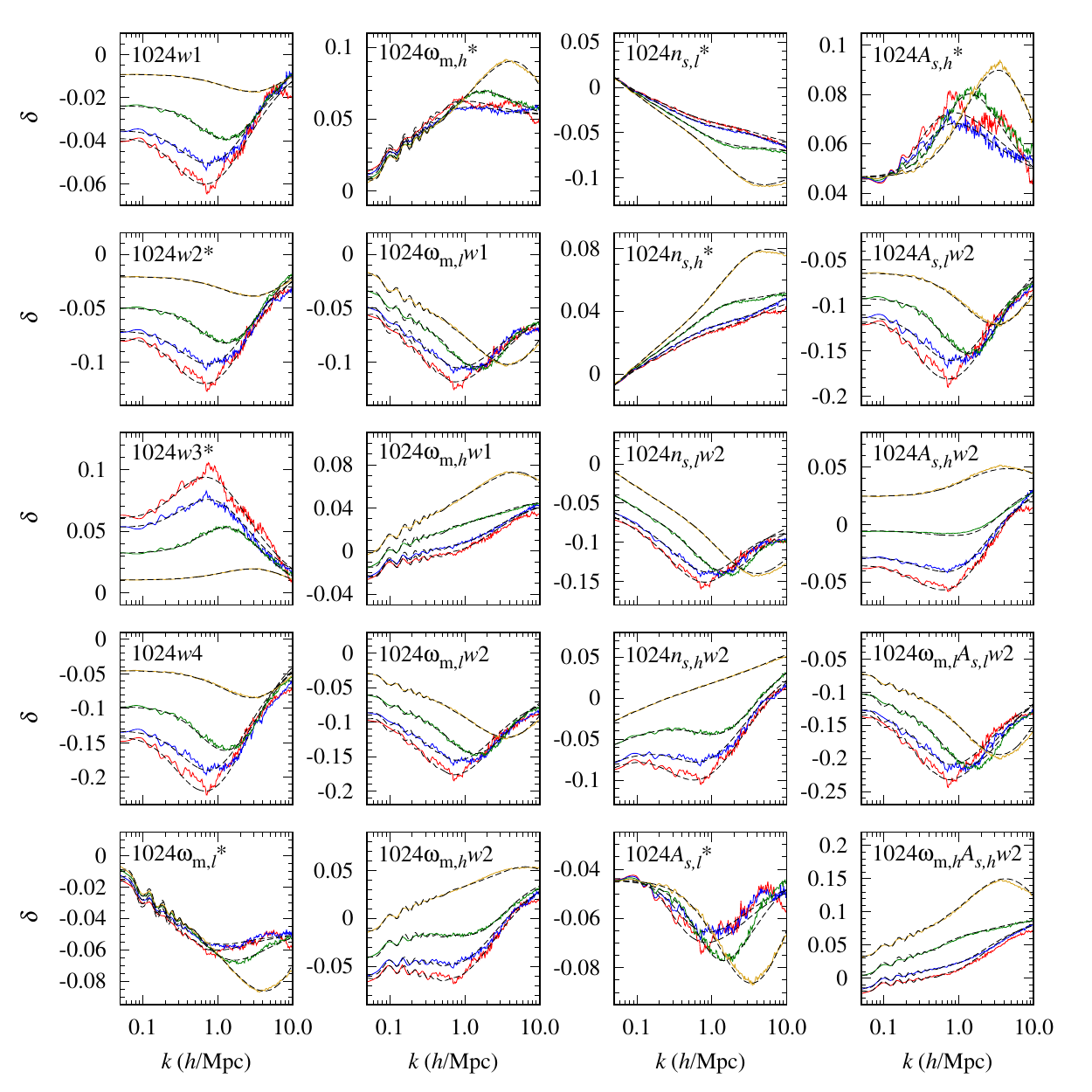}
\end{center}
\caption{Relative matter power spectra constructed from the simulations of table~\ref{tab:runs2}  with respect to the reference simulation {\tt 1024Ref} at $a=0.85$ (red), $0.70$ (blue), $0.50$ (green), and $0.30$ (yellow).  The black dashed lines denote predictions of the restricted form of {\sc RelFit} based on equation~(\ref{eq:formaltns}).  An asterisk denotes a simulation that has been used to calibrate {\sc RelFit}.\label{fig:calibration}}
\end{figure}

We fit each weighted mean $(\delta/\gamma)_X$ using rational functions of quadratic polynomials in $\log_{10} y$, where the fitting coefficients are themselves functions of the scale factor~$a$.
In all cases,  $(\delta/\gamma)_X$ must converge to the predictions of linear theory at $k \to 0$,  a condition we explicitly enforce in all of our fitting functions by tuning down the rational functions with a $1-e^{-y}$ factor.  Specifically,  for variations in $X={\cal N},\omega_{\rm m},D$, we use the functional form
\begin{equation}
\label{eq:functionalform}
(\delta/\gamma)_{X} \simeq 1+ \left(1-e^{-y} \right) \frac{b_0^X + b_1^X \log_{10} y + b_2^X (\log_{10} y)^2}{ 1 + c_1^X \log_{10} y + c_2^X  (\log_{10} y)^2 },
\end{equation}
while variations in $n_s$ are well described by
\begin{equation}
(\delta/\Gamma)_{n_s} \simeq e^{-y} \ln (k/k_{\rm piv}) + \left(1-e^{-y} \right) \frac{b_0^{n_s} + b_1^{n_s} \log_{10} y + b_2^{n_s} (\log_{10} y)^2}{ 1 + c_1^{n_s} \log_{10} y + c_2^{n_s}  (\log_{10} y)^2 }.
\end{equation}
In all cases $X={\cal N}, \omega_{\rm m},D,n_s$, the coefficients  $b_{0,1,2}^X = b_{0,1,2}^X(a)$ and $c_{1,2}^X = c_{0,1,2}^X(a)$ are polynomials of the scale factor $a$ alone given in appendix~\ref{sec:coefficients}, and we remind the reader again that no attempts have been made to model the $k \to \infty$ behaviours of the fitting functions.

\begin{figure}[t]
\begin{center}
\includegraphics[width=11.cm]{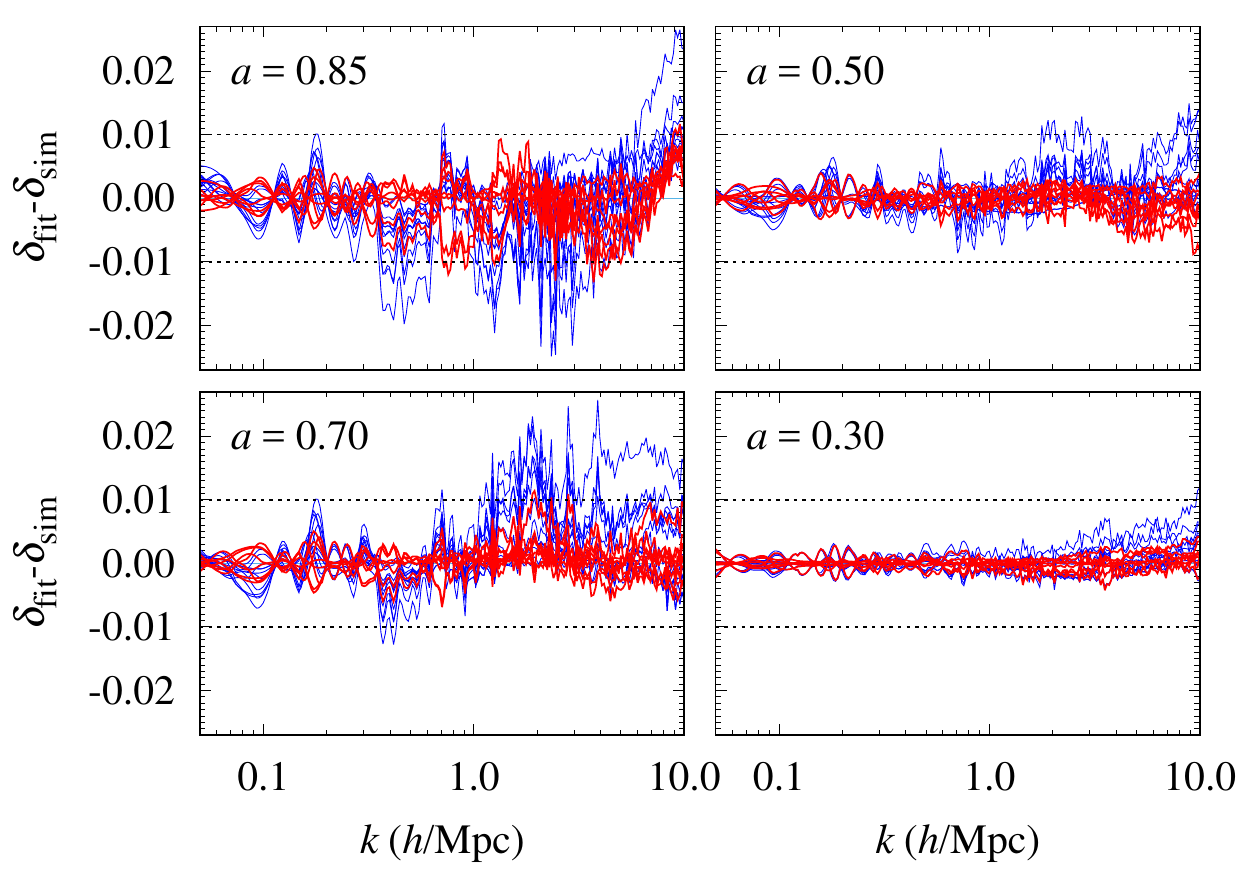}
\end{center}
\caption{Fitting errors at $a=0.85,0.70, 0.50,0.30$. The coloured lines represent the differences between the predictions of the restricted form of {\sc RelFit},~$\delta_{\rm fit}$, based on equation~(\ref{eq:formaltns}), and the relative matter power spectra,~$\delta_{\rm sim}$, formed from the simulations of table~\ref{tab:runs2}.  Red lines denote the subset of eight relative power spectra used to calibrate {\sc RelFit}, while the blue lines denote the remaining 12 relative power spectra not used for  calibration. 
 \label{fig:error}}
\end{figure}

Figure~\ref{fig:calibration} shows the predictions of the restricted form of {\sc RelFit}, $\delta_{\rm fit}$, based on equation~(\ref{eq:formaltns}), against the relative matter power spectra,~$\delta_{\rm sim}$, constructed from the simulations of table~\ref{tab:runs2}; figure~\ref{fig:error} shows the corresponding fitting errors formed from their differences.  The fit is across the board excellent.  At $a=0.85$ and  for the whole range of wavenumbers explored, no individual error exceeds  0.01 in magnitude for the eight calibration models, or exceeds 0.025 for the remaining 12 models not used in the  calibration of {\sc RelFit}.
 The fit improves as we move to smaller scale factors:  at $a=0.30$, the fitting error is always well below 0.01 for the entire $k$-range.  

The reasoning behind {\sc RelFit} together with the parameter dependence of the linear matter power spectrum in $w$CDM cosmologies, equation~(\ref{eq:linearpowerspectrum}), 
also suggests that varying the dimensionless Hubble parameter~$h$ should produce an effect on the nonlinear matter power spectrum identical to varying the linear growth function~$D$.   Likewise, {\sc RelFit} in its present form imposes no restriction on the time-dependence of the dark energy equation of state parameter.  These scenarios will be explored further in sections~\ref{sec:emu} and~\ref{sec:extendedmodels} respectively.

\subsubsection{$X=T$}
\label{sec:transfer}

While none of the pairs of simulations in~table~\ref{tab:runs2} models explicitly a variation in the linear transfer function~$T$ alone, following the arguments of section~\ref{sec:further} it is possible to construct a fitting function for $(\delta/\gamma)_{T}$ using a combination of our set of {\tt 1024}$\omega_{\rm m}$ simulations and the fitting functions derived in section~\ref{sec:restrictedfit}.  With $(\delta/\gamma)_T$ available,   a more general form of {\sc RelFit}  based on equation~(\ref{eq:formns}) could be achieved, potentially widening the applicability of the fitting function also to target cosmologies involving variations in the physical baryon density $\omega_{{\rm b}}$ (section~\ref{sec:emu}) as well as in the effective number of neutrinos~$N_{\rm eff}$ (section~\ref{sec:extendedmodels}).

Recall that varying $\omega_{\rm m}$ changes simultaneously the normalisation ${\cal N}$, the linear transfer function~$T$, and the linear growth function $D$.  Then, beginning with the relative nonlinear matter power spectra formed from the pairs \{{\tt 1024}$\omega_{{\rm m},l}$,  {\tt 1024Ref}\} and  \{{\tt 1024}$\omega_{{\rm m},h}$, {\tt 1024Ref}\}, a simple procedure based on equation~(\ref{eq:matteronly}) can be used to recover $(\delta/\gamma)_{T}$ in each case: 
\begin{enumerate}
\item Compute the variation in $D$ due to the change in $\omega_{\rm m}$ alone, and use it in {\sc RelFit} to calculate the corresponding nonlinear variation;
\item Repeat the above for the nonlinear variation in~${\cal N}$ due to $\omega_{\rm m}$;
\item Remove the $D$ and ${\cal N}$ contributions of steps 1 and 2 from the  simulated relative nonlinear matter power spectrum via equation~(\ref{eq:matteronly}) to form the relative nonlinear power spectrum under variations in the linear transfer function $T$ alone;
\item Divide the relative nonlinear power spectrum of step 3 through by its linear counterpart to form $(\delta/\gamma)_T$.
\end{enumerate}
We have repeated this process for the two pairs of relative power spectra, formed a weighted average as described in section~\ref{sec:restrictedfit}, and fit it using a rational function of the form~(\ref{eq:functionalform}).  The resulting fitting coefficients  $b_{0,1,2}^T = b_{0,1,2}^T(a)$ and $c_{1,2}^T = c_{0,1,2}^T(a)$ can be found in appendix~\ref{sec:coefficients}.

\begin{figure}[t]
\begin{center}
\includegraphics[width=11.cm]{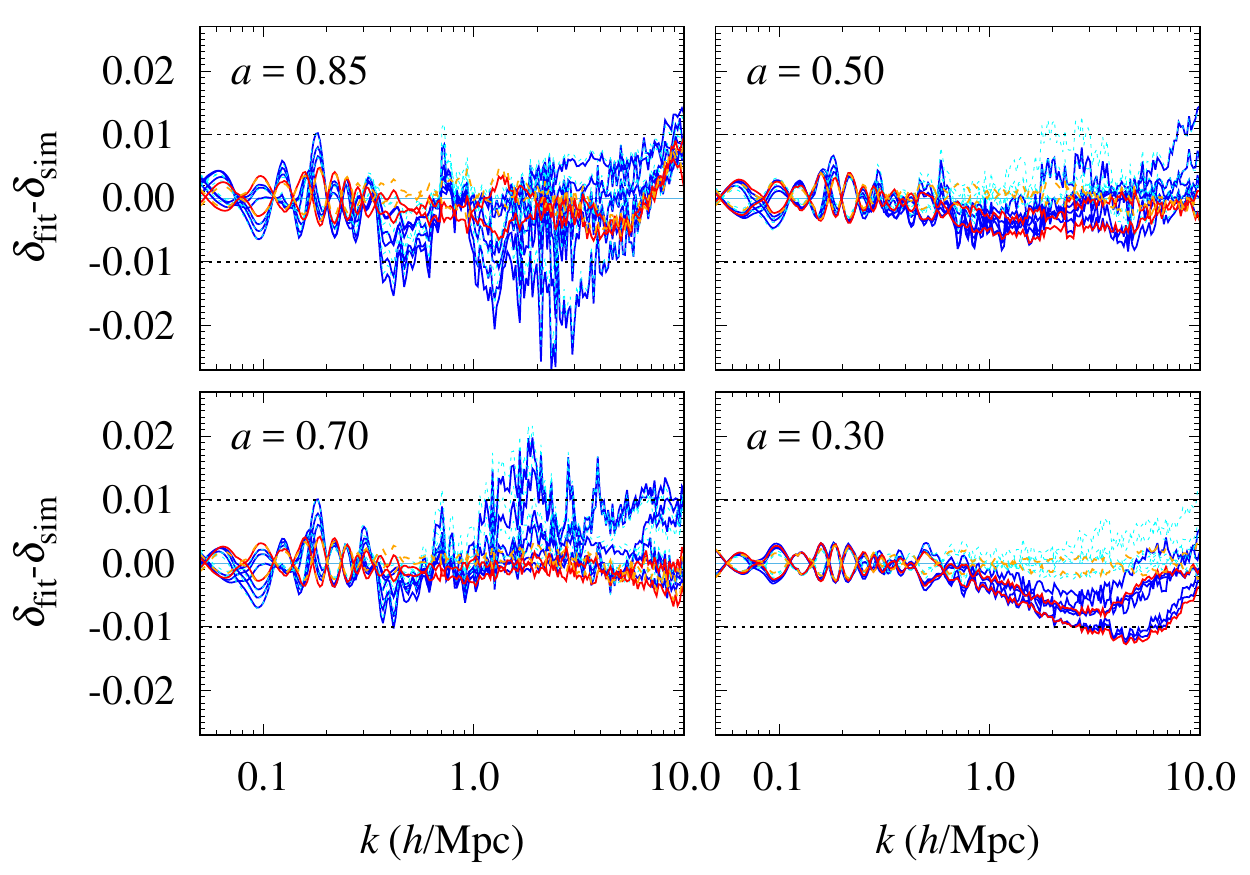}
\end{center}
\caption{Same as figure~\ref{fig:error}, except the predictions of {\sc RelFit}, $\delta_{\rm fit}$, have been computed from its general form~(\ref{eq:formns}).
Red lines denote the relative power spectra formed from {\tt 1024}$\omega_{{\rm m},l}$ and {\tt 1024}$\omega_{{\rm m},h}$ relative to {\tt 1024Ref} used to calibrate {\sc RelFit}, while the blue lines denote the remaining six {\tt 1024}$\omega_{{\rm m}}{\tt XXX}$ relative power spectra not used for  calibration.  For comparison we also show the corresponding fitting errors arising from the restricted form of {\sc RelFit}, equation~(\ref{eq:formaltns}), in orange and cyan.~\label{fig:error-full}}
\end{figure}

Figure~\ref{fig:error-full} shows the fitting errors of the general form of {\sc RelFit}, equation~(\ref{eq:formns}), for the eight {\tt 1024}$\omega_{{\rm m}}${\tt XXX} simulations of table~\ref{tab:runs2} relative to  {\tt 1024Ref}.\footnote{Recall that relative matter power spectra formed from pairs of simulations {\it without} variations in $\omega_{\rm m}$ are not affected by the choice between the general and the restricted form of {\sc RelFit}.}
Again, we see that the fit is across the board excellent, and at $a=0.85, 0.70, 0.50$, comparable to that of the restricted form (figure~\ref{fig:error}).  At $a=0.30$, however, the general form of {\sc RelFit} appears to systematically underestimate the simulated power spectra at $k \gtrsim 1\, h$/Mpc by some 0.005 to 0.01.  This is likely an artefact of the admittedly convoluted method with which we have extracted~$(\delta/\gamma)_{T}$ in this section, and can potentially be improved with calibrations against dedicated simulations in which only the linear transfer function is varied.
 We shall defer this exercise to a later publication.  Suffice it to say here that figure~\ref{fig:error-full} demonstrates the robustness of the general strategy of fitting function construction proposed in this work.


\subsection{Application to extended models}
\label{sec:extendedmodels}

The form of {\sc RelFit}---phrased in terms of variations in the linear transfer function, linear growth function, etc.---suggests that its applicability extends beyond the $w$CDM cosmologies  we have used to calibrate its free parameters.
In order to test this possibility, we have performed an additional set of simulations using {\sc Gadget}-2/{\sc Camb},  detailed in table~\ref{tab:runs3}, that go beyond $w$CDM in two different ways: (i) a time-dependent dark energy equation of state parameter~$w(a)$, which at the linear level affects only the growth function, and (ii) a linear transfer function modified by a non-canonical effective number of neutrinos $N_{\rm eff}$.

\begin{table*}[t]
\begin{center}
{\footnotesize
  \hspace*{0.0cm}\begin{tabular}
  {lccccccc} \hline \hline
  Run & $L (h^{-1}$Mpc) & $N$ & $z_i$ & $r_s(h^{-1}$kpc)  & $w_0$ & $w_a$ & $N_{\rm eff}$ \\       \hline
		{\tt 1024$w2w_a1$} & 256 & $1024^3$ &  49  & 6  & $-0.85$  & $0.3$ & 3.04 \\
	{\tt 1024$w2w_a2$} & 256 & $1024^3$ &  49  & 6   & $-0.85$  & $0.2$ & 3.04\\
	{\tt 1024$w2w_a3$} & 256 & $1024^3$ &  49  & 6 & $-0.85$  & $0.1$ &3.04\\
	{\tt 1024$w2w_a4$} & 256 & $1024^3$ &  49  & 6 & $-0.85$  & $-0.1$ & 3.04\\
\hline
{\tt 1024}$N_{\rm eff}3.3$ & 256 & $1024^3$ &  49  & 6 & $-1.00$  & $0.0$ & 3.34 \\
{\tt 1024}$N_{\rm eff}4.0$ & 256 & $1024^3$ &  49  & 6  & $-1.00$  & $0.0$ & 4.04 \\
  \hline \hline
  \end{tabular}
  }
  \end{center}
    \caption{Additional simulations of extended cosmological models performed using {\sc Gadget}-2/{\sc Camb}, used in section~\ref{sec:extendedmodels} as blind tests (i.e., not calibration) of {\sc RelFit}: $L$ is the simulation box length, $N$ the number of simulation particles, $z_i$ the initial redshift, $r_s$ the gravitational softening length,  $N_{\rm eff}$ is the effective number of neutrinos, while  $\{w_0, w_a \}$ replace $w$ to parameterise a possible time dependence of the dark energy equation of state by way of equation~(\ref{eq:wa}).  All other cosmological parameters not listed here are held at their reference $\Lambda$CDM values given in table~\ref{tab:params}.~\label{tab:runs3}}
\end{table*}

\paragraph{(i) Time-dependent dark energy equation of state} Dynamical dark energy models such as quintessence typically predict effective equations of state for the dark energy component that change with time~(e.g.,~\cite{Copeland:2006wr}).
 The exact time dependence varies from model to model.
Here, we use for simplicity  a time dependence parameterised by~\cite{Chevallier:2000qy,Linder:2002et}
 \begin{equation}
 \label{eq:wa}
 w(a) = w_0 + w_a (1-a),
\end{equation} 
where we fix $w_0=-0.85$, but allow $w_a$ to vary in the interval $w_a \in [-0.1, 0.3]$ in our simulations. Current cosmological measurements do not provide strong constraints on the time dependence of  $w(a)$, and the models represented by our choices of $w_a$ values, while spanning a parameter range comparable to only about 1.5 times the standard deviation inferred from the 2018 Planck+SNe+BAO data~\cite{Aghanim:2018eyx}, do in fact deviate strongly from the reference $\Lambda$CDM cosmology in their matter power spectrum predictions.

Extending {\sc RelFit}  to include a time-dependent dark energy equation of state parameter simply requires that we redefine the linear growth function variations $\Delta D^2$ and $\Delta D_1^2$ that appear in equations~(\ref{eq:formns})~and~(\ref{eq:formaltns}) as
\begin{equation}
\begin{aligned}
\Delta D^2 \to \Delta D^{2} & \equiv D^{2} - \bar{D}^{2} \equiv  D^2(w(a), \omega_{\rm m},h;a)-D^2(\bar{w}(a), \bar{\omega}_{\rm m},h;a) ,\\
\Delta D^2_1 \to \Delta D_1^2 &\equiv D_1^2 - \bar{D}^2  \equiv D^2(\bar{w}(a), \omega_{\rm m},\bar{h};a)-D^2(\bar{w}(a), \bar{\omega}_{\rm m},\bar{h};a) ,
\end{aligned}
\end{equation}
where $\bar{w}(a)$ denotes  the reference $\Lambda$CDM choices of $\bar{w}_0 =-1$ and $\bar{w}_a=0$ in equation~(\ref{eq:wa}).


\paragraph{(ii) Non-canonical effective number of neutrinos}

Any light thermal particle species that decouples while relativistic will behave in the cosmological context essentially like a standard-model neutrino, and contribute to the non-photon radiation energy density, conventionally parameterised as the effective number of thermalised neutrinos $N_{\rm eff}$.  Well-known examples of such particle species 
include sterile neutrinos and axions~(e.g.,~\cite{Hannestad:2015tea,Archidiacono:2015mda}).

Phenomenologically, increasing $N_{\rm eff}$ alone shifts the epoch of matter--radiation equality to a lower redshift according to~\cite{Abazajian:2012ys},
\begin{equation}
1+z_{\rm eq} = \frac{\omega_m}{\omega_\gamma} \frac{1}{1+0.227 N_{\rm eff}},
\end{equation}
where $\omega_\gamma$ is the present-day photon energy density.  
For the linear matter power spectrum, changes in $z_{\rm eq}$ are manifested primarily as a shift in the location of the turning point~$k_{\rm eq}$ according to
\begin{equation}
\begin{aligned}
\label{eq:keq}
k_{\rm eq}  & \equiv a_{\rm eq} H(a_{\rm eq}) \\
& \simeq 4.7 \times 10^{-4} \sqrt{\omega_{\rm m} (1+z_{\rm eq})}\;\; {\rm Mpc}^{-1},
\end{aligned}
\end{equation}
which, within the structure of {\sc RelFit}, is captured by a variation in the linear transfer function. Then,
incorporating $N_{\rm eff}$ into {\sc RelFit} simply requires that we use the general form (\ref{eq:formns}) of the fitting function together with 
\begin{equation}
\begin{aligned}
\Delta T^2 \to \Delta T^{2}  \equiv T^{2} - \bar{T}^{2} \equiv  T^2(\omega_{\rm m},\omega_{\rm b},N_{\rm eff}; k)-T^2(\bar{\omega}_{\rm m}, \bar{\omega}_b,\bar{N}_{\rm eff}; k) ,
\end{aligned}
\end{equation}
where the linear transfer function~$T$ is now a function of three cosmological parameters.

 The 2018 Planck+external data combination currently constrains $N_{\rm eff}$ most tightly to
$N_{\rm eff} = 2.99^{+0.34}_{-0.33}$ (95\% C.I.)~\cite{Aghanim:2018eyx}; 
our two choices of $N_{\rm eff}=3.34$ and $N_{\rm eff}= 4.04$ in table~\ref{tab:runs3} therefore represent respectively a $2\sigma$ and a $20 \sigma$ variation away from the 2018 Planck best-fit.

\bigskip

\begin{figure}[t]
\begin{center}
\includegraphics[width=14.cm]{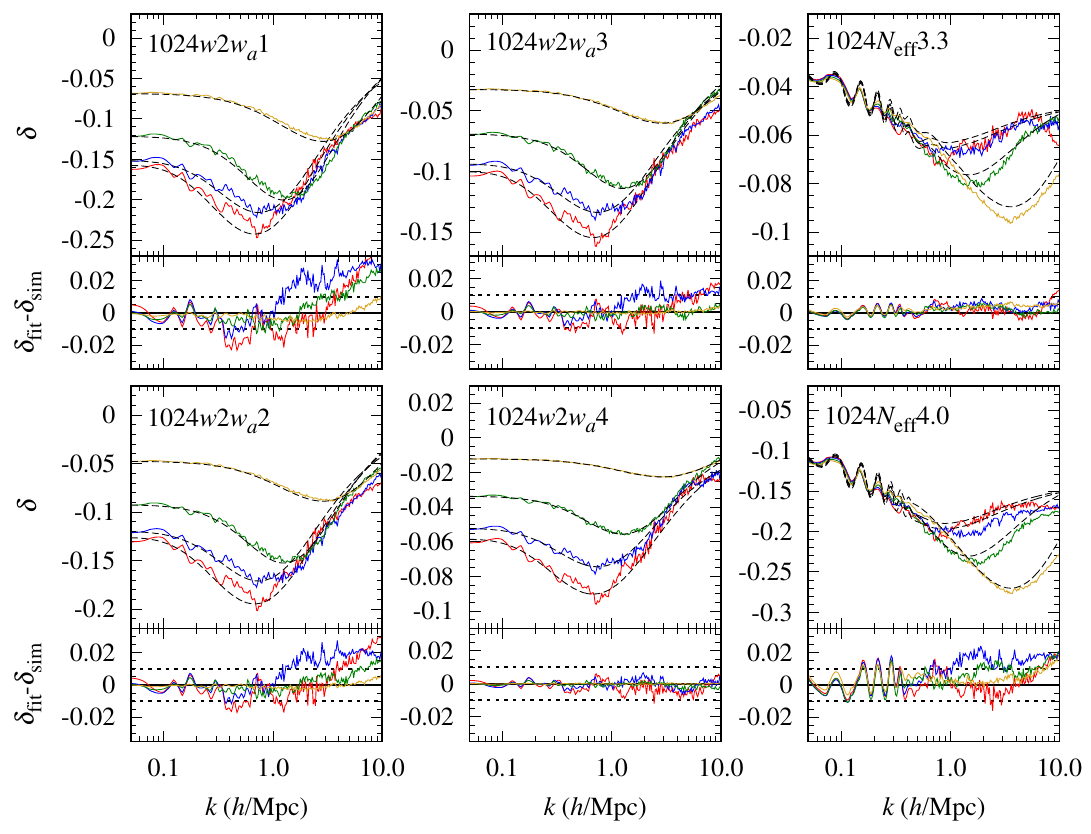}
\end{center}
\caption{Relative matter power spectra constructed from the simulations of table~\ref{tab:runs3}  with respect to the reference simulation {\tt 1024Ref} at $a=0.85$ (red), $0.70$ (blue), $0.50$ (green), and $0.30$ (yellow).  The black dashed lines denote the predictions of {\sc RelFit}.\label{fig:additional}}
\end{figure}

\noindent Figure~\ref{fig:additional} shows the predictions of {\sc RelFit}---as calibrated originally in section~\ref{sec:fitforms}---against the relative matter power spectra constructed from the simulations of table~\ref{tab:runs3}, together with the corresponding fitting errors.  Again, the differences between the predictions of {\sc RelFit} and the simulated relative power spectra up to $k \sim  1\, h$/Mpc generally do not exceed about 0.01; in the case of {\tt 1024}$N_{\rm eff}4.0$, the large fluctuations around zero seen at  $k \sim  0.1 \to 1\, h$/Mpc are a consequence of nonlinear corrections to the baryon acoustic oscillations, which in principle can be modelled approximately using a suppression factor (as has been implemented in, e.g., {\sc HMCode}~\cite{Mead:2015yca}, but not in {\sc RelFit}).  

Beyond $k \gtrsim  1\, h$/Mpc the fitting errors tend to increase, although for most $w(a)$ and $N_{\rm eff}$ cosmologies tested here the {\sc RelFit} predictions  still fall within 0.02 of the simulation results.
 The only exception is  the case of {\tt 1024}$w2w_{a}1$, where at $k \gtrsim  4\, h$/Mpc the deviation is up to~0.03.  We note however that the particular $w(a)$ cosmology represented by this simulation is fairly far away from the $\Lambda$CDM reference cosmological model in terms of the deviation of its linear matter power spectrum from the reference case ($\gtrsim15$\% at $a \geq 0.70$).   Given the ``perturbative'' nature of {\sc RelFit}, it is perhaps not surprising that its simple linear prescription should break down at large wavenumbers.  
 
We conclude section~\ref{sec:extendedmodels} with the emphasis  that none of the simulations of table~\ref{tab:runs3} has been used to calibrate {\sc RelFit}.  In particular, the  $(\delta/\gamma)_{T}$ fitting function that forms the basis of the {\sc RelFit} predictions in the two $N_{\rm eff}$ scenarios has  been extracted from a combination of target cosmology simulations that have nothing to do with varying $N_{\rm eff}$ at face value.   That {\sc RelFit} is still capable of predicting to $0.01 \to 0.02$ the relative power spectra of these target cosmologies speaks again for the general soundness of our strategy.


\subsection{Comparison with {\sc CosmicEmu}, {\sc Halofit},  and {\sc HMCode}}
\label{sec:emu}

\subsubsection{Single-parameter variations}
\label{sec:singleparametervariation}

The essence of {\sc RelFit} is a set of first-order logarithmic functional derivatives of the nonlinear matter power spectrum $P({\bf \Theta}; k; a)$ with respect to variations in the linear matter power spectrum $P_{\rm L}({\bf \Theta}; k; a)$ evaluated at the reference cosmology ${\bf \Theta} = {\bf \Theta}_0$.  Predicting a target nonlinear $P({\bf \Theta}; k; a)$ relative to the reference $P({\bf \Theta}_0; k; a)$ simply consists in multiplying these derivatives with the relevant variations in the linear $P_{\rm L}({\bf \Theta}; k; a)$ away from the reference $P_{\rm L}({\bf \Theta}_0; k; a)$.  One immediately concludes that the smaller the linear variations a target cosmology produces, the higher the fidelity of {\sc RelFit} in predicting its nonlinear variations.

We take as a formal assessment of ``smallness''  the fractional variation in the linear matter power spectrum at the ``peak'' wavenumber $k_{\rm peak}$, defined in equation~(\ref{eq:kpeak}), of the reference cosmology. Then, at $a=0.85$ and for single-parameter variations, a maximum 10\% (15\%) variation corresponds to target cosmological parameter values falling in the region
\begin{equation}
\begin{aligned}
\label{eq:singleparametermodels}
(0.1322)\; 0.1351 &\leq \omega_{\rm m} \leq 0.1493 \; (0.1522), \\ 
(1.868)\;  1.978 & \leq 10^9 A_s \leq 2.418  \; (2.495), \\
 [(0.7784) \; 0.8010 &  \leq \sigma_8  \leq 0.8855 \; (0.9)], \\
(0.90) \; 0.92 & \leq n_s \leq 1.00 \; (1.02), \\
(-1.3) \; -1.26 & \leq w \leq -0.82\;  (-0.75), \\
(0.55) \; 0.585 & \leq h \leq 0.775 \; (0.83), \\
0.0215 & \leq \omega_{\rm b} \leq 0.0235,
\end{aligned}
\end{equation} 
where the equivalent $\sigma_8$ range assumes all parameters but $A_s$ held  fixed at their reference values, and we have included in this list the (as-yet-unexplored) Hubble parameter~$h$ and  physical baryon density~$\omega_{{\rm b}}$.  Where applicable the parameter region~(\ref{eq:singleparametermodels}) is larger than that of equation~(\ref{eq:calibrationmodels})  used to establish  {\sc RelFit}, while the $\omega_{\rm b}$ range, representing $2\to -5$\% variations in the linear matter power spectrum, has been chosen so as to stay within the confines of  the Mira--Titan  simulations~\cite{Heitmann:2015xma,Lawrence:2017ost}.%
\footnote{In the same vein,  $10^9 A_s=2.495$, or equivalently, $\sigma_8 = 0.9$, represents only a 13.4\% variation from the reference cosmology, and $w=-1.3$  only  11.4\%  at $a=0.85$.}
Simple power counting then suggests that the output of {\sc RelFit} in the region~(\ref{eq:singleparametermodels}) should be accurate to $ \lesssim0.01$ ($ \lesssim 0.02$).

\begin{figure}[!t]
	\begin{center}
		\includegraphics[width=6.9cm]{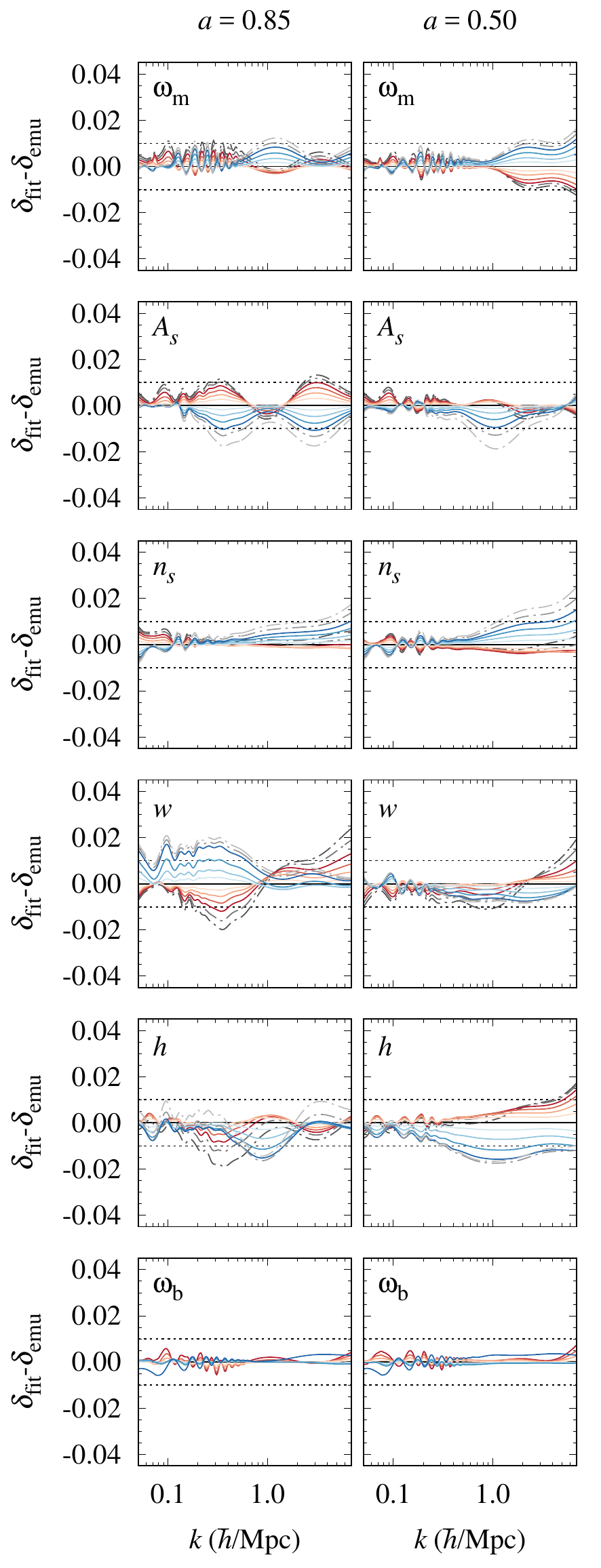}
		\includegraphics[width=6.9cm]{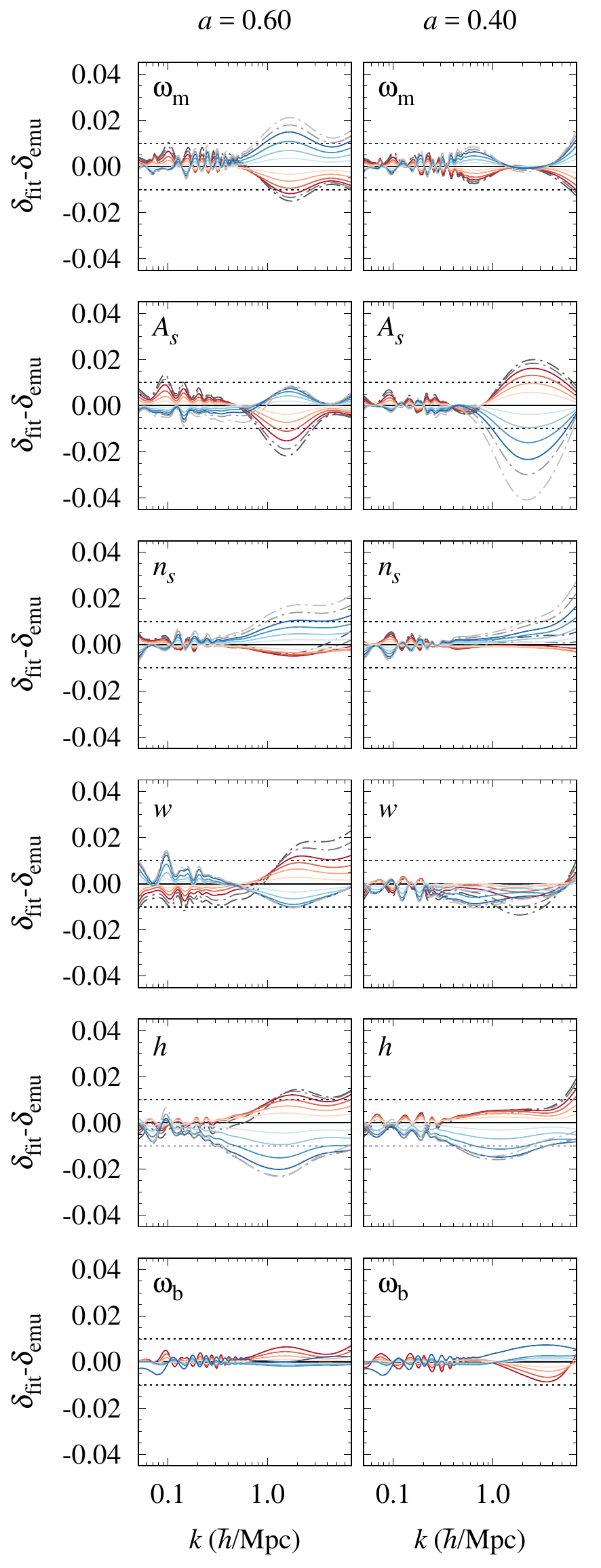}
	\end{center}
	\caption{Differences between the predictions of {\sc RelFit} $\delta_{\rm fit}$ and {\sc CosmicEmu} $\delta_{\rm emu}$ in single-parameter variations  at the calibration scale factors $a=0.85, 0.50$ (left) and  at $a=0.60, 0.40$ {\it not} used for calibration (right).
		Solid lines on the blue-to-red spectrum correspond to the low-to-high end of the ``10\%-variation'' single-parameter ranges of equation (\ref{eq:singleparametermodels}), while the dot-dash lines represent additional target cosmologies encompassed by the ``15\%-variation'' ranges.  Note that wavenumbers and the absolute power spectra 
		have units of $\bar{h}$/Mpc and $({\rm Mpc}/\bar{h})^3$ respectively, where $\bar{h}=0.673$.\label{fig:single}}
\end{figure}

The left panel of  figure~\ref{fig:single} compares  the output of {\sc RelFit} in the parameter region~(\ref{eq:singleparametermodels})
at the calibration scale factors $a=0.85,0.50$, with the predictions of the {\sc CosmicEmu} emulator trained on the Mira--Titan simulations~\cite{Heitmann:2015xma,Lawrence:2017ost}. 
For comparable cosmological parameters, {\sc CosmicEmu} interpolates in the parameter region
 \begin{equation}
\begin{aligned}
\label{eq:singleparametermodelsemu}
0.12 &\leq \omega_{\rm m} \leq 0.155, \\
 0.7 &  \leq \sigma_8  \leq 0.9, \\
0.85 & \leq n_s \leq 1.05, \\
-1.3 & \leq w \leq -0.7, \\
0.55 & \leq h \leq 0.85, \\
0.0215 & \leq \omega_{\rm b} \leq 0.0235,
\end{aligned}
\end{equation} 
which, with the exception of $\omega_{\rm m}$ and $\sigma_8$, is marginally larger than the ``15\%-variation'' parameter region defined in equation~(\ref{eq:singleparametermodels}).
As can be seen, the agreement between {\sc RelFit} and {\sc CosmicEmu} in the region~(\ref{eq:singleparametermodels}) is remarkable: with few exceptions, the two sets of predictions agree to 0.01 (0.02) or better across the whole wavenumber range tested.

The same comparison at the ``off-calibration'' scale factors $a=0.60,0.40$ is shown  in the right panel of figure~\ref{fig:single},
which serves to test the $a$-dependence of the fitting coefficients presented in appendix~\ref{sec:coefficients}.
At $a=0.60$ the agreement between {\sc RelFit} and {\sc CosmicEmu} is as good as or at marginally worse than the ``on-calibration'' comparisons discussed above.  The $a=0.40$ results are likewise concordant for variations in $\omega_{\rm m}$, $n_s$,  $w$, $h$, and $\omega_{\rm b}$ across the whole $k$-range, but appear to diverge  at $k \sim 2\ h$/Mpc by as much as 4\% for variations in $A_s$.  This may be an error of interpolation in {\sc RelFit} consequent to a sparsely sampled $a$-space---recall that we have calibrated {\sc RelFit} at only four instances ($a=0.85, 0.70, 0.50, 0.30$).  Interestingly, however, while 
 {\sc CosmicEmu} uses eight samples in a similar timeframe ($a= 1.0, 0.91, 0.81, 0.70, 0.60, 0.50, 0.38, 0.33$), the particular instance of $a=0.40$ is likewise off-calibration.  To pin down the exact source of discrepancy would require new simulations, which we defer to a later publication.

Lastly, while it may be tempting to interpret figure~\ref{fig:single} as an accuracy test of {\sc RelFit}, it must be kept in mind that {\sc CosmicEmu} itself has a claimed error margin of  4\% on the absolute power spectrum~\cite{Lawrence:2017ost}.   Likewise, relative power spectra formed from its output are in some cases---particularly when the target and reference cosmologies are far apart---demonstrably erroneous by up to 2\% as $k \to 0$, due to convergence to linear perturbation theory not having been explicitly enforced in the emulation process (in contrast to the calibration of {\sc RelFit}, which does respect convergence to linear theory).  Nonetheless, it is encouraging that agreement to $0.01 \to 0.02$ or better can be achieved in a fairly broad parameter region, especially given that {\sc RelFit} and {\sc CosmicEmu} have been calibrated against completely independent simulations  generated from two different $N$-body codes.


\subsubsection{Multi-parameter variations}
\label{sec:multiemu}

\begin{table*}[t]
	\begin{center}
		{\footnotesize
			\hspace*{0.0cm}\begin{tabular}
				{lcccccc|c} \hline \hline
				Model & $\omega_{\rm m}$ & $A_s$ & $n_s$ & $w$ & $h$ & $\omega_{\rm b}$ & $\sigma_8$   \\ 
				\hline	
				M01 & 0.140092 &  2.13003 & 0.914590	 & $-0.858985$ & 0.683514 & 0.022283 &	 0.7725 \\ 
				M02 & 0.145117 & 	2.19429 & 0.918589 & $-1.13693$ &	0.615461 & 0.022034	& 0.8451 \\ 
				M03 & 0.140688 &  2.24912 & 0.993800 & $-0.815920$ & 0.651725 & 0.022436 & 0.8008 \\  
				M04 & 0.143777 & 2.23732 &  0.915382 & $-0.973216$ & 0.606349 & 0.021910 & 0.8206 \\  
				M05 & 0.145488 & 2.35429 &	0.968273 & $-0.849820$	&  0.615829 & 0.022298 &	0.8372 \\  
				M06 & 0.145904 & 2.12986 &	0.990991 & 	$-1.18144$ & 0.698259	& 0.022046	&  0.8976 \\ 
				M07 & 0.139331 & 2.13040 & 	0.911948 &	$-0.889670$	& 0.669114 & 0.022372 & 0.7734 \\  
				M08 & 0.141269 &  2.00676	& 0.966808 & $-0.879664$ & 0.638494 & 0.022906 & 0.7600 \\  
				M09 & 0.148683 & 2.22116 & 0.957208 & $-0.975085$ & 0.588572 & 0.022523 & 0.8362 \\  
				M10 & 0.138605 & 1.99789 & 0.941659 & 	$-1.16323$ & 0.672825 & 0.021899 & 0.8125  \\ 
				\hline \hline
			\end{tabular}
		}
	\end{center}
	\caption{Randomly sampled cosmologies on the ``surface of 15\%-variation'' defined by the parameter ranges~(\ref{eq:singleparametermodels}).\label{tab:random}}
\end{table*}
 
Next we test {\sc RelFit} against {\sc CosmicEmu} in the full  6-parameter space of equation~(\ref{eq:singleparametermodels}).
 To do so we draw 10 sets of six random numbers on a 5-sphere of unit radius centred on the origin, where each  axis represents a cosmological parameter direction.  These random numbers are then rescaled according to the parameter ranges of equation~(\ref{eq:singleparametermodels}), assuming that the reference $\Lambda$CDM cosmology sits at the centre of the sphere.
Table~\ref{tab:random} shows the 10 target cosmologies sampled in this manner.  
 The sampling procedure ensures that all 10 target cosmologies reside on the ``surface of 15\% variation'' away from the reference, where we generically expect the fitting error of {\sc RelFit} to be the largest---up to $\sim 0.02$ by the arguments of section~\ref{sec:singleparametervariation}; by the same token we can expect the errors of {\sc RelFit} to be smaller than $\sim 0.02$ for those cosmologies contained within the surface.  

\begin{figure}[!t]
	\begin{center}
		\includegraphics[width=11.5cm]{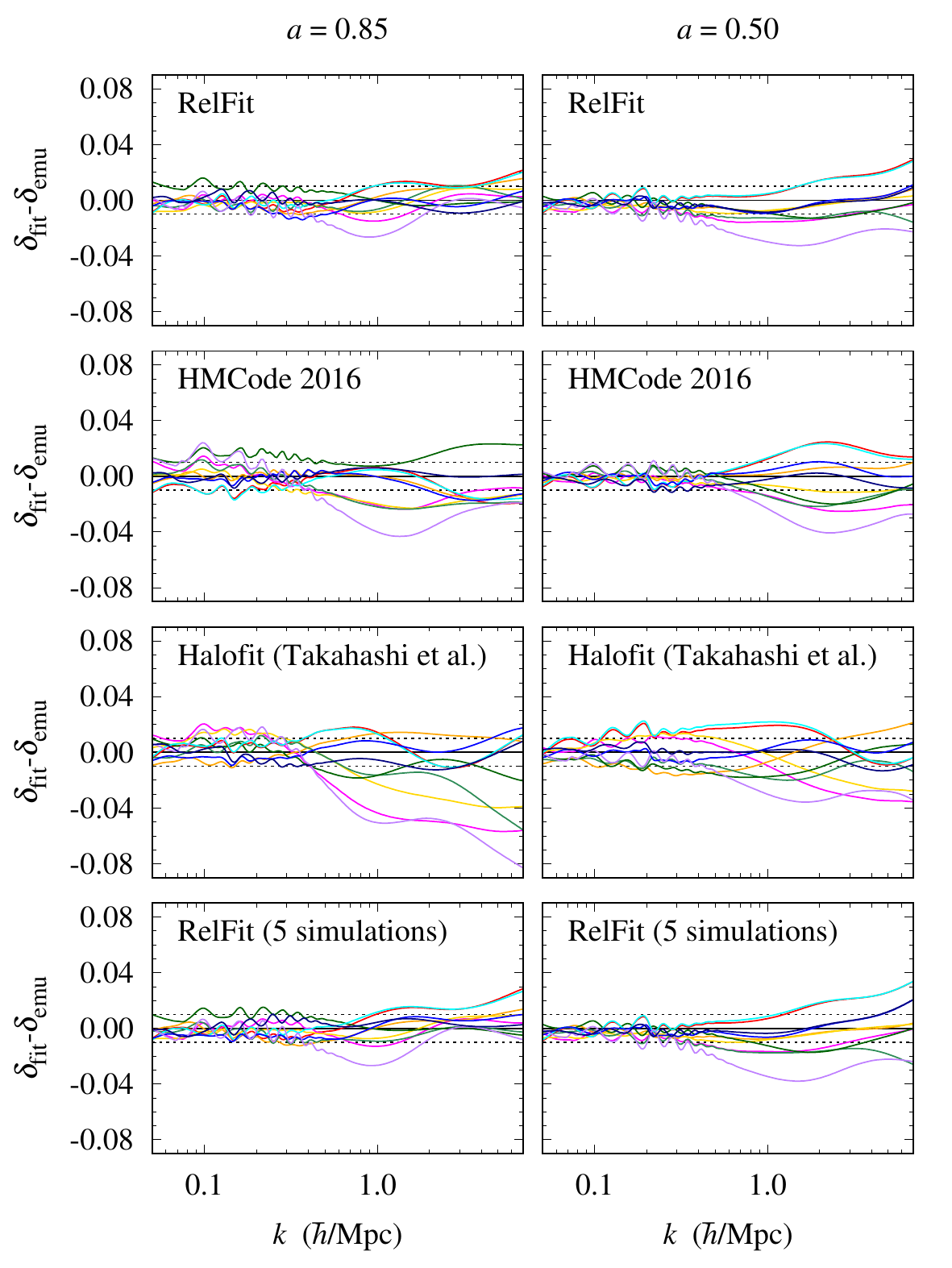}
	\end{center}
	\caption{Differences between the predictions of various fitting formulae $\delta_{\rm fit}$ and {\sc CosmicEmu} $\delta_{\rm emu}$ at  $a=0.85, 0.50$ for the 10 random models of table~\ref{tab:random}.  $\delta_{\rm fit}$ has been computed using {\sc RelFit} of this work (top),  {\sc HMCode} 2016 version~\cite{hmcode2016} as implemented in {\sc Camb} version 1.0.4~\cite{Lewis:1999bs} (second from top), {\sc Halofit}~\cite{halofit} including the updates of~\cite{Takahashi:2012em} as implemented in {\sc Camb} (third from top), and  an alternative calibration of {\sc RelFit} against  the minimum five simulations (bottom).	 Note that wavenumbers and the absolute power spectra 
		have units of $\bar{h}$/Mpc and $({\rm Mpc}/\bar{h})^3$ respectively, where $\bar{h}=0.673$.\label{fig:multi}}
\end{figure}

Figure~\ref{fig:multi} compares the output of {\sc RelFit} for the 10 target cosmologies of table~\ref{tab:random} at $a = 0.85, 0.50$ with the predictions of {\sc CosmicEmu}.  As with the single-parameter comparisons of section~\ref{sec:singleparametervariation}, the consistency between the two sets of predictions is remarkable:  for the most part the output of {\sc RelFit} is within 0.01 of  the {\sc CosmicEmu} predictions for the whole range of wavenumbers tested, and offers a clearly better concordance than can be achieved with 
{\sc HMCode} (2016 version)~\cite{hmcode2016} and especially {\sc Halofit}~\cite{halofit} as updated in~\cite{Takahashi:2012em} for the same set of target cosmologies,  also shown in figure~\ref{fig:multi}.

The ``worst-performing'' cosmology of the {\sc RelFit} set has a maximum deviation from the {\sc CosmicEmu} predictions of 0.03 at $k \sim 1.5\, \bar{h}$/Mpc and $a=0.50$.  Interestingly, the same cosmology also exhibits the largest deviation from {\sc CosmicEmu} under corrections with both  {\sc HMCode} and {\sc Halofit}.  Bearing in mind that {\sc CosmicEmu} has a claimed accuracy of 4\%~\cite{Lawrence:2017ost}, these deviations could  suggest that the inaccuracy lies with {\sc CosmicEmu} {\it  itself} rather than with the three fitting functions. It is likewise intriguing  that except at $k \lesssim 0.5 \, \bar{h}$/Mpc, the agreement between {\sc RelFit} and {\sc CosmicEmu} does not improve with a decreasing scale factor~$a$, in contrast to the fidelity of  {\sc RelFit} to simulation results, which, as shown in figures~\ref{fig:error} and~\ref{fig:error-full}, does improve significantly from $a=0.85$ to  $a=0.30$ across the board.
 Further investigation of these oddities is however beyond the scope of the present work. 

We conclude our study with a comparison of an alternative calibration of {\sc RelFit}---against only five simulations,%
\footnote{The simulations used are {\tt 1024}$A_{s,l}$,  {\tt 1024}$w4$,  {\tt 1024}$\omega_{{\rm m},l}$, {\tt 1024}$n_{s,{\rm h}}$, {\tt 1024Ref} of table~\ref{tab:runs2}.}
 the minimum required to map out the four approximate-universal forms~$(\delta/\gamma)_X$---to  {\sc CosmicEmu},  using again the 10 target cosmologies of table~\ref{tab:random}.  This comparison is shown in the bottom panels of figure~\ref{fig:multi} as ``{\sc RelFit} (5 simulations)''.
Evidently, this even ``cheaper'' version of {\sc RelFit} agrees with {\sc CosmicEmu} almost as well as the default version (calibrated against nine simulations), with only marginal deteriorations (and possibly a hint of  systematic bias) in the agreement at $k \gtrsim 1 \, \bar{h}$/Mpc  and still outperforming both {\sc HMCode} and {\sc Halofit}.  
While we do not advocate this alternative calibration  because of potential biases introduced by the one-sided derivative estimates  (see section~\ref{sec:fitforms}), this exercise serves to illustrate succinctly the power of the {\sc RelFit} method, and supports our thesis that an accurate fitting function to the relative nonlinear matter power spectrum can indeed be obtained very cheaply.


\section{Conclusions}
\label{sec:conc}

The central message of this work is twofold:  (i) The relative matter power spectrum, defined as the fractional deviation in the absolute matter power spectrum produced by a target cosmology away from a  reference $\Lambda$CDM prediction, is fairly insensitive to the specifics of a simulation and can be computed to 1\%-level accuracy at a much lower computational cost than can the absolute matter power spectrum itself.  (ii) Within the $w$CDM class of models tested, the relative nonlinear power spectrum has the interesting property that when divided through by its linear counterpart under single-parameter variations, the result exhibits an approximate universality for each class of variations at the onset of nonlinearity.  Exploiting this and the property of multiplicability of the relative power spectrum, it is possible to construct full fitting functions to any cosmology in the vicinity of $\Lambda$CDM in a piece-wise manner, whereby component fitting functions are sought for single-parameter variations and then multiplied together to form the full fitting function.

Point 1 offers an advantage in the exploration of the nonlinear matter power spectrum under variation of cosmology, in that once an ultra-precise reference absolute matter power spectrum has been computed, variations away from the reference cosmology can be investigated at a relatively low cost, enabling a larger swath of parameter space to be explored or a particular parameter region of interest to be more densely sampled. Point 2 enables independent, piece-wise studies of cosmological models on nonlinear scales, by which we mean a fitting function for variations in, e.g., the primordial curvature power spectrum can be constructed independently of that for variations in, e.g., the dark energy equation of state.
Both have particular implications for the investigation of ``non-standard'' or ``exotic'' cosmologies: 
 Because computational costs have been significantly reduced, the task of exploring exotic model parameter spaces is now  possible for a much wider section of the scientific community. Computing the nonlinear matter power spectrum at 1\%-level accuracy can be made a far more egalitarian exercise than is currently feasible with conventional methods.

As an  illustration of the approach, we have used nine relatively inexpensive $w$CDM simulations  (box length $L=256 \, h^{-1}$Mpc and $N=1024^3$ particles, initialised at $z_{i}=49$)
spanning the parameter directions  $\{\omega_{\rm m}, A_{s}, n_{s},w\}$ 
to construct the fitting function {\sc RelFit} that is able to reproduce to $0.01 \to 0.02$ accuracy or better the relative nonlinear matter power spectra of 20-odd  $w$CDM cosmologies at $0.85 \geq a \geq 0.30$ up to $k \simeq 10 \, h$/Mpc.  {\sc RelFit} is likewise  applicable---without modification and to the same accuracy---to cosmologies in which  $w(a)= w_{0} + w_{a} (1-a)$ parameterises 
a time-dependent dark energy equation of state,  and where  $N_{\rm eff}$ may deviate from the canonical $N_{\rm eff}=3.04$.

Testing {\sc RelFit} against the output of the {\sc CosmicEmu} emulator trained on the Mira--Titan simulations~\cite{Heitmann:2015xma,Lawrence:2017ost}, we find again consistency at better than $0.01 \to 0.02$  in a large region of the 6-parameter space $\{\omega_{\rm m},A_{s}, n_{s}, ,w,\omega_{b},h\}$, despite {\sc RelFit} not having been calibrated against the same  simulations---or any high-quality simulation for that matter.   For the set of 10 randomly selected cosmologies examined, the ability of {\sc RelFit} to replicate the {\sc CosmicEmu} predictions surpasses that of both {\sc Halofit}~\cite{halofit} (with updates~\cite{Takahashi:2012em}) and {\sc HMCode} (2016 version)~\cite{hmcode2016}.  The same success can be reproduced even with only five calibrating simulations, although for reasons of minimising potential systematic biases,   the nine-simulation calibration is preferable---this version of {\sc RelFit} is summarised in appendix~\ref{sec:coefficients}.

To conclude, the relative matter power spectrum is an inexpensive and democratically accessible route to fulfilling the 1\%-level accuracy demands of the forthcoming generation of large-scale structure probes.  Our prototype  fitting function~{\sc RelFit} for $w(a)$CDM+$N_{\rm eff}$  cosmologies, which takes the linear matter power spectrum as an input, can be readily implemented in publicly available linear Boltzmann codes such as {\sc Camb}~\cite{Lewis:1999bs} and {\sc Class}~\cite{Blas:2011rf} together with, e.g., an output of {\sc CosmicEmu} as a placeholder for the ultra-precise reference absolute power spectrum yet to come.  In the future we shall extend the approach to  cosmologies including massive neutrinos, as well as more ``exotic'' scenarios such as decaying dark matter, interacting dark matter, and dark energy perturbations.

\acknowledgments

 Y$^3$W thanks Amol Upadhye for useful discussions about {\sc CosmicEmu}.
STH is supported by a grant from the Villum Foundation. 
Y$^3$W is supported in part by the Australian Research Council's Discovery Project (project DP170102382) and Future Fellowship (project FT180100031) funding schemes. 


\appendix

\section{Connection to the stable clustering ansatz}
\label{sec:pd}

The Peacock--Dodds (PD) fitting formula renders the dimensionless nonlinear matter power spectrum, $\Delta_{\rm NL}^2 ({\bf \Theta}; k;a)  \equiv k^3 P({\bf \Theta}; k;a)/(2 \pi^2)$, in terms of a simple function of the dimensionless linear power spectrum, $\Delta_{\rm L}^2 ({\bf \Theta}; k;a)  \equiv k^3 P_{\rm L}({\bf \Theta}; k;a)/(2 \pi^2)$,  and the $k$-independent linear growth function~$D$~\cite{PD}:
\begin{equation}
\Delta^2_{\rm NL}({\bf \Theta}; k_{\rm NL}; a) = f(\Delta_{\rm L}^2({\bf \Theta}; k_{\rm L}; a),D({\bf \Theta}; a)).
\end{equation}
Here, the nonlinear and linear wavenumbers, $k_{\rm NL}$ and $k_{\rm L}$, satisfy a scaling relation
\begin{equation}
\label{eq:mapping}
k_{\rm L} = \frac{k_{\rm NL}}{[1+\Delta^2_{\rm NL}({\bf \Theta};k_{\rm NL}; a)]^{1/3}}
\end{equation}
following from the stable clustering ansatz~\cite{Hamilton:1991es,PD}.  The dependences of $f$ on $\Delta_{\rm L}^2$ and $D$ on strongly nonlinear scales are likewise stipulated by stable clustering.   The general form of $f$, however, must be determined from and calibrated against simulations, which introduces an additional, empirical dependence on the effective linear spectral index 
\begin{equation}
n(k_{\rm L}) \equiv \frac{\partial \log \Delta^2_{\rm L}({\bf \Theta}; k_{\rm L}; a)}{\partial \log k_{\rm L}}
\end{equation}
 in the fitting coefficients of the PD formula that we shall ignore for now.

Then, perturbing around ${\bf \Theta}_0$  it is  straightforward to establish that
\begin{equation}
\delta({\bf \Theta}, {\bf \Theta}_0;k_{\rm NL}) \simeq \frac{R\, \gamma({\bf \Theta}, {\bf \Theta}_0; k_{\rm L})  + V \, \Delta D/\bar{D} }{1 + R \, U\, n(k_{\rm L}) }
\label{eq:delta}
\end{equation}
to leading order in $\delta, \gamma, \Delta D/\bar{D}$, with expansion coefficients
\begin{eqnarray}
\label{eq:R}
R &\equiv& \left.  \frac{\partial \log f(\Delta^2_{\rm L}(k_{\rm L}),D,n(k_{\rm L}))}{\partial \log \Delta^2_{\rm L}(k_{\rm L})} \right|_{{\bf \Theta} = {\bf \Theta}_0}, \\
V & \equiv & \left. \frac{\partial \log f(\Delta^2_{\rm L}(k_{\rm L}),D,n(k_{\rm L}))}{\partial \log D}  \right|_{{\bf \Theta} = {\bf \Theta}_0},\\
U  &\equiv& \left. \frac{1}{3} \frac{\Delta^2_{\rm NL}(k_{\rm NL})}{1+\Delta^2_{\rm NL}(k_{\rm NL})}  \right|_{{\bf \Theta} = {\bf \Theta}_0}.
\end{eqnarray}
Note that with the exception of $U$, all quantities on the RHS of equation~(\ref{eq:delta}) are technically functions of $k_{\rm L}$, and must be mapped to $k_{\rm NL}$ via equation~(\ref{eq:mapping}) before they can be meaningfully interpreted.

\begin{figure}[t]
\begin{center}
\includegraphics[width=6cm]{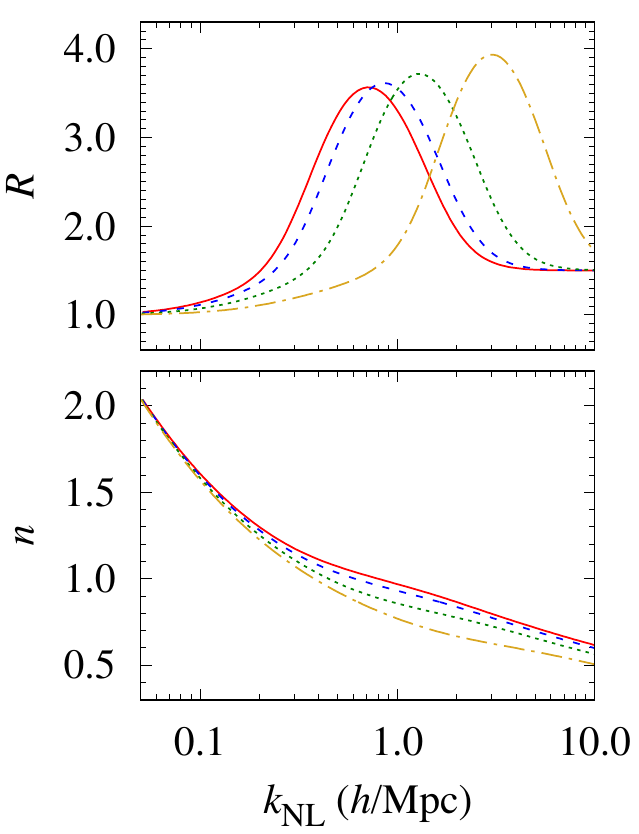}
\includegraphics[width=6cm]{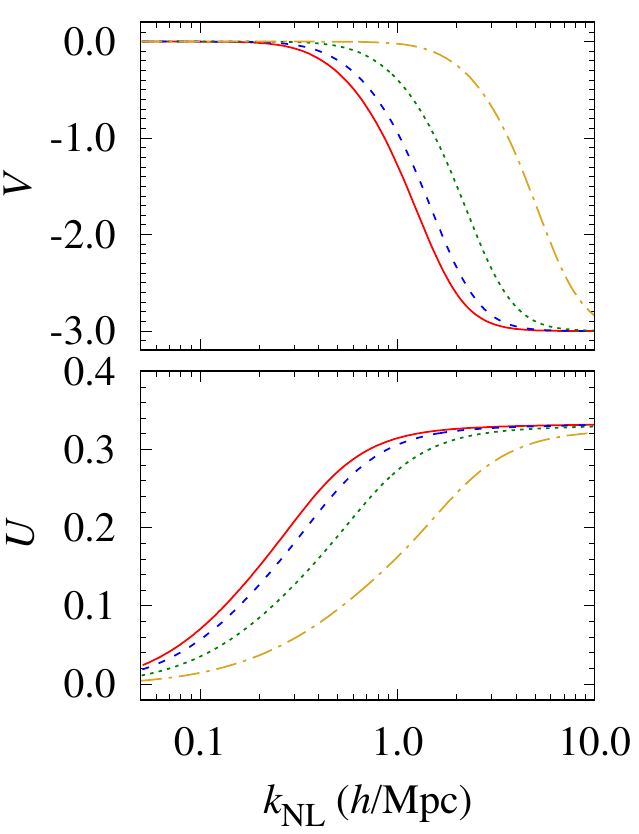}
\end{center}
\caption{The coefficients~$R$, $V$, $n$, and $U$ as they appear in equation~(\ref{eq:delta}) as functions of the nonlinear wavenumber~$k_{\rm NL}$ at $a=0.85$ (red/solid), $0.70$ (blue/dash), $0.50$ (green/dotted), and $0.30$ (gold/dot-dash).  All have been computed using the PD fitting formula applied to a dewiggled linear power spectrum of the canonical reference $\Lambda$CDM model of table~\ref{tab:params}.\label{fig:rrstv}}
\end{figure}

Figure~\ref{fig:rrstv} shows the expansion coefficients as functions of $k_{\rm NL}$ for the canonical reference~$\Lambda$CDM model at $a=0.85,0.70,0.50, 0.30$.  We have used the (cosmology- and simulation-dependent) PD fitting formula~\cite{PD} to make this plot.%
\footnote{Note that in evaluating the expansion coefficients in figure~\ref{fig:rrstv}, we have applied the~PD formula to a ``dewiggled'' linear power spectrum, in order to show more clearly the coefficients' broad-band characteristics.}  However, a number of cosmology-  and simulation-independent features can still be identified:
\begin{enumerate}
\item At any given epoch, the function $R$ generally evolves from unity in the linear regime to a peak around $k_{\rm peak}$ (see equation~(\ref{eq:kpeak}))
 at which structures begin to acquire nonlinearities.
  It then drops to a constant asymptotic value $R_\infty$;    
    here, the stable clustering ansatz stipulates that 
\begin{equation}
\label{eq:stable}
f(\Delta^2_{\rm L}, D) \propto (\Delta^2_{\rm L})^{3/2} (D/a)^{-3}
\end{equation}
on very nonlinear scales, leading to $R_\infty = 3/2$.

\item In the linear regime the function $f$ approaches $\Delta^2_{\rm L}$ and does not depend on the growth factor $D$; in this limit  $V$~asymptotes to  zero.  In the nonlinear limit the assumption of stable clustering and hence equation~(\ref{eq:stable}) lead to $V_\infty= -3$.  The transition between the two regimes again takes place around  $k_{\rm peak}$.   Note that $V$ is always negative in the PD~fitting formula.

\item 
Barring baryon acoustic oscillations, the effective linear spectral index  $n(k_{\rm L})$~is  always positive for all cosmologies consistent with current observations; it takes on a value  $\sim 4$ at wavenumbers $k \lesssim k_{\rm eq}$ defined in equation~(\ref{eq:keq}), and decreases monotonically beyond $k \sim k_{\rm eq}$ to zero on strongly nonlinear scales.

\item The $U$ term traces its origin to the cosmology-dependent mapping of the nonlinear wavenumber~$k_{\rm NL}$ to the linear wavenumber~$k_{\rm L}$ via equation~(\ref{eq:mapping}).  For $\Lambda$CDM-type cosmologies, however, we expect $U$ to increase monotonically from zero in the linear regime to $U_\infty=1/3$ in the strongly nonlinear regime.

\end{enumerate}

With these considerations in mind it is straightforward to establish the qualitative behaviours of some of the approximate-universal forms~$(\delta/\gamma)_X$.
For $X={\cal N}$, e.g., from varying only the primordial fluctuation amplitude~$A_s$ between the reference and target cosmologies, the relative linear power spectrum $\gamma$ is $k$-independent and  $\Delta D/\bar{D}=0$.  Then, we find from equation~(\ref{eq:delta})
\begin{equation}
\begin{aligned}
(\delta/\gamma)_{\cal N}
 &\simeq \frac{R  }{1 + R \, U\, n(k_{\rm L}) }\\
 &\xrightarrow[\rm stable]{}  
  \frac{3}{2+ n(k_{\rm L})},
 \label{eq:npd}
 \end{aligned}
\end{equation}
where we can immediately conclude from the first approximate equality that 
the relative nonlinear power spectrum $\delta$ at $k_{\rm NL}$ is always larger in magnitude than the corresponding  $\gamma$ at $k_{\rm L}$,  i.e., $(\delta/\gamma)_{\cal N} \gtrsim 1$.
 In the stable clustering limit we expect $(\delta/\gamma)_{\cal N}$ to tend to the expression in the second line of equation~(\ref{eq:npd}), 
which for our reference $\Lambda$CDM cosmology evaluates numerically to $(\delta/\gamma)_{\cal N}\simeq 1.2$ at $k \sim 10\, h$/Mpc, a prediction that 
appears to be borne out by our simulation results at $a=0.85, 0.50$ shown in figures~\ref{fig:asomegamns} and~\ref{fig:asomegamnshighz}.

In the case of $X = D$, e.g., from varying the dark energy equation of state parameter~$w$, we have again a $k$-independent $\gamma  = 2 \,\Delta D/\bar{D}$, such that   
 equation~(\ref{eq:delta}) becomes
\begin{equation}
(\delta/\gamma)_D \simeq \frac{2 R+ V}{2 + 2 R\, U\, n(k_{\rm L})}.
\end{equation}
Recall that on linear scales $V$ evaluates to zero.  Thus, we expect $(\delta/\gamma)_D$ to evolve on mildly nonlinear scales approximately like $(\delta/\gamma)_{\cal N} \gtrsim 1$.
As we move further  into the nonlinear regime, however, $V(g)$ becomes increasingly negative; when the condition $ V < 2 (1 - R + R\, U\, n)$ is satisfied,  $(\delta/\gamma)_D$ flips from $\gtrsim 1$ to $\lesssim 1$, also observed in figures~\ref{fig:w} and~\ref{fig:wtheta}.   In the stable clustering limit, $(\delta/\gamma)_D$ should vanish exactly (to all orders): our simulation results in  figures~\ref{fig:w} and~\ref{fig:wtheta} appear to suggest this behaviour at large wavenumbers $k$.  The simulations do not however have the necessary dynamical range to fully resolve this limit.


\section{{\sc RelFit} fitting coefficients}
\label{sec:coefficients}

The fitting function {\sc RelFit} for the relative nonlinear matter power spectrum  in $w$CDM cosmologies comes in a general form,   
\begin{equation}
\begin{aligned}
&1+ \delta({\bf \Theta}, {\bf \Theta}_0; k;a)  \\
&\simeq 
\left[ 1+  (\delta/\gamma)_{\cal N}  \frac{\Delta {\cal N}}{\bar{\cal N}} \right]
  \left[1+  (\delta/\Gamma)_{n_s} \Gamma \right]
\left[ 1+  (\delta/\gamma)_{D}  \frac{\Delta D^2}{\bar{D}^2}  \right]
\left[ 1+  (\delta/\gamma)_{T}  \frac{\Delta T^2}{\bar{T}^2}  \right],
  \end{aligned}
  \end{equation}
and a restricted form,
\begin{equation}
\begin{aligned}
1+ \delta({\bf \Theta}, {\bf \Theta}_0; k;a)  
\simeq \frac{ 1+  (\delta/\gamma)_{\cal N}  \frac{\Delta {\cal N}}{\bar{\cal N}} }{ 1+  (\delta/\gamma)_{\cal N}  \frac{\Delta {\cal N}_1}{\bar{\cal N}} }
  \left[1+  (\delta/\Gamma)_{n_s} \Gamma \right]
\left[1+  (\delta/\gamma)_{\omega_{\rm m}}  \gamma_{\omega_{\rm m}} \right]
\frac{ 1+  (\delta/\gamma)_{D}  \frac{\Delta D^2}{\bar{D}^2}  }{ 1+  (\delta/\gamma)_{D}  \frac{\Delta D_1^2}{\bar{D}^2} },
  \end{aligned}
  \end{equation}
 the latter of which applies to a restricted set of $w$CDM parameters.

Here, the target and reference $\Lambda$CDM cosmologies are specified respectively by
\begin{equation}
\begin{aligned}
&{\bf \Theta} =  \{\theta_{w(a),\omega_{\rm m}, \omega_{\rm b},h,A_s,n_s}=\bar{\theta}_{w(a),\omega_{\rm m},\omega_{\rm b}, h, A_s,n_s}; \\
& \hspace*{25mm} {\cal N} (A_s, \omega_{ \rm m}), n_s, D(w(a),\omega_{\rm m},h;a),T(\omega_{\rm m},\omega_{\rm b},N_{\rm eff};k) \}, \\
&{\bf \Theta}_0 =  \{\theta_{w(a), \omega_{\rm m},\omega_{\rm b},h,A_s,n_s}= \bar{\theta}_{w(a), \omega_{\rm m},\omega_{\rm b},h,A_s,n_s}; \\
& \hspace*{25mm}
\bar{\cal N} \equiv {\cal N} (\bar{A}_s,\bar{\omega}_{ \rm m}), n_{s},  \bar{D}\equiv D(\bar{w}(a),\bar{\omega}_{\rm m},\bar{h};a), \bar{T}\equiv T(\bar{\omega}_{\rm m},\bar{\omega}_{\rm b},\bar{N}_{{\rm eff}};k ) \},
\end{aligned}
\end{equation}
where
${\cal N} \equiv A_{s}/\omega_{\rm m}^{2}$ is the overall normalisation of the linear matter power spectrum, $D$ is the linear growth function, $T$  the linear transfer function, and 
$\Delta X \equiv X - \bar{X}$ denotes the variation in  $X = {\cal N}, D, T$ between the target and reference cosmologies.    
The function 
\begin{equation}
\Gamma (k) \equiv \Delta n_s \sum_{i=0}^\infty  \frac{\left[\Delta n_s \ln (k/k_{\rm piv})\right]^{i}}{(i+1)!}
\end{equation}
specifies the variation in the shape of the primordial curvature power spectrum, taken to be of a power-law form, with $\Delta n_{s} \equiv n_{s}-\bar{n}_{s}$ for a $k$-independent $n_{s}$.
In the case of the restricted form of {\sc RelFit}, which applies if the only parameter varied in the linear transfer function is the physical matter density $\omega_{\rm m}$,  we require also the
auxiliary definitions 
\begin{equation}
\begin{aligned}
\Delta {\cal N}_1 &\equiv {\cal N}_1 - \bar{\cal N} \equiv {\cal N} (\bar{A}_s, \omega_m)- \bar{\cal N},\\
\Delta D_1^2 &\equiv D_1^2 - \bar{D}^2  \equiv D^2(\bar{w}(a), \omega_{\rm m},\bar{h};a)- \bar{D}^2, \\
 \gamma_{\omega_{\rm m}} & \equiv \gamma({\bf \Theta}=\{\bar{\theta}_{\omega_{\rm m}}; \omega_{\rm m}\}, {\bf \Theta}_0=\{\bar{\theta}_{\omega_{\rm m}};\bar{ \omega}_{\rm m}\}),
\end{aligned}
\end{equation} 
where the last entry denotes the relative linear  power spectrum in which only the physical matter density~$\omega_{\rm m}$ is varied away from its reference $\Lambda$CDM value.

The crux of {\sc RelFit} are the approximate-universal forms $(\delta/\gamma)_{X}$ and $(\delta/\Gamma)_{n_{s}}$.  For $X={\cal N},\omega_{\rm m},D,T$, these are given by
\begin{equation}
(\delta/\gamma)_{X} \simeq 1+ \left(1-e^{-y} \right) \frac{b_0^X + b_1^X \log_{10} y + b_2^X (\log_{10} y)^2}{ 1+ c_1^X \log_{10} y + c_2^X  (\log_{10} y)^2 },
\end{equation}
while for $X=n_s$ we have
\begin{equation}
(\delta/\Gamma)_{n_s} \simeq e^{-y} \ln (k/k_{\rm piv}) + \left(1-e^{-y} \right) \frac{b_0^{n_s} + b_1^{n_s} \log_{10} y + b_2^{n_s} (\log_{10} y)^2}{ 1 + c_1^{n_s} \log_{10} y + c_2^{n_s}  (\log_{10} y)^2 }
\end{equation}
 with the $\Gamma(k)$ series truncated at $i=2$.  Here, the independent variable is
\begin{equation}
y(k,a) \equiv \left[\frac{k}{h/{\rm Mpc}} \right] \big /  \left[\frac{k_\sigma (a) }{h/{\rm Mpc}} \right]^{0.65},
\end{equation}
where $k_\sigma\equiv 1/x$ is defined by the condition
\begin{equation}
\label{eq:sigma1}
\sigma^2(x=k_\sigma^{-1},a) = \frac{1}{2 \pi^2} \int d \ln k \ k^3 P_{\rm L}({\bf \Theta}_0; k;a) \ e^{-k^2 x^2} = 1.
\end{equation}
The corresponding $a$-dependent coefficients, calibrated against nine simulations at output scale factors $a = 0.85, 0.70, 0.50, 0.30$  are as follows:
\begin{itemize}
\item $X={\cal N}$:
\begin{equation}
\begin{aligned}
b_0^{\cal N} &= -1.27262 a^{-2} + 8.49321 a^{-1} -15.6289 + 9.75478 a,\\
b_1^{\cal N} &=0.383462 a^{-2} - 2.75936 a^{-1} +3.86886 -1.00869 a,\\
b_2^{\cal N} &= 0.88578 a^{-2} - 6.35666 a^{-1}  +12.7673 - 8.39676 a,\\
c_1^{\cal N} & = 0.971655 a^{-2}  -6.42766 a^{-1} + 13.1775 -6.13817 a,\\
c_2^{\cal N} &= -0.861939 a^{-2} + 3.71565 a^{-1}  -2.15088 + 0.782779 a.
\end{aligned}
\end{equation}
\item $X=\omega_{\rm m}$: 
\begin{equation}
\begin{aligned}
b_0^{\omega_{\rm m}} &=-0.701687 a^{-2} +4.83922 a^{-1} -9.3655+5.42385 a,\\
b_1^{\omega_{\rm m}} &=-0.934228 a^{-2} +5.5201 a^{-1}- 10.3304 +5.62847 a,\\
b_2^{\omega_{\rm m}} &=1.56111 a^{-2} -10.3063 a^{-1} + 19.1107-12.0445 a, \\
c_1^{\omega_{\rm m}} &=  0.27832 a^{-2} -3.2378 a^{-1} + 9.54675 -4.63525 a,\\
c_2^{\omega_{\rm m}} &= -2.11622 a^{-2} + 11.4841 a^{-1}-18.0275 + 11.0678 a.
\end{aligned}
\end{equation}

\item $X=D$:
\begin{equation}
\begin{aligned}
b_0^D &= -1.20647 a^{-2} + 8.05854 a^{-1}  -14.9436 + 9.2222 a,\\
b_1^D &= 0.331031 a^{-2}  -2.39088 a^{-1}  + 2.30353 -1.03277 a,\\
b_2^D &= 1.39963 a^{-2} - 8.95383 a^{-1} +16.0116 -10.8365 a,\\
c_1^D &= 0.431109 a^{-2} - 2.67855 a^{-1} + 5.22015 -1.89007 a,\\
c_2^D &=-0.644279 a^{-2} + 3.11752 a^{-1} -2.6547 + 1.02493 a.\\
\end{aligned}
\end{equation}

\item $X=T$:
\begin{equation}
\begin{aligned}
b_0^{T} & = -1.00004 a^{-2} + 6.72007 a^{-1} -12.5078 + 7.50648 a, \\
b_1^{T} &=  -0.12367 a^{-2}  +0.48026 a^{-1} - 1.48121 +1.3353 a,\\ 
b_2^{T}& =  0.702664 a^{-2}  -5.01405 a^{-1} +9.20829 -5.95565 a, \\
c_1^{T} & =1.52249 a^{-2}  -10.1414 a^{-1} +20.8795  -10.7079 a,\\
c_2^{T} & = -0.128336 a^{-2}  -1.23333 a^{-1} +7.52606 -4.59947 a.
\end{aligned}
\end{equation}

\item $X=n_s$:
\begin{equation}
\begin{aligned}
b_0^{n_s}&=-0.356502 a^{-2} + 3.94524 a^{-1}  -6.66721 + 3.61934 a,\\
b_1^{n_s}&= -1.6224 a^{-2} + 10.7091 a^{-1} -16.2796 + 10.9023 a,\\
b_2^{n_s}& = -0.407575 a^{-2}+ 1.40031 a^{-1} + 0.350623+ 0.73293 a,\\
c_1^{n_s} & = -0.439116 a^{-2} + 2.62068 a^{-1}  -3.487+ 2.56173 a,\\
c_2^{n_s} & =-0.908028 a^{-2} + 6.05265 a^{-1}  -11.5555 + 7.49648 a.
\end{aligned}
\end{equation}
\end{itemize}


\bibliographystyle{utcaps}

\bibliography{newrefs}

\providecommand{\href}[2]{#2}\begingroup\raggedright\begin{thebibliography}{10}

\bibitem{euclid}
{\bfseries EUCLID} Collaboration, R.~Laureijs {\em et al.}, ``{Euclid
  Definition Study Report},''
\href{http://arxiv.org/abs/1110.3193}{{\ttfamily arXiv:1110.3193
  [astro-ph.CO]}}.

\bibitem{lsst}
{\bfseries LSST Science, LSST Project} Collaboration, P.~A. Abell {\em et al.},
  ``{LSST Science Book, Version 2.0},''
\href{http://arxiv.org/abs/0912.0201}{{\ttfamily arXiv:0912.0201
  [astro-ph.IM]}}.

\bibitem{Bernardeau:2013oda}
F.~Bernardeau, ``{The evolution of the large-scale structure of the universe:
  beyond the linear regime},'' in {\em {100e Ecole d'Ete de Physique:
  Post-Planck Cosmology Les Houches, France, July 8-August 2, 2013}}.
\newblock 2013.
\newblock \href{http://arxiv.org/abs/1311.2724}{{\ttfamily arXiv:1311.2724
  [astro-ph.CO]}}.
\newblock
\url{http://inspirehep.net/record/1264888/files/arXiv:1311.2724.pdf}.
\newblock

\bibitem{Heitmann:2008eq}
K.~Heitmann, M.~White, C.~Wagner, S.~Habib, and D.~Higdon, ``{The Coyote
  Universe I: Precision Determination of the Nonlinear Matter Power
  Spectrum},'' \href{http://dx.doi.org/10.1088/0004-637X/715/1/104}{{\em
  Astrophys. J.} {\bfseries 715} (2010)  104--121},
\href{http://arxiv.org/abs/0812.1052}{{\ttfamily arXiv:0812.1052 [astro-ph]}}.

\bibitem{Schneider:2015yka}
A.~Schneider, R.~Teyssier, D.~Potter, J.~Stadel, J.~Onions, D.~S. Reed, R.~E.
  Smith, V.~Springel, F.~R. Pearce, and R.~Scoccimarro, ``{Matter power
  spectrum and the challenge of percent accuracy},''
  \href{http://dx.doi.org/10.1088/1475-7516/2016/04/047}{{\em JCAP} {\bfseries
  1604} (2016) no.~04, 047},
\href{http://arxiv.org/abs/1503.05920}{{\ttfamily arXiv:1503.05920
  [astro-ph.CO]}}.

\bibitem{Heitmann:2015xma}
K.~Heitmann {\em et al.}, ``{The Mira-Titan Universe: Precision Predictions for
  Dark Energy Surveys},''
  \href{http://dx.doi.org/10.3847/0004-637X/820/2/108}{{\em Astrophys. J.}
  {\bfseries 820} (2016) no.~2, 108},
\href{http://arxiv.org/abs/1508.02654}{{\ttfamily arXiv:1508.02654
  [astro-ph.CO]}}.

\bibitem{Lawrence:2017ost}
E.~Lawrence, K.~Heitmann, J.~Kwan, A.~Upadhye, D.~Bingham, S.~Habib, D.~Higdon,
  A.~Pope, H.~Finkel, and N.~Frontiere, ``{The Mira-Titan Universe II: Matter
  Power Spectrum Emulation},''
  \href{http://dx.doi.org/10.3847/1538-4357/aa86a9}{{\em Astrophys. J.}
  {\bfseries 847} (2017) no.~1, 50},
\href{http://arxiv.org/abs/1705.03388}{{\ttfamily arXiv:1705.03388
  [astro-ph.CO]}}.

\bibitem{halofit}
{\bfseries VIRGO Consortium} Collaboration, R.~E. Smith, J.~A. Peacock,
  A.~Jenkins, S.~D.~M. White, C.~S. Frenk, F.~R. Pearce, P.~A. Thomas,
  G.~Efstathiou, and H.~M.~P. Couchmann, ``{Stable clustering, the halo model
  and nonlinear cosmological power spectra},''
  \href{http://dx.doi.org/10.1046/j.1365-8711.2003.06503.x}{{\em Mon. Not. Roy.
  Astron. Soc.} {\bfseries 341} (2003)  1311},
\href{http://arxiv.org/abs/astro-ph/0207664}{{\ttfamily arXiv:astro-ph/0207664
  [astro-ph]}}.

\bibitem{Takahashi:2012em}
R.~Takahashi, M.~Sato, T.~Nishimichi, A.~Taruya, and M.~Oguri, ``{Revising the
  Halofit Model for the Nonlinear Matter Power Spectrum},''
  \href{http://dx.doi.org/10.1088/0004-637X/761/2/152}{{\em Astrophys. J.}
  {\bfseries 761} (2012)  152},
\href{http://arxiv.org/abs/1208.2701}{{\ttfamily arXiv:1208.2701
  [astro-ph.CO]}}.

\bibitem{Mead:2015yca}
A.~Mead, J.~Peacock, C.~Heymans, S.~Joudaki, and A.~Heavens, ``{An accurate
  halo model for fitting non-linear cosmological power spectra and baryonic
  feedback models},'' \href{http://dx.doi.org/10.1093/mnras/stv2036}{{\em Mon.
  Not. Roy. Astron. Soc.} {\bfseries 454} (2015) no.~2, 1958--1975},
\href{http://arxiv.org/abs/1505.07833}{{\ttfamily arXiv:1505.07833
  [astro-ph.CO]}}.

\bibitem{hmcode2016}
A.~Mead, C.~Heymans, L.~Lombriser, J.~Peacock, O.~Steele, and H.~Winther,
  ``{Accurate halo-model matter power spectra with dark energy, massive
  neutrinos and modified gravitational forces},''
  \href{http://dx.doi.org/10.1093/mnras/stw681}{{\em Mon. Not. Roy. Astron.
  Soc.} {\bfseries 459} (2016) no.~2, 1468--1488},
\href{http://arxiv.org/abs/1602.02154}{{\ttfamily arXiv:1602.02154
  [astro-ph.CO]}}.

\bibitem{Heitmann:2013bra}
K.~Heitmann, E.~Lawrence, J.~Kwan, S.~Habib, and D.~Higdon, ``{The Coyote
  Universe Extended: Precision Emulation of the Matter Power Spectrum},''
  \href{http://dx.doi.org/10.1088/0004-637X/780/1/111}{{\em Astrophys. J.}
  {\bfseries 780} (2014)  111},
\href{http://arxiv.org/abs/1304.7849}{{\ttfamily arXiv:1304.7849
  [astro-ph.CO]}}.

\bibitem{Heitmann:2009cu}
K.~Heitmann, D.~Higdon, M.~White, S.~Habib, B.~J. Williams, and C.~Wagner,
  ``{The Coyote Universe II: Cosmological Models and Precision Emulation of the
  Nonlinear Matter Power Spectrum},''
  \href{http://dx.doi.org/10.1088/0004-637X/705/1/156}{{\em Astrophys. J.}
  {\bfseries 705} (2009)  156--174},
\href{http://arxiv.org/abs/0902.0429}{{\ttfamily arXiv:0902.0429
  [astro-ph.CO]}}.

\bibitem{Lawrence:2009uk}
E.~Lawrence, K.~Heitmann, M.~White, D.~Higdon, C.~Wagner, S.~Habib, and
  B.~Williams, ``{The Coyote Universe III: Simulation Suite and Precision
  Emulator for the Nonlinear Matter Power Spectrum},''
  \href{http://dx.doi.org/10.1088/0004-637X/713/2/1322}{{\em Astrophys. J.}
  {\bfseries 713} (2010)  1322--1331},
\href{http://arxiv.org/abs/0912.4490}{{\ttfamily arXiv:0912.4490
  [astro-ph.CO]}}.

\bibitem{Dakin:2019dxu}
J.~Dakin, S.~Hannestad, and T.~Tram, ``{Fully relativistic treatment of
  decaying cold dark matter in $N$-body simulations},''
  \href{http://dx.doi.org/10.1088/1475-7516/2019/06/032}{{\em JCAP} {\bfseries
  1906} (2019) no.~06, 032},
\href{http://arxiv.org/abs/1904.11773}{{\ttfamily arXiv:1904.11773
  [astro-ph.CO]}}.

\bibitem{Diacoumis:2018ezi}
J.~A.~D. Diacoumis and Y.~Y. Wong, ``{On the prior dependence of cosmological
  constraints on some dark matter interactions},''
  \href{http://dx.doi.org/10.1088/1475-7516/2019/05/025}{{\em JCAP} {\bfseries
  1905} (2019) no.~05, 025},
\href{http://arxiv.org/abs/1811.11408}{{\ttfamily arXiv:1811.11408
  [astro-ph.CO]}}.

\bibitem{Dakin:2019vnj}
J.~Dakin, S.~Hannestad, T.~Tram, M.~Knabenhans, and J.~Stadel, ``{Dark energy
  perturbations in $N$-body simulations},''
  \href{http://dx.doi.org/10.1088/1475-7516/2019/08/013}{{\em JCAP} {\bfseries
  1908} (2019)  013},
\href{http://arxiv.org/abs/1904.05210}{{\ttfamily arXiv:1904.05210
  [astro-ph.CO]}}.

\bibitem{McDonald:2005gz}
P.~McDonald, H.~Trac, and C.~Contaldi, ``{Dependence of the non-linear mass
  power spectrum on the equation of state of dark energy},''
  \href{http://dx.doi.org/10.1111/j.1365-2966.2005.09881.x}{{\em Mon. Not. Roy.
  Astron. Soc.} {\bfseries 366} (2006)  547--556},
\href{http://arxiv.org/abs/astro-ph/0505565}{{\ttfamily arXiv:astro-ph/0505565
  [astro-ph]}}.

\bibitem{Aad:2014rta}
{\bfseries ATLAS} Collaboration, G.~Aad {\em et al.}, ``{A measurement of the
  ratio of the production cross sections for $W$ and $Z$ bosons in association
  with jets with the ATLAS detector},''
  \href{http://dx.doi.org/10.1140/epjc/s10052-014-3168-9}{{\em Eur. Phys. J.}
  {\bfseries C74} (2014) no.~12, 3168},
\href{http://arxiv.org/abs/1408.6510}{{\ttfamily arXiv:1408.6510 [hep-ex]}}.

\bibitem{Song:2008qt}
Y.-S. Song and W.~J. Percival, ``{Reconstructing the history of structure
  formation using Redshift Distortions},''
  \href{http://dx.doi.org/10.1088/1475-7516/2009/10/004}{{\em JCAP} {\bfseries
  0910} (2009)  004},
\href{http://arxiv.org/abs/0807.0810}{{\ttfamily arXiv:0807.0810 [astro-ph]}}.

\bibitem{Villante:1998pe}
F.~L. Villante, G.~Fiorentini, and E.~Lisi, ``{Solar neutrino interactions:
  Using charged currents at SNO to tell neutral currents at
  Super-Kamiokande},'' \href{http://dx.doi.org/10.1103/PhysRevD.59.013006}{{\em
  Phys. Rev.} {\bfseries D59} (1999)  013006},
\href{http://arxiv.org/abs/hep-ph/9807360}{{\ttfamily arXiv:hep-ph/9807360
  [hep-ph]}}.

\bibitem{Ren:2017xov}
{\bfseries MINERvA} Collaboration, L.~Ren {\em et al.}, ``{Measurement of the
  antineutrino to neutrino charged-current interaction cross section ratio in
  MINERvA},'' \href{http://dx.doi.org/10.1103/PhysRevD.97.019902,
  10.1103/PhysRevD.95.072009}{{\em Phys. Rev.} {\bfseries D95} (2017) no.~7,
  072009}, \href{http://arxiv.org/abs/1701.04857}{{\ttfamily arXiv:1701.04857
  [hep-ex]}}.
[Addendum: Phys. Rev.D97,no.1,019902(2018)].

\bibitem{Cataneo:2018cic}
M.~Cataneo, L.~Lombriser, C.~Heymans, A.~Mead, A.~Barreira, S.~Bose, and B.~Li,
  ``{On the road to percent accuracy: non-linear reaction of the matter power
  spectrum to dark energy and modified gravity},''
  \href{http://dx.doi.org/10.1093/mnras/stz1836}{{\em Mon. Not. Roy. Astron.
  Soc.} {\bfseries 488} (2019) no.~2, 2121--2142},
\href{http://arxiv.org/abs/1812.05594}{{\ttfamily arXiv:1812.05594
  [astro-ph.CO]}}.

\bibitem{Giblin:2019iit}
B.~Giblin, M.~Cataneo, B.~Moews, and C.~Heymans, ``{On the road to percent
  accuracy II: calibration of the non-linear matter power spectrum for
  arbitrary cosmologies},'' \href{http://dx.doi.org/10.1093/mnras/stz2659}{{\em
  Mon. Not. Roy. Astron. Soc.} {\bfseries 490} (2019) no.~4, 4826--4840},
\href{http://arxiv.org/abs/1906.02742}{{\ttfamily arXiv:1906.02742
  [astro-ph.CO]}}.

\bibitem{Cataneo:2016suz}
M.~Cataneo, S.~Foreman, and L.~Senatore, ``{Efficient exploration of cosmology
  dependence in the EFT of LSS},''
  \href{http://dx.doi.org/10.1088/1475-7516/2017/04/026}{{\em JCAP} {\bfseries
  1704} (2017) no.~04, 026},
\href{http://arxiv.org/abs/1606.03633}{{\ttfamily arXiv:1606.03633
  [astro-ph.CO]}}.

\bibitem{Ade:2015xua}
{\bfseries Planck} Collaboration, P.~A.~R. Ade {\em et al.}, ``{Planck 2015
  results. XIII. Cosmological parameters},''
  \href{http://dx.doi.org/10.1051/0004-6361/201525830}{{\em Astron. Astrophys.}
  {\bfseries 594} (2016)  A13},
\href{http://arxiv.org/abs/1502.01589}{{\ttfamily arXiv:1502.01589
  [astro-ph.CO]}}.

\bibitem{Aghanim:2018eyx}
{\bfseries Planck} Collaboration, N.~Aghanim {\em et al.}, ``{Planck 2018
  results. VI. Cosmological parameters},''
\href{http://arxiv.org/abs/1807.06209}{{\ttfamily arXiv:1807.06209
  [astro-ph.CO]}}.

\bibitem{Springel:2005mi}
V.~Springel, ``{The Cosmological simulation code GADGET-2},''
  \href{http://dx.doi.org/10.1111/j.1365-2966.2005.09655.x}{{\em Mon. Not. Roy.
  Astron. Soc.} {\bfseries 364} (2005)  1105--1134},
\href{http://arxiv.org/abs/astro-ph/0505010}{{\ttfamily arXiv:astro-ph/0505010
  [astro-ph]}}.

\bibitem{Lewis:1999bs}
A.~Lewis, A.~Challinor, and A.~Lasenby, ``{Efficient computation of CMB
  anisotropies in closed FRW models},''
  \href{http://dx.doi.org/10.1086/309179}{{\em Astrophys. J.} {\bfseries 538}
  (2000)  473--476},
\href{http://arxiv.org/abs/astro-ph/9911177}{{\ttfamily arXiv:astro-ph/9911177
  [astro-ph]}}.

\bibitem{Crocce:2006ve}
M.~Crocce, S.~Pueblas, and R.~Scoccimarro, ``{Transients from Initial
  Conditions in Cosmological Simulations},''
  \href{http://dx.doi.org/10.1111/j.1365-2966.2006.11040.x}{{\em Mon. Not. Roy.
  Astron. Soc.} {\bfseries 373} (2006)  369--381},
\href{http://arxiv.org/abs/astro-ph/0606505}{{\ttfamily arXiv:astro-ph/0606505
  [astro-ph]}}.

\bibitem{Potter:2016ttn}
D.~Potter, J.~Stadel, and R.~Teyssier, ``{PKDGRAV3: Beyond Trillion Particle
  Cosmological Simulations for the Next Era of Galaxy Surveys},''
  \href{http://dx.doi.org/10.1186/s40668-017-0021-1}{{\em
  Comput.~Astrophys.~Cosmol.} {\bfseries 4} (2017)  2},
\href{http://arxiv.org/abs/1609.08621}{{\ttfamily arXiv:1609.08621
  [astro-ph.IM]}}.

\bibitem{Blas:2011rf}
D.~Blas, J.~Lesgourgues, and T.~Tram, ``{The Cosmic Linear Anisotropy Solving
  System (CLASS) II: Approximation schemes},''
  \href{http://dx.doi.org/10.1088/1475-7516/2011/07/034}{{\em JCAP} {\bfseries
  1107} (2011)  034},
\href{http://arxiv.org/abs/1104.2933}{{\ttfamily arXiv:1104.2933
  [astro-ph.CO]}}.

\bibitem{Dakin:2017idt}
J.~Dakin, J.~Brandbyge, S.~Hannestad, T.~Haugbølle, and T.~Tram,
  ``{$\nu$CO$N$CEPT: Cosmological neutrino simulations from the non-linear
  Boltzmann hierarchy},''
  \href{http://dx.doi.org/10.1088/1475-7516/2019/02/052}{{\em JCAP} {\bfseries
  1902} (2019)  052},
\href{http://arxiv.org/abs/1712.03944}{{\ttfamily arXiv:1712.03944
  [astro-ph.CO]}}.

\bibitem{Dodelson:2003ft}
S.~Dodelson, {\em {Modern Cosmology}}.
\newblock Academic Press, Amsterdam, 2003.
\newblock
\url{http://www.slac.stanford.edu/spires/find/books/www?cl=QB981:D62:2003}.
\newblock

\bibitem{Linder:2003dr}
E.~V. Linder and A.~Jenkins, ``{Cosmic structure and dark energy},''
  \href{http://dx.doi.org/10.1046/j.1365-2966.2003.07112.x}{{\em Mon. Not. Roy.
  Astron. Soc.} {\bfseries 346} (2003)  573},
\href{http://arxiv.org/abs/astro-ph/0305286}{{\ttfamily arXiv:astro-ph/0305286
  [astro-ph]}}.

\bibitem{Chevallier:2000qy}
M.~Chevallier and D.~Polarski, ``{Accelerating universes with scaling dark
  matter},'' \href{http://dx.doi.org/10.1142/S0218271801000822}{{\em Int. J.
  Mod. Phys.} {\bfseries D10} (2001)  213--224},
\href{http://arxiv.org/abs/gr-qc/0009008}{{\ttfamily arXiv:gr-qc/0009008
  [gr-qc]}}.

\bibitem{Linder:2002et}
E.~V. Linder, ``{Exploring the expansion history of the universe},''
  \href{http://dx.doi.org/10.1103/PhysRevLett.90.091301}{{\em Phys. Rev. Lett.}
  {\bfseries 90} (2003)  091301},
\href{http://arxiv.org/abs/astro-ph/0208512}{{\ttfamily arXiv:astro-ph/0208512
  [astro-ph]}}.

\bibitem{Linder:2005in}
E.~V. Linder, ``{Cosmic growth history and expansion history},''
  \href{http://dx.doi.org/10.1103/PhysRevD.72.043529}{{\em Phys. Rev.}
  {\bfseries D72} (2005)  043529},
\href{http://arxiv.org/abs/astro-ph/0507263}{{\ttfamily arXiv:astro-ph/0507263
  [astro-ph]}}.

\bibitem{Hamilton:1991es}
A.~J.~S. Hamilton, A.~Matthews, P.~Kumar, and E.~Lu, ``{Reconstructing the
  primordial spectrum of fluctuations of the universe from the observed
  nonlinear clustering of galaxies},''
\href{http://dx.doi.org/10.1086/186057}{{\em Astrophys. J.} {\bfseries 374}
  (1991)  L1}.

\bibitem{PD}
J.~A. Peacock and S.~J. Dodds, ``{Nonlinear evolution of cosmological power
  spectra},'' \href{http://dx.doi.org/10.1093/mnras/280.3.L19}{{\em Mon. Not.
  Roy. Astron. Soc.} {\bfseries 280} (1996)  L19},
\href{http://arxiv.org/abs/astro-ph/9603031}{{\ttfamily arXiv:astro-ph/9603031
  [astro-ph]}}.

\bibitem{Copeland:2006wr}
E.~J. Copeland, M.~Sami, and S.~Tsujikawa, ``{Dynamics of dark energy},''
  \href{http://dx.doi.org/10.1142/S021827180600942X}{{\em Int. J. Mod. Phys.}
  {\bfseries D15} (2006)  1753--1936},
\href{http://arxiv.org/abs/hep-th/0603057}{{\ttfamily arXiv:hep-th/0603057
  [hep-th]}}.

\bibitem{Hannestad:2015tea}
S.~Hannestad, R.~S. Hansen, T.~Tram, and Y.~Y.~Y. Wong, ``{Active-sterile
  neutrino oscillations in the early Universe with full collision terms},''
  \href{http://dx.doi.org/10.1088/1475-7516/2015/08/019}{{\em JCAP} {\bfseries
  1508} (2015) no.~08, 019},
\href{http://arxiv.org/abs/1506.05266}{{\ttfamily arXiv:1506.05266 [hep-ph]}}.

\bibitem{Archidiacono:2015mda}
M.~Archidiacono, T.~Basse, J.~Hamann, S.~Hannestad, G.~Raffelt, and Y.~Y.~Y.
  Wong, ``{Future cosmological sensitivity for hot dark matter axions},''
  \href{http://dx.doi.org/10.1088/1475-7516/2015/05/050}{{\em JCAP} {\bfseries
  1505} (2015) no.~05, 050},
\href{http://arxiv.org/abs/1502.03325}{{\ttfamily arXiv:1502.03325
  [astro-ph.CO]}}.

\bibitem{Abazajian:2012ys}
K.~N. Abazajian {\em et al.}, ``{Light Sterile Neutrinos: A White Paper},''
\href{http://arxiv.org/abs/1204.5379}{{\ttfamily arXiv:1204.5379 [hep-ph]}}.

\end{thebibliography}\endgroup

\end{document}